\documentclass{aastex631}

 \usepackage{subfigure}
\usepackage{appendix}

\newcommand{\hcfn}{H$^{13}$CO$^+$ }

\newcommand{\cfh}{C$_4$H}

\newcommand{\cctht}{$c$-C$_3$H$_2$ }

 \usepackage{ulem}
 
\usepackage{mhchem}
\usepackage{rotating}
\usepackage{booktabs}

\usepackage{multirow}
\usepackage{appendix}
\shorttitle{ C$_4$H 9-8 and  $c$-C$_3$H$_2$ 2-1}
\shortauthors{Yijia Liu et al.}
\graphicspath{{./}{figures/}}

\begin{document}

\title{Spatial distribution of C$_4$H and $c$-C$_3$H$_2$ in cold molecular cores}

\correspondingauthor{Ningyu Tang}
\email{nytang@ahnu.edu.cn}

\author[0009-0001-0943-1195]{Yijia LIu}
\affiliation{Department of Physics, Anhui Normal University, Wuhu, Anhui 241002, People’s Republic of China}

\author[0000-0001-6106-1171]{Junzhi Wang}
\affiliation{Guangxi Key Laboratory for Relativistic Astrophysics, Department of Physics, Guangxi University, Nanning 530004, PR China}

\author[0000-0001-6016-5550]{Shu Liu}
\affiliation{National Astronomical Observatories, Chinese Academy of Sciences, Beijing 100101, People’s Republic of China}

\author[0000-0002-2169-0472]{Ningyu Tang}
\affiliation{Department of Physics, Anhui Normal University, Wuhu, Anhui 241002, People’s Republic of China}

\author[0000-0002-3866-414X]{Yan Gong}
\affiliation{Max-Planck-Institut f\"ur Radioastronomie, Auf dem Hügel 69, 53121, Bonn, Germany}

\author[0000-0002-2243-6038]{Yuqiang Li}
\affiliation{Shanghai Astronomical Observatory, 80 Nandan Road, Shanghai 200030, China}

\author[0000-0003-3520-6191]{Juan Li}
\affiliation{Shanghai Astronomical Observatory, 80 Nandan Road, Shanghai 200030, China}

\author[0009-0001-2414-7221]{Rui Luo}
\affiliation{Guangxi Key Laboratory for Relativistic Astrophysics, Department of Physics, Guangxi University, Nanning 530004, PR China}

\author[0009-0009-6136-0417]{Yani Xu}
\affiliation{Guangxi Key Laboratory for Relativistic Astrophysics, Department of Physics, Guangxi University, Nanning 530004, PR China}






\begin{abstract}

C$_4$H and $c$-C$_3$H$_2$, as unsaturated hydrocarbon molecules, are important for forming large organic molecules in the interstellar medium. We present mapping observations of C$_4$H ($N$=9$-8$) lines,  $c$-C$_3$H$_2$ ($J_{Ka,Kb}$=2$_{1,2}$-1$_{0,1}$) 
 and H$^{13}$CO$^+$ ($J$=1$-0$) 
 toward 19 nearby cold molecular cores in the Milky Way with the IRAM 30m telescope. C$_4$H 9--8 was detected in 13  sources, while $c$-C$_3$H$_2$ was detected in 18 sources. The widely existing C$_4$H and $c$-C$_3$H$_2$ molecules in cold cores provide material to form large organic molecules. Different spatial distributions between C$_4$H 9--8 and $c$-C$_3$H$_2$ 2--1 were found. The relative abundances of these three molecules were obtained under the assumption of local thermodynamic equilibrium conditions with a fixed excitation temperature. The abundance ratio of C$_4$H to $c$-C$_3$H$_2$ ranged from 0.34 $\pm$ 0.09 in G032.93+02 to 4.65 $\pm$ 0.50 in G008.67+22. A weak correlation between C$_4$H/H$^{13}$CO$^+$ and $c$-C$_3$H$_2$/H$^{13}$CO$^+$ abundance ratios was  found, with a correlation coefficient of 0.46, which indicates that there is no tight astrochemical connection between  C$_4$H and $c$-C$_3$H$_2$ molecules.


\end{abstract}

\keywords{Astrochemistry (75) --- Interstellar molecules (849)  --- Chemical abundances (224) --- Abundance ratios(11)
 }


\section{Introduction} \label{sec:intro}

Organic molecules are important  complex molecules in the interstellar medium (ISM). 
 Hydrocarbons, the simplest organic molecules, have been recognized as ubiquitous in the ISM since  the 1970s \citep{1974ApJ...193L.115T,1985ApJ...299L..63T,1987ApJ...323L.149Y}. Small unsaturated hydrocarbon molecules, such as C$_2$H, C$_2$H$_2$, C$_4$H, C$_6$H, $l$-C$_3$H$_2$  and $c$-C$_3$H$_2$, are very important for the formation of large organic molecules. 
 Due to their permanent dipole moments and relatively high abundances, C$_4$H and $c$-C$_3$H$_2$, which can form in cold molecular cores, have been detected in cold dark clouds such as TMC-1  \citep{1981ApJ...248L.113I,1985ApJ...298L..61M,1989AJ.....97.1403M}.

C$_4$H is the simplest example of a polyyne, an organic substance with alternating single- and triple-bond positions in the molecular structure \citep{2015A&A...575A..82C}. It was first detected in the circumstellar envelopes of carbon-rich evolved stars \citep{1978ApJ...224L..27G} and then in  very different environments, eg., cold dark clouds \citep{1981ApJ...248L.113I,1985ApJ...294L..55G} and diffuse medium \citep{1983A&A...127..241B}.
Based on several chemical networks, C$_4$H can be formed by some routes for the C$_n$H family  and  through the reaction of C with $l$-C$_3$H$_2$ \citep{2014MNRAS.437..930L,2023ApJ...944L..45R}. Meanwhile, C$_4$H can  undergo destruction to form  C$_4$H$^-$ through the radiation absorption of an electron \citep{2008ApJ...685..272H,2016JPhB...49t4003G}.  $c$-C$_3$H$_2$  is a cyclic structure  with an isomer,  the linear  $l$-C$_3$H$_2$.  The cyclic isomer $c$-C$_3$H$_2$ exhibits greater stability and higher abundance compared to the  linear counterpart $l$-C$_3$H$_2$.
As the first organic ring in the  ISM, $c$-C$_3$H$_2$ was first detected in Sgr B2 (OH) \citep{1985ApJ...299L..63T}  and subsequently found in other sources \citep{2001ApJ...552..168F,2010ApJ...722.1633S,2012A&A...548A..68P}. 
$c$-C$_3$H$_2$ can be formed through  dissociative recombination of C$_3$H$_3^+$  \citep{1984ApJ...285..618H,1987IJMSI..76..307S,2006A&A...449..631P} and the reaction between $l$-C$_3$H$_2$ and H \citep{2017MNRAS.470.4075L}. In dense molecular clouds, the reaction of $c$-C$_3$H$_2$  with O forms HC$_3$O \citep{2017MNRAS.470.4075L}.

Although hydrogen is much more abundant than carbon in interstellar clouds, the abundance of  C$_4$H  ($\sim  2.8  \times 10^{-9}$) and $c$-C$_3$H$_2$   ($\sim  2.8  \times 10^{-10}$)  compared to molecular hydrogen is relatively high \citep{2021A&A...648A..83Z}, suggesting that interstellar chemistry is not in thermodynamic equilibrium \citep{2008CP....343..292V}. The abundance ratios of C$_4$H to $c$-C$_3$H$_2$ may give us unique information on the physics and chemistry of hydrocarbon molecules.  The distribution of such  small unsaturated hydrocarbon molecules in cold cores is important  for understanding the formation of other organic molecules. However,    there is a paucity of studies about the abundances as well as spatial distributions of C$_4$H and $c$-C$_3$H$_2$  in cold cores. 
 To further understand the relationship between C$_4$H and  $c$-C$_3$H$_2$, mapping surveys that consider    sources with varying C$_4$H  to  $c$-C$_3$H$_2$ abundances are necessary.

As a good dense gas tracer, H$^{13}$CO$^+$ 1--0 is used  to investigate the location of cold cores   {\citep{2018A&A...614A..26A}.   In this work, we investigated  the spatial distributions and relative abundances of C$_4$H, $c$-C$_3$H$_2$, and H$^{13}$CO$^+$  toward the 19 cold cores.
  A comprehensive survey of these cores was conducted to determine the connection between C$_4$H and $c$-C$_3$H$_2$ in these sources. 
 The organization of this paper is as follows.  In Section \ref{sec:obs},  we describe  the mapping observations  and data reduction.  In Section  \ref{sec:results},   we show the results of  our observations.  The discussions are presented in Section \ref{sec:discussion} and a summary is given in Section   \ref{sec:summary}. 

\section{Observation and date reduction}
 \label{sec:obs}


The on-the-fly (OTF) mapping observations  of  a sample of  19 Planck Cold Dust Clumps as cold  dust clumps as cold  molecular cores , which were selected with strong CO 1--0 emission \citep{2012ApJ...756...76W}, were made  with the  IRAM 30m telescope on Pico Veleta,  Spain,  in 2020  September  (project number: 016-20, PI: Shu Liu). The data were taken with the 3 mm (E0) band of the Eight Mixer Receiver   and the fast Fourier transform spectrometers  backend  covering two intermediate frequencies with a 2 GHz bandwidth and 48.8 kHz spectral resolution  in dual polarization. The beam size of the IRAM 30m telescope is $\sim$ 24$^{''}$ at 85\,GHz. The typical system temperatures were around 150 K in the 3 mm band. Pointing was checked every 2 hours with nearby strong quasars. Focus was checked and corrected at the beginning of each run and during sunsets/sunrises. The antenna temperature ($T_{\rm A}^{\ast}$) was converted to the main beam brightness temperature ($T_{\rm mb}$), using $T_{\rm mb}$=$T_{\rm A}^{\ast}\cdot F_{\rm eff}/B_{\rm eff}$, where the forward efficiency $F_{\rm eff}$ is 0.95 and beam efficiency $B_{\rm eff}$ is 0.81 for the 3 mm band. The pixel size of the final regridded maps was  9$''$.


 C$_4$H (9$-8$), $c$-C$_3$H$_2$ ($J_{Ka,Kb}$=2$_{1,2}$-1$_{0,1}$) (hereafter $c$-C$_3$H$_2$ 2--1)  and H$^{13}$CO$^+$ ($J$=1$-0$) (hereafter H$^{13}$CO$^+$ 1--0)  lines were included  in the observations.  Detailed information on the physical parameters of the molecular lines was obtained from  the Cologne Database for Molecular Spectroscopy  ({\sc CDMS})\footnote{https://cdms.astro.uni-koeln.de/classic} \citep{2005JMoSt.742..215M} and is listed in  Table \ref{table:Physical parameters}.  All data processing was conducted for the regridded maps  using the {\sc CLASS} package, which is a part of the {\sc GILDAS}\footnote{http://www.iram.fr/IRAMFR/GILDAS} software. For each line observed for each source, we first took a quick look at the velocity ranges in the spectral emission region. These velocity ranges  were used as ‘mask' with  ‘set window' in CLASS  when the first-order baseline was removed. Then we used  ‘print area' in  {\sc CLASS}  to get the velocity-integrated fluxes for each  pixel, which provided spatial distribution maps for each line.

Since  C$_4$H 9--8 ($J$=19/2$-17/2$)  is strongly blended  with  C$_4$H 9--8 ($J$=19/2$-19/2$),  which cannot be separated due to line broadening, both of them are marked as  C$_4$H 9--8 ($J$=19/2$-17/2$) later in this paper. There similar cases are for C$_4$H 9--8 ($J$=17/2$-15/2$)   and C$_4$H 9--8 ($J$=17/2$-17/2$).  The information of the observations for each source is shown in Table \ref{table:source}.

\section{Results}
 \label{sec:results}

With the mapping observations of 19 cold cores, H$^{13}$CO$^+$ 1--0 was detected in all sources. $c$-C$_3$H$_2$ 2--1  emission was detected in 18 sources, except for G008.52+21. The emission of C$_4$H 9--8 ($J$=19/2$-17/2$) and C$_4$H 9--8 ($J$=17/2$-15/2$) was detected in 13 sources while remaining undetected in G003.73+16, G006.32+20, G006.41+20, G008.52+21, G031.44+04,  and G032.93+02  (see Table \ref{table:distribution}).

\subsection{Spatial distribution of  C$_4$H, $c$-C$_3$H$_2$ and H$^{13}$CO$^+$ lines}
\label{sec:spatial distribution}


Considering the comparable weakness of the individual signals of C$_4$H 9-8 (J=19/2--17/2) and C$_4$H 9-8 (J=17/2--15/2), along with the nearly identical Einstein coefficients and upper-level energies,  combining them into one total velocity-integrated intensity is a reasonable approach, which offers a higher fidelity for analyzing the C$_4$H distribution compared with one single line.    The averaged value of the two C$_4$H 9--8 lines will be used  for the discussion of the spatial distribution of C$_4$H. 

The velocity-integrated intensity maps of C$_4$H 9--8 (magenta contours), $c$-C$_3$H$_2$ 2--1 (gray scale and blue contours) and H$^{13}$CO$^+$ 1--0 (green contours)   of six sources are shown in   Figure  \ref{map-1} (G001.38+20, G001.84+16,  G006.04+36, G021.20+04,  and G021.66+03) and Figure \ref{map-2} (G032.93+22), while other sources are displayed in   Figure \ref{appendix}.   Two velocity components with different  spatial distributions derived from a  velocity-integrated  map  were found in G032.93+22. The two velocity ranges were  from 10.8 to 12.2 km s$^{-1}$ and 12.2 to 13.4 km s$^{-1}$, respectively, for deriving velocity-integrated maps (see Figure \ref{map-2}).


The spatial distribution information of C$_4$H 9--8, $c$-C$_3$H$_2$ 2--1 and H$^{13}$CO$^+$ 1--0 for each source is listed in Table \ref{table:distribution}. C$_4$H 9--8 spatial distributions could  be clearly found in 10 out of 13 sources. However, our current sensitivity and angular resolution are insufficient to unveil the spatial distributions of C$_4$H 9--8 in the remaining three sources (G001.84+16,  G007.14+05,  and G021.20+04). $c$-C$_3$H$_2$ 2--1 was only marginally detected in  G006.32+20 and G006.41+20,  while  clear  spatial distribution could be found in the  other 16 sources. 
H$^{13}$CO$^+$ 1--0 emission could be resolved  in all 19 sources.

Among the 13 sources with detection of C$_4$H 9--8,  different spatial distribution  between C$_4$H 9--8 and $c$-C$_3$H$_2$ 2--1 could  be found in 4 sources (G001.84+16, G007.14+05, G021.20+04, and G021.66+03). Even though  C$_4$H 9--8 in G007.14+05 and G021.20+04  could  not be resolved mainly due to weak emission,  a clear spatial offset between the  peak  of C$_4$H 9--8 and  $c$-C$_3$H$_2$ 2--1 could  be found.  Different spatial distributions  between C$_4$H 9--8 and $c$-C$_3$H$_2$ 2--1 could  not be found in the  other 9 sources  (G001.38+20,   G006.04+36,  G008.67+22,   G025.48+06,  G026.85+06,  G028.45-06,  G028.71+03,   G030.78+05,  and G058.16+03).


Notes for individual sources with useful spatial information are presented below:

$\bf G001.38+20$:  A 220$''$  $\times$  220$''$ map of C$_4$H 9--8, $c$-C$_3$H$_2$ 2--1 and H$^{13}$CO$^+$ 1--0 was obtained. Strong emissions of $c$-C$_3$H$_2$ 2--1 and H$^{13}$CO$^+$ 1--0 were detected with similar spatial distribution, which shows a banded structure  extended over 200$''$ in the east-to-west direction. The three lines peak at the same position approximately at (-90$''$, 10$''$).

$\bf G001.84+16$:  The OTF mode was used to cover 120$''$ $\times$ 120$''$ for C$_4$H 9--8, $c$-C$_3$H$_2$ 2--1 and H$^{13}$CO$^+$ 1--0. Both $c$-C$_3$H$_2$ 2--1 and H$^{13}$CO$^+$ 1--0  were detected and were seen with  significantly different spatial distributions. $c$-C$_3$H$_2$ 2--1 distributes in a north-to-south direction at the east of the mapping area, extending over 120$''$, while H$^{13}$CO$^+$ 1--0  shows a northwest-to-southeast direction distribution. Weak C$_4$H 9--8 emission is mainly located at the northeast of the mapping area.

$\bf  G006.04+36$:   The mapping size of C$_4$H 9--8, $c$-C$_3$H$_2$ 2--1 and H$^{13}$CO$^+$ 1--0 is 240$''$ $\times$ 240$''$.   Two strong $c$-C$_3$H$_2$ 2--1 components were evident, with one centered at about  (-50$''$, -50$''$) and the other at  (0$''$, 50$''$).  In contrast, two H$^{13}$CO$^+$ 1--0 components were observed, one predominantly centered around (-150 $''$, -130$''$) without $c$-C$_3$H$_2$ 2--1 emission, and the other centered at about (-50$''$, -80$''$) in close proximity to $c$-C$_3$H$_2$ 2--1 emission.
 No H$^{13}$CO$^+$ 1--0 was detected near the component centered at about (0$''$, 50$''$), where both $c$-C$_3$H$_2$ 2--1 and C$_4$H 9--8 were detected. The three components present different properties based on the emission of these three lines.


$\bf  G021.20+04$:  The mapping size of C$_4$H 9--8,  $c$-C$_3$H$_2$ 2--1 and H$^{13}$CO$^+$ 1--0 is 220$''$ $\times$ 100$''$.  All three lines were detected, showing different spatial distributions.  Strong $c$-C$_3$H$_2$ 2--1 and H$^{13}$CO$^+$ 1--0 emissions exhibit a north-to-south distribution centered at about (30$''$,0$''$). The emission peaks of all three lines are different.  The peak coordinates of $c$-C$_3$H$_2$ 2--1 and H$^{13}$CO$^+$ 1--0 are about (30$''$, 0$''$).  C$_4$H 9--8 shows two peaks in the northeast and southwest directions of $c$-C$_3$H$_2$ 2--1  approximately at (40$''$, 25$''$) and  (25$''$, --20$''$).

$\bf G021.66+03$:  A  220$''$ $\times$ 220$''$ size map of C$_4$H 9--8, $c$-C$_3$H$_2$ 2--1 and H$^{13}$CO$^+$ 1--0 was obtained. No clear difference in the spatial distributions  between  $c$-C$_3$H$_2$ 2--1 and H$^{13}$CO$^+$ 1--0 was found,  with a semielliptical distribution  in the north-to-south direction. The $c$-C$_3$H$_2$ 2--1 and H$^{13}$CO$^+$ 1--0 emission peak is at about (-5$''$, 45$''$). The C$_4$H 9--8 emission presents a slightly different spatial distribution from  that of $c$-C$_3$H$_2$ 2--1 and H$^{13}$CO$^+$ 1--0.

$\bf G032.93+22$:  Two velocity components with different spatial distributions were found for both $c$-C$_3$H$_2$ 2--1 and H$^{13}$CO$^+$ 1--0 (see Figure 1),  while C$_4$H 9--8 was not detected in the velocity-integrated map. The spatial distribution of $c$-C$_3$H$_2$ 2--1 is similar to that of H$^{13}$CO$^+$ 1--0 for each velocity component.  After  being spatially averaged with several pixels,  C$_4$H 9--8 was detected for both velocity components (see Figure \ref{map-2}).



 \subsection{Column densities and relative abundances  } 
\label{sec:column densities and relative abundance } 

Based on the velocity-integrated maps of C$_4$H 9--8, $c$-C$_3$H$_2$ 2--1 and H$^{13}$CO$^+$ 1--0,  there was more than one core in  some sources, so 25 regions from the 19 mapping sources were  selected to  calculate the relative abundance of the three molecules.  The  spatially averaged spectra of these three lines were used to derive velocity-integrated intensities with a single-component  Gaussian fitting. The regions used for obtaining the spatially averaged spectra are marked with yellow  boxes, while black boxes are also used if the  second core is used in one map.

Both $c$-C$_3$H$_2$ 2--1 and H$^{13}$CO$^+$ 1--0 were  detected in  25 regions. C$_4$H 9--8 ($J$=19/2$-17/2$) and C$_4$H 9--8 ($J$=17/2$-15/2$) emission was  detected in 21 regions except for G006.32+20 (yellow), G006.41+20 (yellow), G008.52+21 (yellow), and G032.93+02 (black). Detailed information on the spectra obtained from the 25 regions is shown in Table \ref{table:Observed date}, including source names, molecular line names, velocity-integrated intensities, FWHM and peak temperatures. The strongest emissions of C$_4$H 9--8 ($J$=19/2$-17/2$) and C$_4$H 9--8 ($J$=17/2$-15/2$) were observed in G028.71+03 (yellow) with integrated intensities of  0.93 $\pm$ 0.03  and 0.88 $\pm$ 0.02 K km\,s$^{-1}$,   respectively.  Meanwhile,  the strongest emissions of $c$-C$_3$H$_2$ 2--1 and H$^{13}$CO$^+$ 1--0 were identified in G030.78+05 (yellow) with integrated intensities of 2.75 $\pm$ 0.02 and 2.12 $\pm$ 0.03 K km\,s$^{-1}$,  respectively.


For sources where  C$_4$H 9--8  was not detected, 3$\sigma$  upper limits for $\int$$T_{\rm mb}\rm dv$  were calculated:
\begin{equation}
	\int  T_{\rm mb}dv=3 rms \sqrt{\delta v \cdot \Delta v}~ (\rm K~km\; s^{-1}),
\end{equation}
    where $\delta v$ is the velocity resolution,  $\Delta v$ is the   line width in km s$^{-1}$, and $rms$  is the root mean square value per channel of the spectrum.

Even though  local thermodynamic equilibrium (LTE)  is hard to be reached for molecules, such as C$_4$H and \ce{c-C_3H_2},   with dipole moment higher than that of CO, LTE assumption is still a reasonable  approximation for deriving column density with only one transition of each molecule.  Assuming LTE and that all three lines are optically thin, the formula for calculating the column density of these three molecules can be given by 
\begin{equation}
	N_{\rm tot}=\frac{8\pi k\nu^2}{hc^3A_{ul}}\frac{Q(T_{\rm ex})}{\rm g_u}e^{E_u/kT_{\rm ex}}\int T_{\rm mb}\rm dv (\rm cm^{-2})
\end{equation} 
where  $k$ is the Boltzmann constant, $\nu$ is the frequency of the molecular spectral line transition, $h$ is the Planck constant, $c$ is the speed of light, $A_{\rm ul}$ is the Einstein emission coefficient, $\rm g_u$ is the upper-level degeneracy, and $E_u$ is the energy of the upper level above the ground state. The frequency $\nu$, $\rm g_u$, $A_{\rm ul}$, and  $E_u$  for six emission lines (C$_4$H 9--8 ($J$=19/2$-17/2$), C$_4$H 9--8 ($J$=19/2$-19/2$), C$_4$H 9--8 ($J$=17/2$-15/2$), C$_4$H 9--8 ($J$=17/2$-17/2$), $c$-C$_3$H$_2$ ($J_{Ka,Kb}$=2$_{1,2}$-1$_{0,1}$), H$^{13}$CO$^+$ ($J$=1$-0$)) were taken from the  CDMS. $Q$($T_{\rm ex}$) is the partition function, which is depends on the excitation temperature $T_{\rm ex}$. Since all the sources  were selected from cold cores, the excitation temperature $T_{\rm ex}$ = 9.375 K was taken for the calculations in this work. Detailed information is shown in Table \ref{table:Physical parameters} from the CDMS.

The frequency difference among C$_4$H 9--8 ($J$=19/2$-17/2$) and C$_4$H 9--8 ($J$=19/2$-19/2$) as well as C$_4$H 9--8 ($J$=17/2$-15/2$) and C$_4$H 9--8 ($J$=17/2$-17/2$) is very small, which leads to the blending of these lines. We calculate the column densities of C$_4$H 
with

\begin{equation}
\label{lias:3}
	N_{\rm tot}=\frac{8\pi k\nu^2}{hc^3}\frac{Q(T_{\rm ex})}{A_{ul_1}\rm g_{u_1}+A_{ul_2}\rm g_{u_2}}e^{E_u/kT_{\rm ex}}\int T_{\rm mb}\rm dv (\rm cm^{-2})
\end{equation} 
where  $A_{ul_1}$, $g_{u_1}$,  and $A_{ul_2}$, $g_{u_2}$ are the values corresponding to the two blended lines, respectively. Since the difference in  $E_u$ between the two blended molecular lines is also very small, we take the  $E_u$  values of C$_4$H 9--8 ($J$=19/2$-17/2$) and C$_4$H 9--8 ($J$=17/2$-15/2$). 

 The relative abundances of C$_4$H and $c$-C$_3$H$_2$ to H$^{13}$CO$^+$, and the column densities of the three molecules are presented in Table \ref{table_Colunm density}. In this table, columns (2)--(4) present the derived column densities of the three molecules, while Column (5)--(7) display their relative abundances.

 The derived column densities of the three molecules are  in Columns (2)-(4) and the  relative abundances in Columns (5)-(7). The abundance ratio of C$_4$H to $c$-C$_3$H$_2$ ranges from 0.34 $\pm$ 0.09 in G032.93+02 (yellow) to 4.65 $\pm$ 0.53 in G008.67+22 (yellow) (see Table \ref{table_Colunm density}).

\section{Discussion} \label{sec:discussion}

With similar Einstein A coefficients and upper-level energies (see Table \ref{table:Physical parameters}), $c$-C$_3$H$_2$ 2--1 and H$^{13}$CO$^+$ 1--0 do represent the abundances of $c$-C$_3$H$_2$ and H$^{13}$CO$^+$ molecules instead of the excitation conditions caused by volume densities and the  kinetic temperature of molecular hydrogen. On the other hand, given a higher $E_u$$\sim$20.5 K than that of $c$-C$_3$H$_2$ 2--1 ($\sim$6.4 K) and H$^{13}$CO$^+$ 1--0 ($\sim$4.2 K), the excitation of C$_4$H 9--8 demands a high kinetic temperature.   Significantly different spatial distributions between  $c$-C$_3$H$_2$ 2--1 and H$^{13}$CO$^+$ 1--0 are seen in several sources (G001.84+16,  G006.04+36,  G006.32+20, G006.41+20,  and  G008.52+21),  while no clear difference can be found in other sources.  Such differences indicate that  the enhancement of $c$-C$_3$H$_2$ can exist even in the cold cores. On the other hand, the absence of C$_4$H in some cores can be explained by their low kinetic temperatures of molecular gas.

The detected abundances of C$_4$H and $c$-C$_3$H$_2$ relative to those  of H$^{13}$CO$^+$ in the 25 regions are plotted in Figure \ref{C4H_C3H2-2}. Only upper limits of C$_4$H column densities are obtained in four regions, while $c$-C$_3$H$_2$ and H$^{13}$CO$^+$ column densities are obtained in all 25 regions.    The detection of  C$_4$H and C$_3$H$_2$ in different cores may vary due to the diverse chemical reaction pathways associated with these two molecules, as well as variations in excitation temperatures among different sources.   The abundance ratio of C$_4$H/H$^{13}$CO$^+$ varies from about 3 to 50, while it is about 2 to 18 for the ratio of $c$-C$_3$H$_2$/H$^{13}$CO$^+$. Even though C$_4$H 9--8 lines are normally weaker than those of $c$-C$_3$H$_2$ 2--1 in these regions,  the relative abundances of C$_4$H are comparable or slightly higher than those of $c$-C$_3$H$_2$ in most of them (see Figure \ref{C4H_C3H2-3}), which is caused by the higher energy levels of C$_4$H than those of $c$-C$_3$H$_2$. Higher abundance of C$_4$H than that of $c$-C$_3$H$_2$ were also seen in the  ``starless'' core  L1521F  \citep{2011ApJ...728..101T}, Orion Bar PDR  \citep{2015A&A...575A..82C}, and four low-mass molecular outflow sources \citep{2021A&A...648A..83Z}. 
Different excitation conditions for molecules with different dipole moments and  critical densities can cause uncertainties in estimating column densities and relative abundances with only one transition for each molecule.  If the excitation temperature of $c$-C$_3$H$_2$ is lower than that of C$_4$H, it will result in a decrement  in  column density of  $c$-C$_3$H$_2$ and an increment in the abundance ratio of C$_4$H to $c$-C$_3$H$_2$. 
Further  observations of multiple transitions will  help us to better determine the abundance of each molecule. For example, combination of C$_4$H 5--4 lines with existing 9--8  lines, and  combination of   $c$-C$_3$H$_2$ 3--2 with   2--1,   can be used to derive a more accurate  relative abundance of $c$-C$_3$H$_2$ and  C$_4$H. Improving the signal-to-noise ratio of such a  weak emission can also be necessary to derive an accurate  abundance ratio.

A very weak trend of increment of  C$_4$H and/or H$^{13}$CO$^+$  with an  increasing $c$-C$_3$H$_2$/H$^{13}$CO$^+$ abundance ratio is observed. The  result is $lg$(C$_4$H/H$^{13}$CO$^+$)=3.35$lg$($c$-C$_3$H$_2$/H$^{13}$CO$^+$)-2.20, with a correlation coefficient of 0.46, using least-square fitting (see Figure \ref{C4H_C3H2-2}). The four upper limits are not used for the fitting. Even though the detected C$_4$H 9--8 emissions are mainly associated with $c$-C$_3$H$_2$ 2--1 emission based on the mapping results, it seems that there is no strong astrochemical connection between C$_4$H and $c$-C$_3$H$_2$ molecules in cold cores. This is supported by  the observed large variation in the C$_4$H/$c$-C$_3$H$_2$ abundance ratio and the astrochemical modeling results \citep{2008ApJ...685..272H,2017MNRAS.470.4075L}.
Based on the Kinetic Database for Astrochemistry network, C$_4$H was thought to be formed through some routes for the C$_n$H family from n=2 to n=10  \citep{2023ApJ...944L..45R}. C$_4$H can also be formed through the reaction of C with $l$-C$_{3}$H$_2$ \citep{2014MNRAS.437..930L}, which is less abundant than $c$-C$_{3}$H$_2$ in molecular clouds. On the other hand,  C$_4$H can be destroyed by  the radiative absorption of an electron to form C$_4$H$^-$ \citep{2008ApJ...685..272H,2016JPhB...49t4003G}. $c$-C$_3$H$_2$ was thought to be formed by the  dissociative recombination of C$_3$H$_3^+$  \citep{1984ApJ...285..618H,1987IJMSI..76..307S,2006A&A...449..631P}, with $c$-C$_3$H$_3^+$+$e^-$ to produce $c$-C$_3$H$_2$ and H. Another pathway for the formation of $c$-C$_3$H$_2$ involves the reaction between $l$-C$_3$H$_2$ and H \citep{2017MNRAS.470.4075L}, which can result in a weak astrochemical connection between $c$-C$_3$H$_2$ and C$_4$H.  The efficient destruction pathway for $c$-C$_3$H$_2$ involves the reaction with O, producing HC$_3$O in dense molecular clouds \citep{2017MNRAS.470.4075L}. 
The comparable abundances of C$_4$H and $c$-C$_3$H$_2$  found in these sources (see Table \ref{table_Colunm density}), should be considered in further  astrochemical  network pathways for these two molecules in  cold molecular cores. However, due to the limited observational results of C$_4$H and $c$-C$_3$H$_2$ in varying  conditions of the ISM, it is hard to make a conclusion about  the dominant chemical network of  these two molecules.
The widely existing $c$-C$_3$H$_2$ and C$_4$H in cold molecular cores provide the building blocks for the formation of large organic molecules in the ISM. 
Further high-resolution  observations  with millimeter interferometers  will be  useful  to obtain a more comprehensive understanding of C$_4$H and $c$-C$_3$H$_2$ chemistry in molecular clouds.


 \section{summary}
  \label{sec:summary}

With the mapping observations of 19 cold molecular cores for C$_4$H (N=9-8 J=19/2-17/2, N=9-8  J=19/2-19/2, N=9-8  J=17/2--15/2, N=9-8 J=17/2--17/2), $c$-C$_3$H$_2$ ($J_{Ka,Kb}$=2$_{1,2}$--1$_{0,1}$) and H$^{13}$CO$^+$ (J=1--0), using the IRAM 30m telescope, the main results are:

1.  H$^{13}$CO$^+$ 1--0 emission was detected in all 19 sources, while $c$-C$_3$H$_2$ 2--1 was detected in 18 sources and C$_4$H 9--8 was detected in 13 sources. The high detection rate of $c$-C$_3$H$_2$ and C$_4$H lines indicates that such unsaturated  hydrocarbon molecules can be formed in cold core or even earlier phases.

2. The emission of $c$-C$_3$H$_2$ 2--1 and H$^{13}$CO$^+$ 1--0, with similar Einstein A coefficients and upper-level energies, in some sources presents different  spatial distributions, which implies that $c$-C$_3$H$_2$ has a distinct chemistry compared to H$^{13}$CO$^+$.

3. The abundance ratio of C$_4$H to $c$-C$_3$H$_2$ spans a wide range  from 0.34 $\pm$ 0.09 in G032.93+02  to 4.65 $\pm$ 0.50 in G008.67+22.

4. The maximum abundance ratio of C$_4$H to H$^{13}$CO$^+$ is 51.48 $\pm$ 2.90 in G028.71+03, and the maximum abundance ratio of $c$-C$_3$H$_2$ to H$^{13}$CO$^+$ is 16.48 $\pm$ 3.50 in G007.14+05.

5. Only a very weak trend between the C$_4$H/H$^{13}$CO$^+$ and $c$-C$_3$H$_2$/H$^{13}$CO$^+$ abundance ratio  was  found,  which is consistent with the astrochemical networks of C$_4$H and $c$-C$_3$H$_2$ in which there is no close connection between these two molecules.

\begin{acknowledgements}

This work has been supported by the University Annual Scientific Research Plan of Anhui Province (No. 2023AH030052, No. 2022AH010013), Zhejiang Lab Open Research Project (No. K2022PE0AB01), the National Natural Science Foundation of China grants 11988101,  the China Manned Space Program through its Space Application System,  the science research    grants from the China Manned Space Project with numbers CMS-CSST grant and Cultivation Project for FAST Scientific Payoff and Research Achievement of CAMS-CAS. This study is based on observations carried out under project 016-12 with the IRAM 30m telescope. IRAM is supported by INSU/CNRS (France), MPG (Germany) and IGN (Spain).

 \end{acknowledgements}

%




\begin{figure*}
\centering

\centering 
\subfigure[]{ \label{fig1:a} 
\includegraphics[height=2.30in,width=0.4\columnwidth]{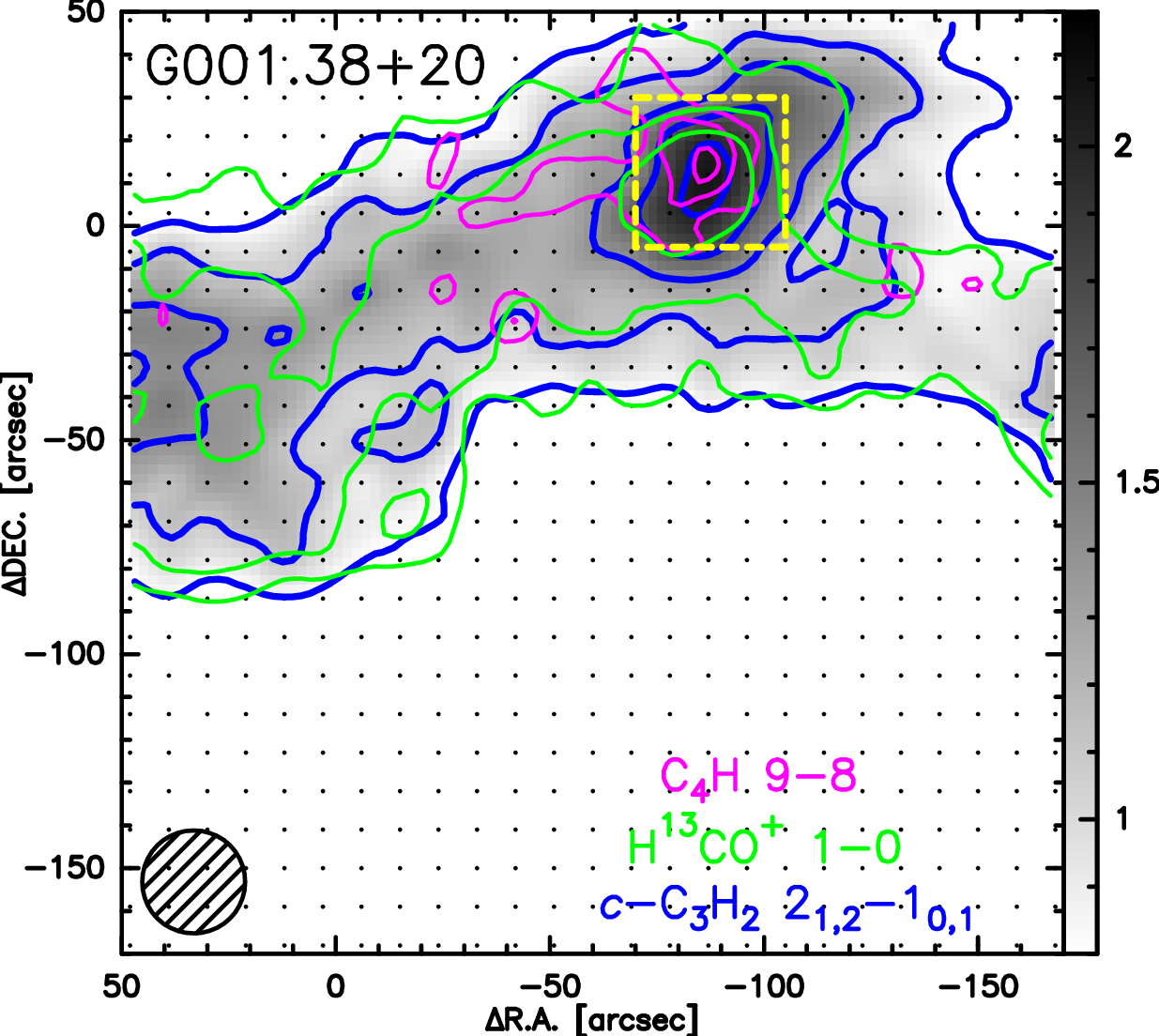} 
} 
\subfigure[]{ \label{fig1:b} 
\includegraphics[height=2.3in,width=0.4\columnwidth]{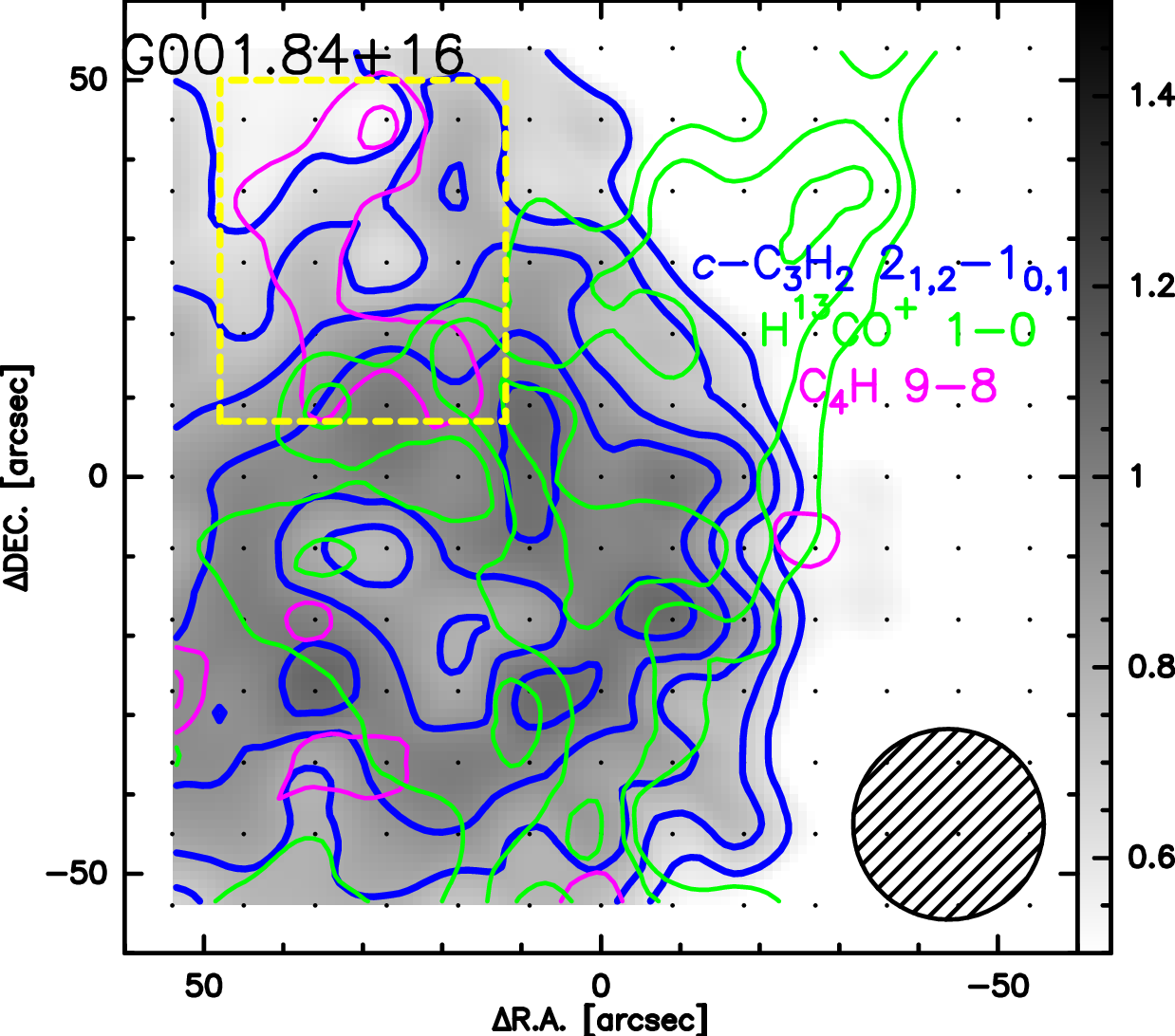} 
} 

\subfigure[] { \label{fig1:c}
\includegraphics[height=1.8in,width=0.4\columnwidth]{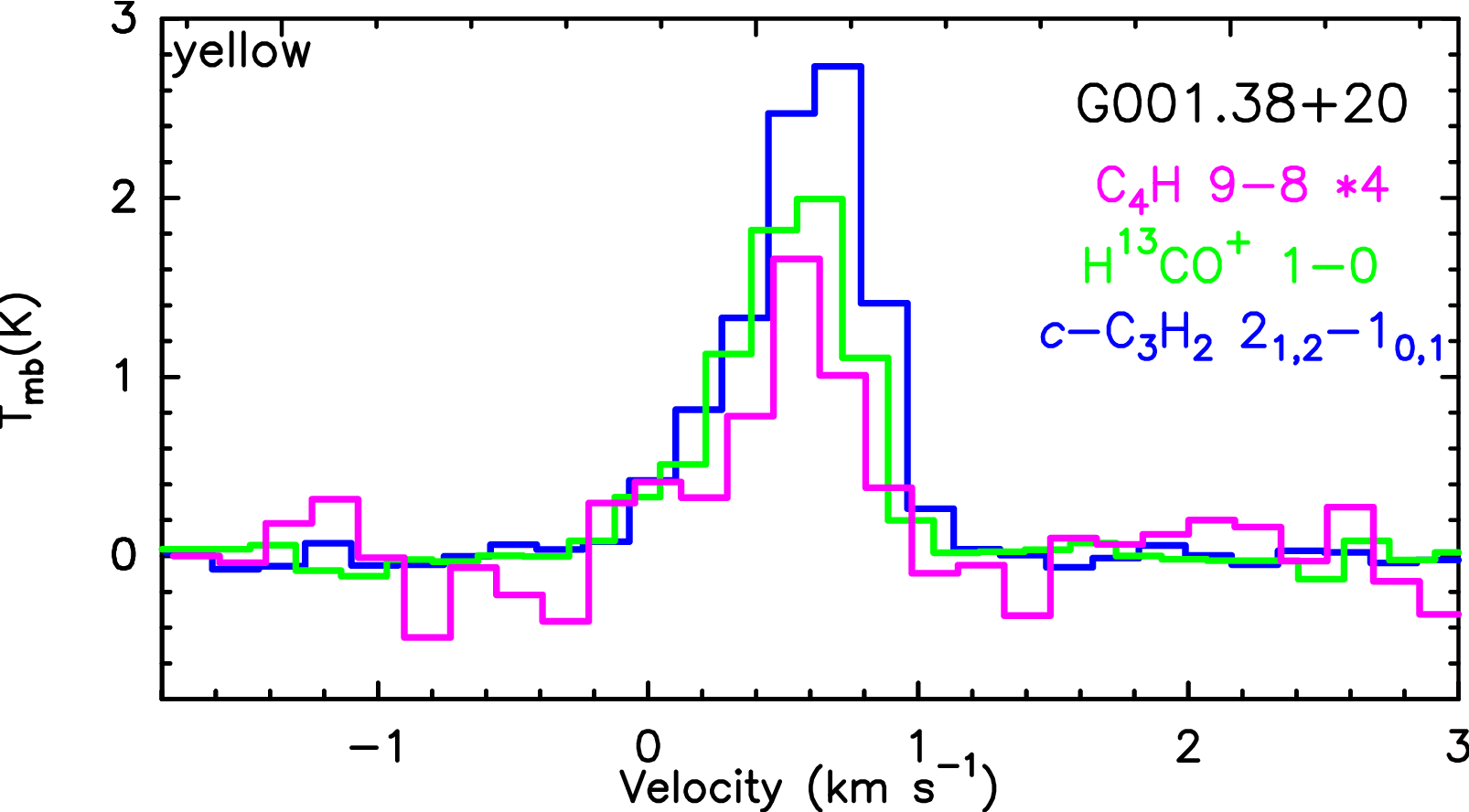} 
} 
\subfigure[]{ \label{fig1:d} 
\includegraphics[height=1.8in,width=0.4\columnwidth]{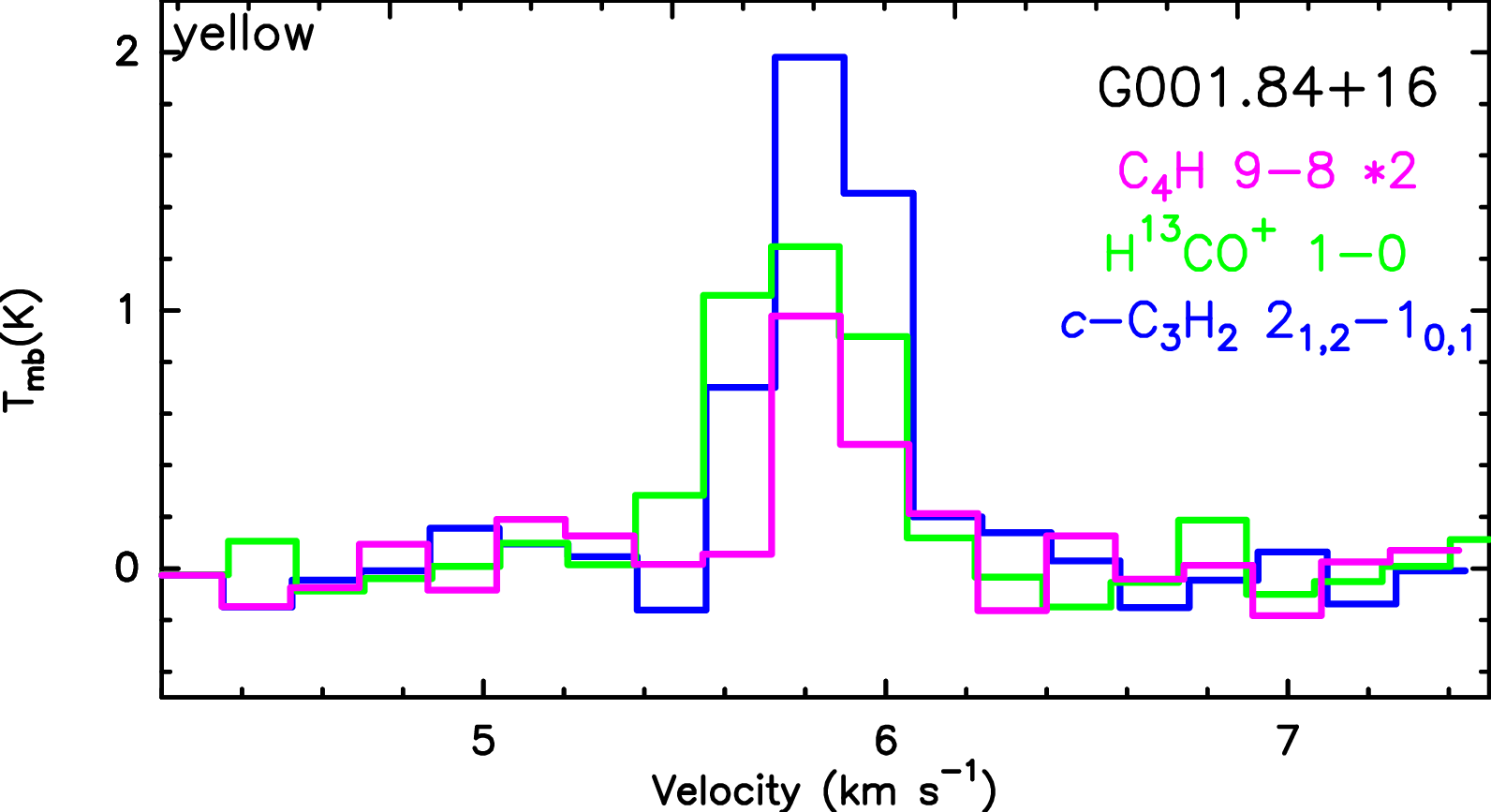} 
}

\caption{The velocity integrated maps and spatial averaged spectra of C$_4$H 9--8, $c$-C$_3$H$_2$ 2--1 and H$^{13}$CO$^+$ 1--0. The source names are presented in the maps and spectra. The gray scale color at the right is in units of K km s$^{-1}$.  
(a): The velocity integrated intensity maps for C$_4$H 9--8 (magenta contours), $c$-C$_3$H$_2$ 2--1 (blue contours) and H$^{13}$CO$^+$ 1--0 (green contours) of G001.38+20. 
(b): The same as  Figure \ref{fig1:a}  but for  G001.84+16. 
(c): Spectra of C$_4$H at 85672.5793 MHz, $c$-C$_3$H$_2$ at 85338.8940 MHz and H$^{13}$CO$^+$ at 86754.2884 MHz in the yellow box of G001.38+20. 
(d): The same as  Figure \ref{fig1:c}   but for  G001.84+16. The  detailed mapping information of all sources for C$_4$H 9--8, $c$-C$_3$H$_2$ 2--1 and H$^{13}$CO$^+$ 1--0  are listed in Table \ref{table:step}. }

 \label{map-1}	 
 \end{figure*}

\clearpage
\addtocounter{figure}{-1}
\centering 

\begin{figure*}
\gridline{
           \fig{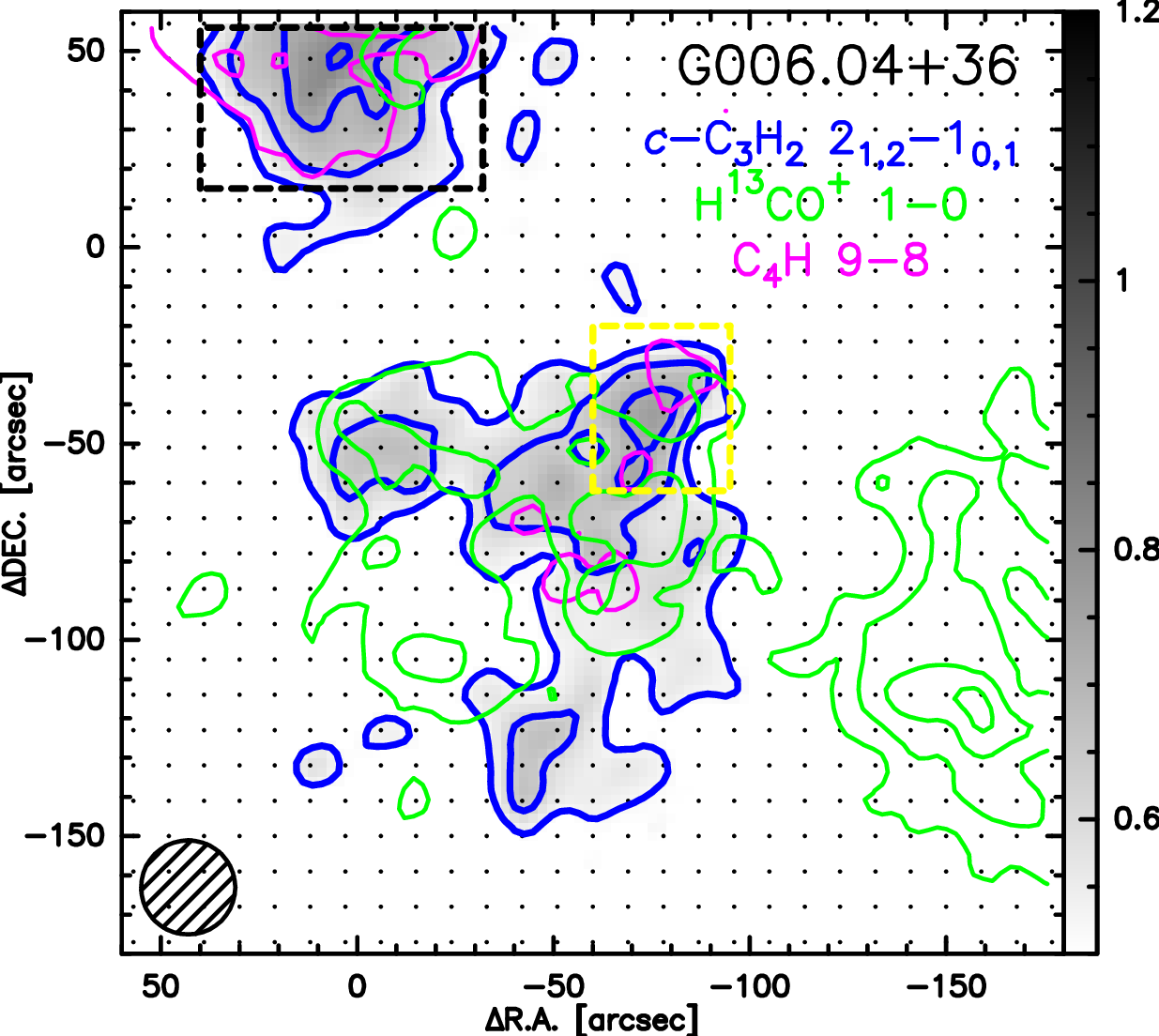}{0.45\textwidth}{(e)}
           }
\gridline{          
          \fig{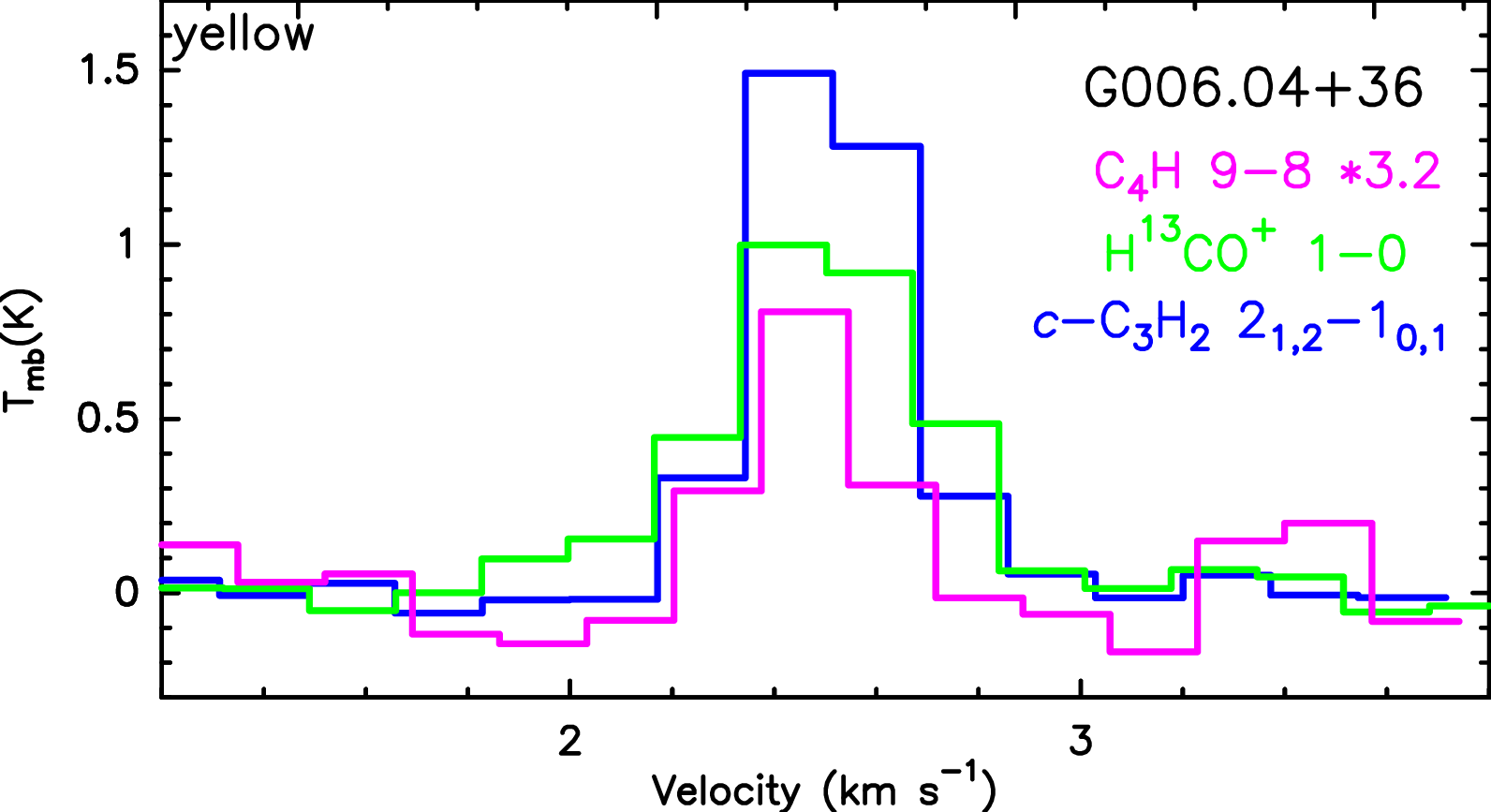}{0.4\textwidth}{(f)}
          \fig{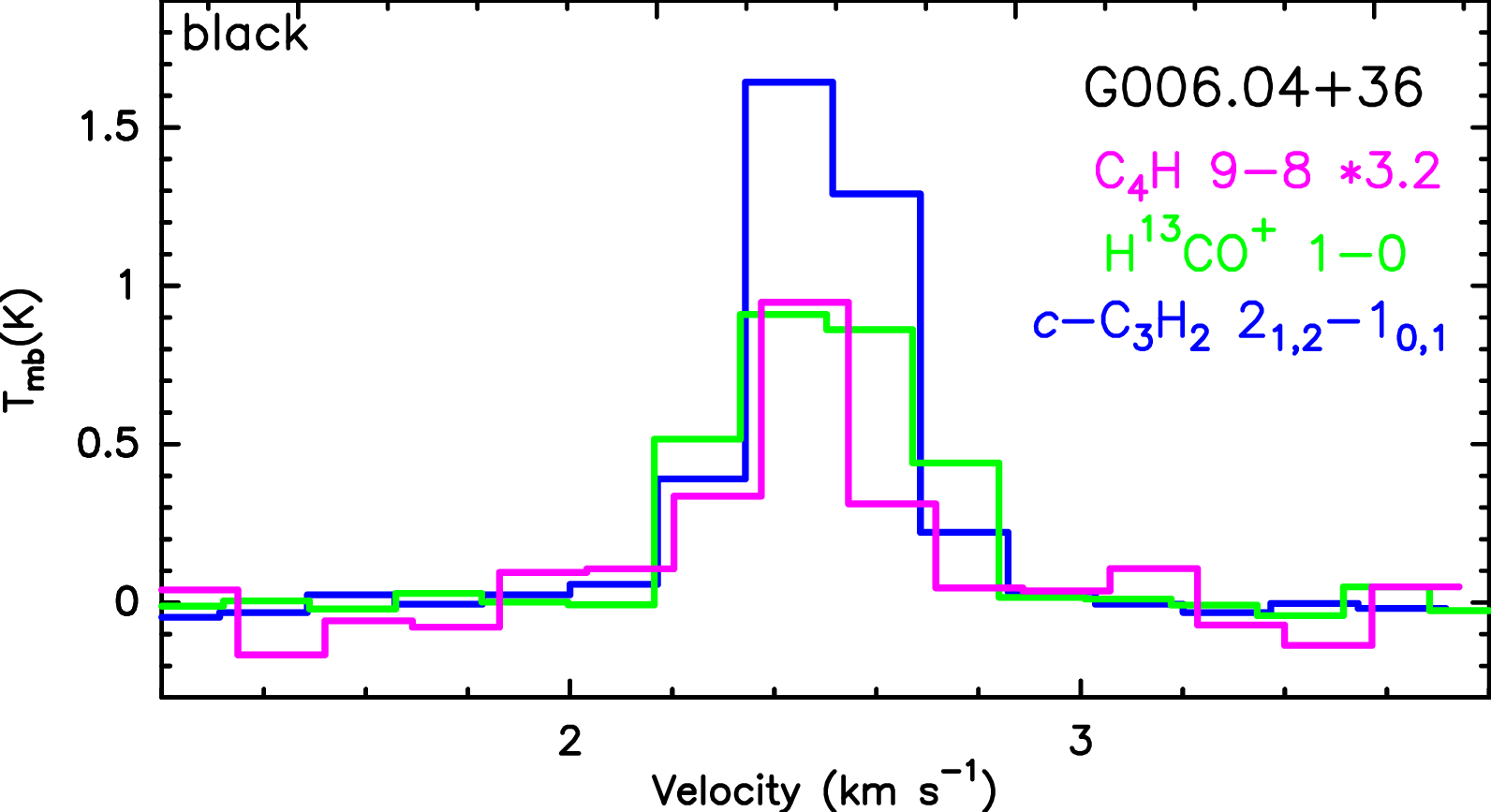}{0.4\textwidth}{(g)}
          }
          
gridline{
           \fig{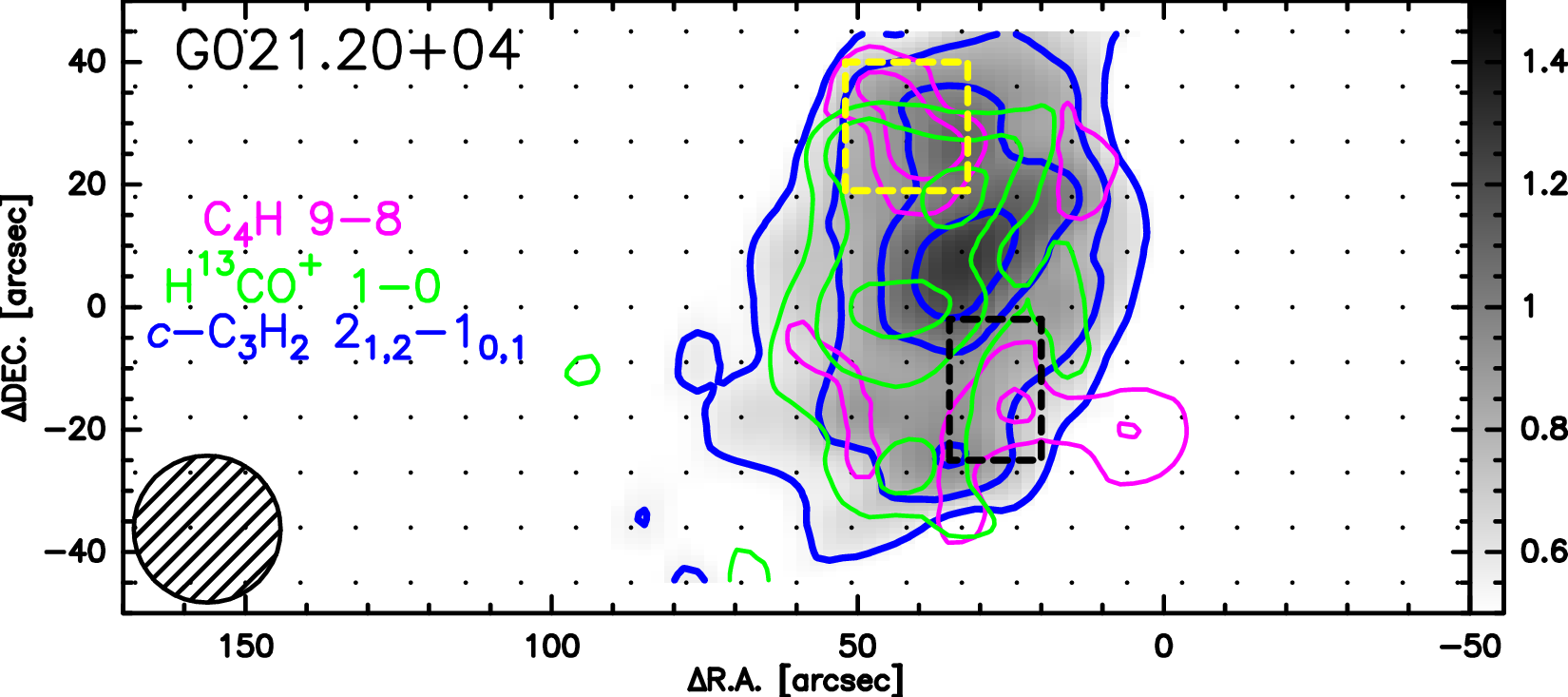}{0.55\textwidth}{(h)}
           }
\gridline{          
          \fig{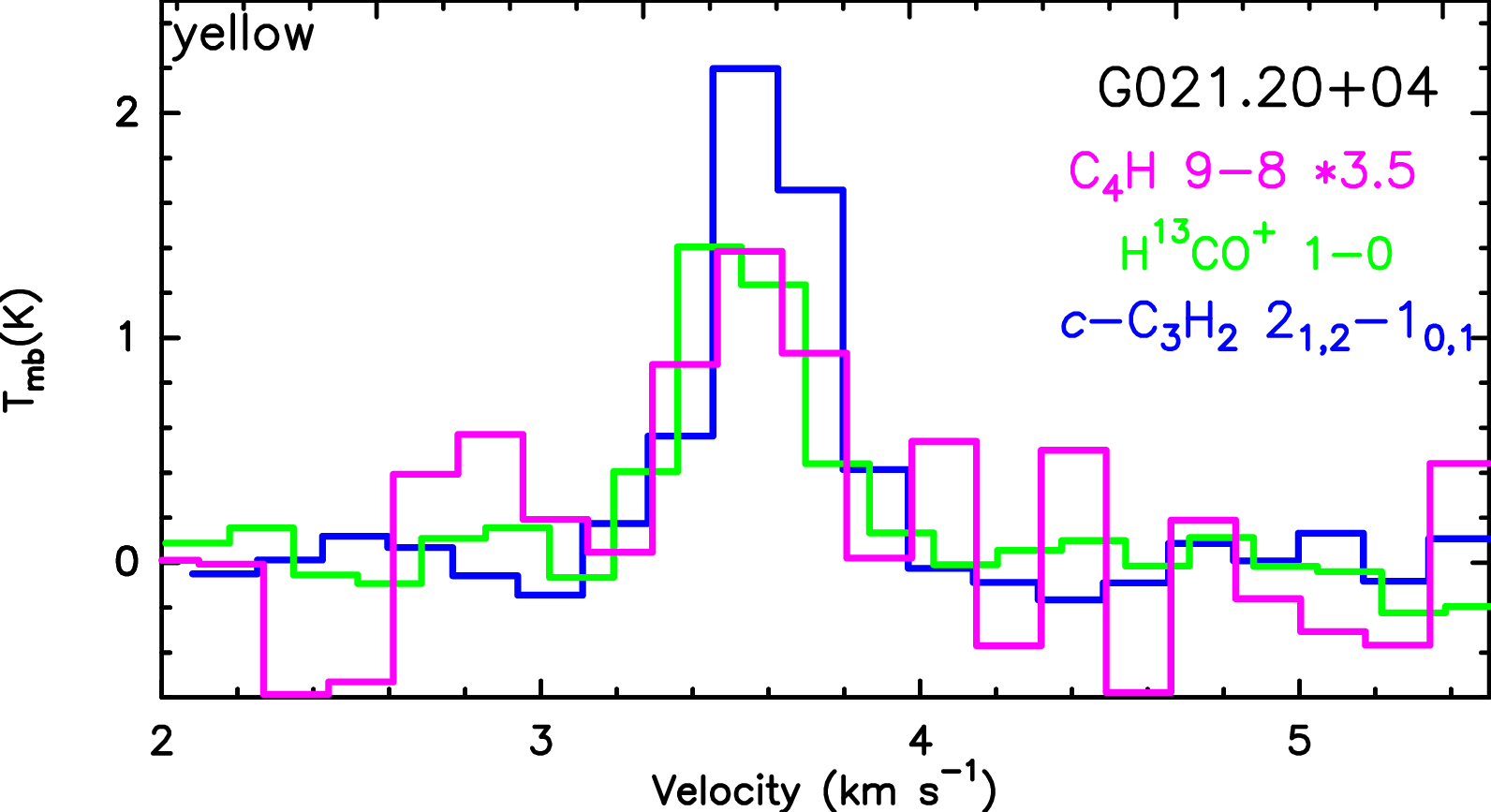}{0.4\textwidth}{(i)}
          \fig{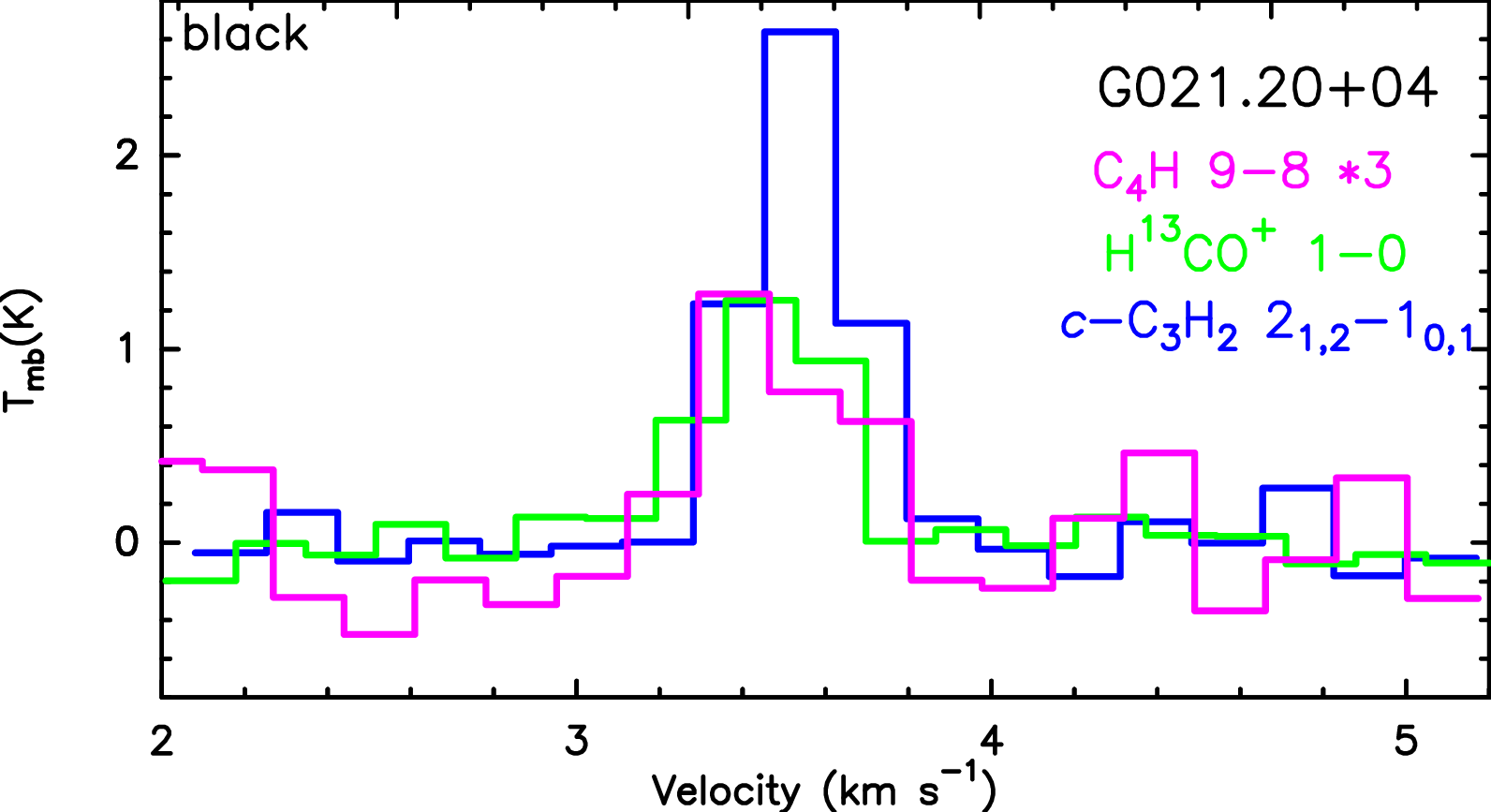}{0.4\textwidth}{(j)}
          }

\caption{Continued.}	
\label{map-1}
\end{figure*}

\clearpage
\addtocounter{figure}{-1}
\centering 

\begin{figure*}

\gridline{          
          \fig{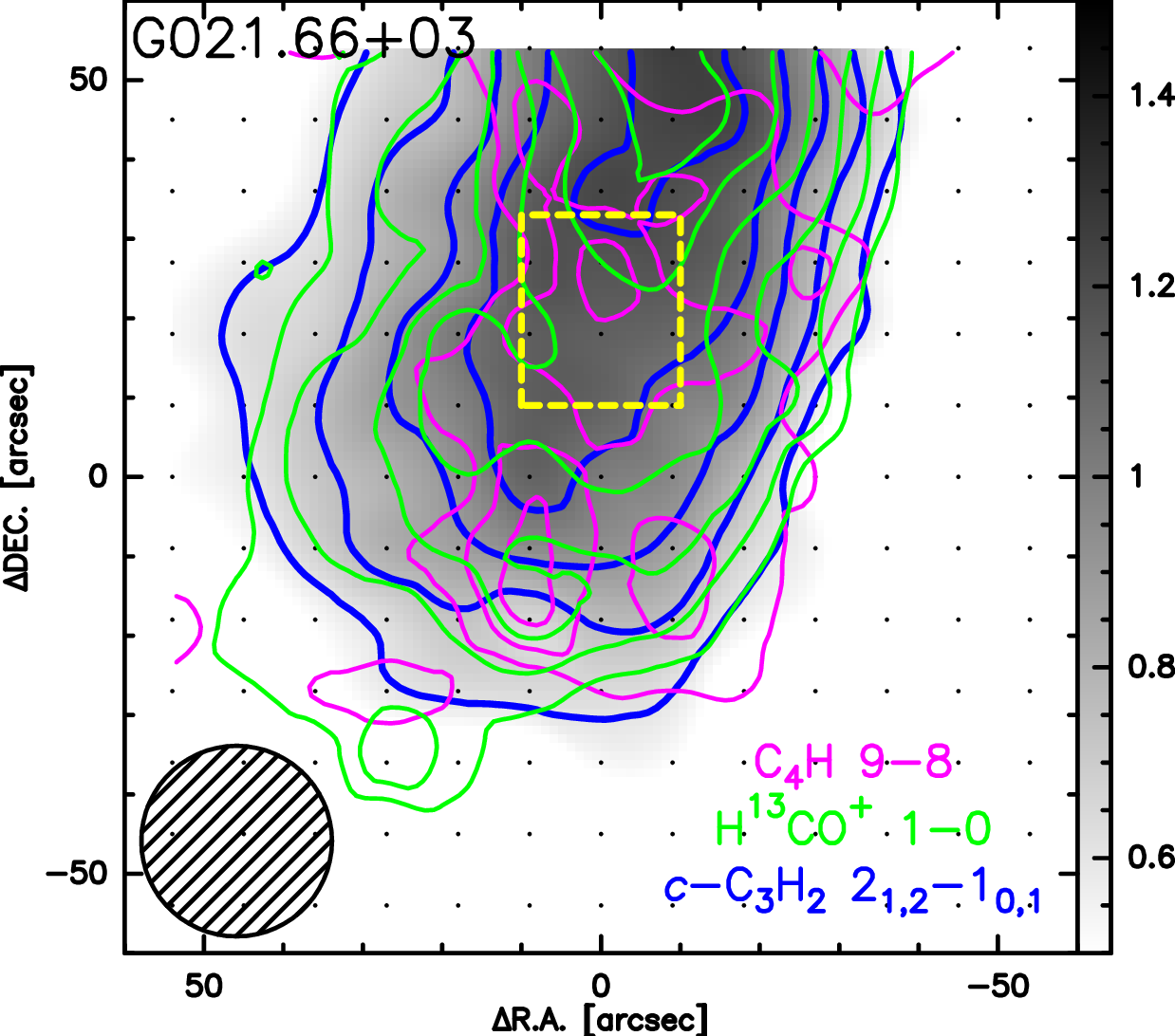}{0.45\textwidth}{(k)}
          \fig{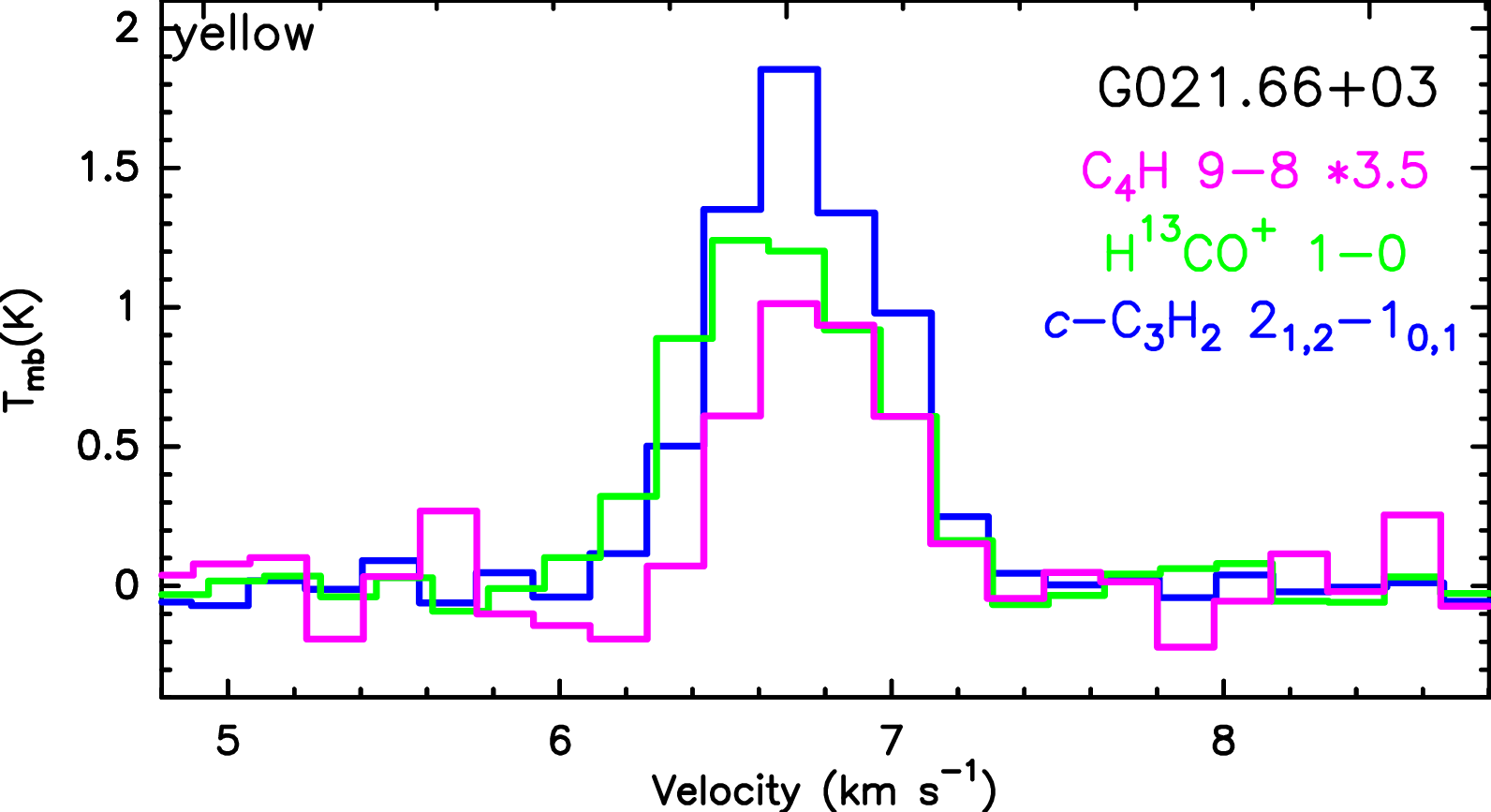}{0.45\textwidth}{(l)}
          }

\caption{Continued.}	
\label{map-1}
\end{figure*}

\centering
\begin{figure*}
\addtocounter{figure}{0}
\centering 
\centering 
\subfigure[]{ \label{fig10:a} 
\includegraphics[width=0.45\columnwidth]{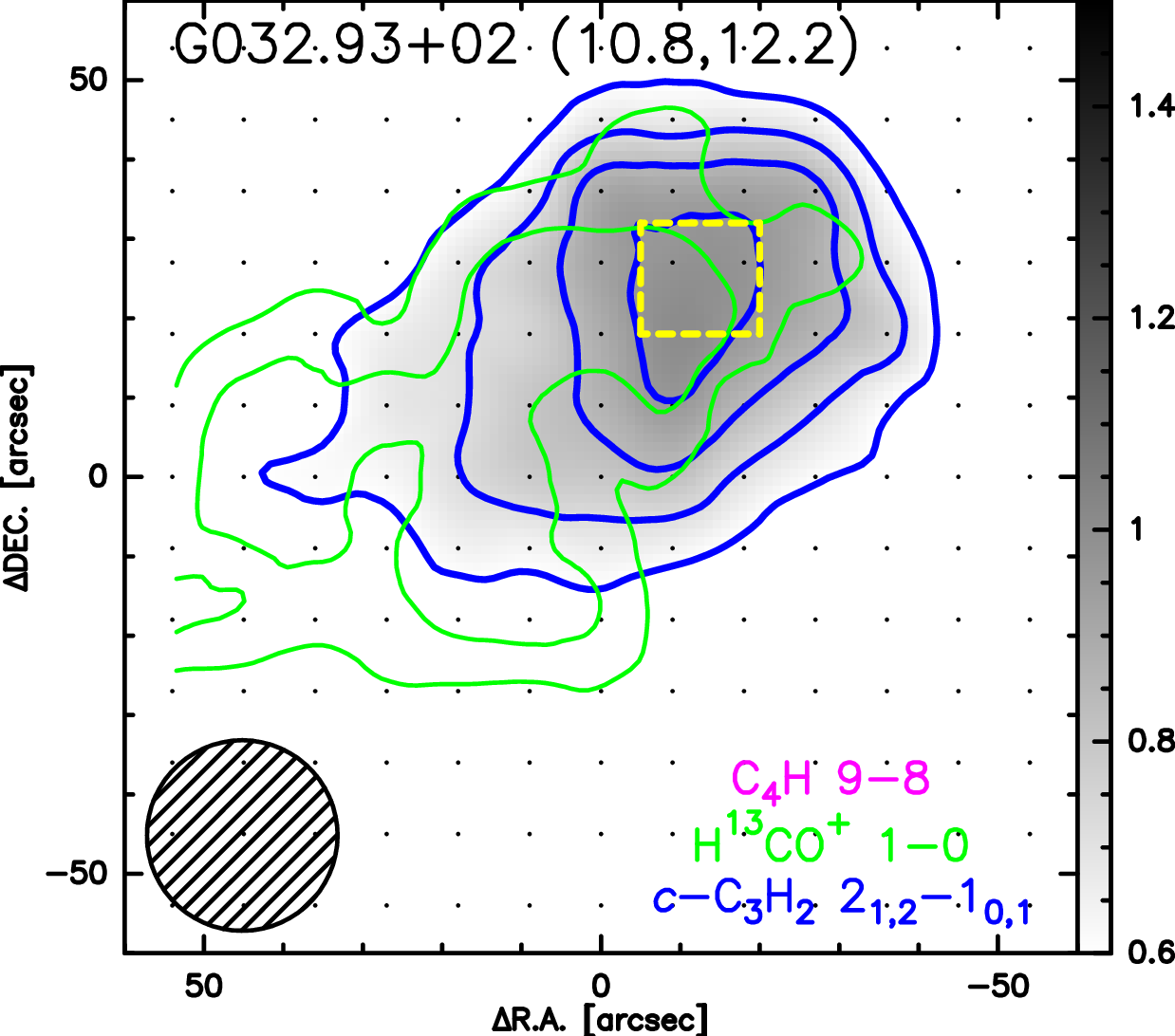} 
} 
\subfigure[]{ \label{fig10:b} 
\includegraphics[width=0.45\columnwidth]{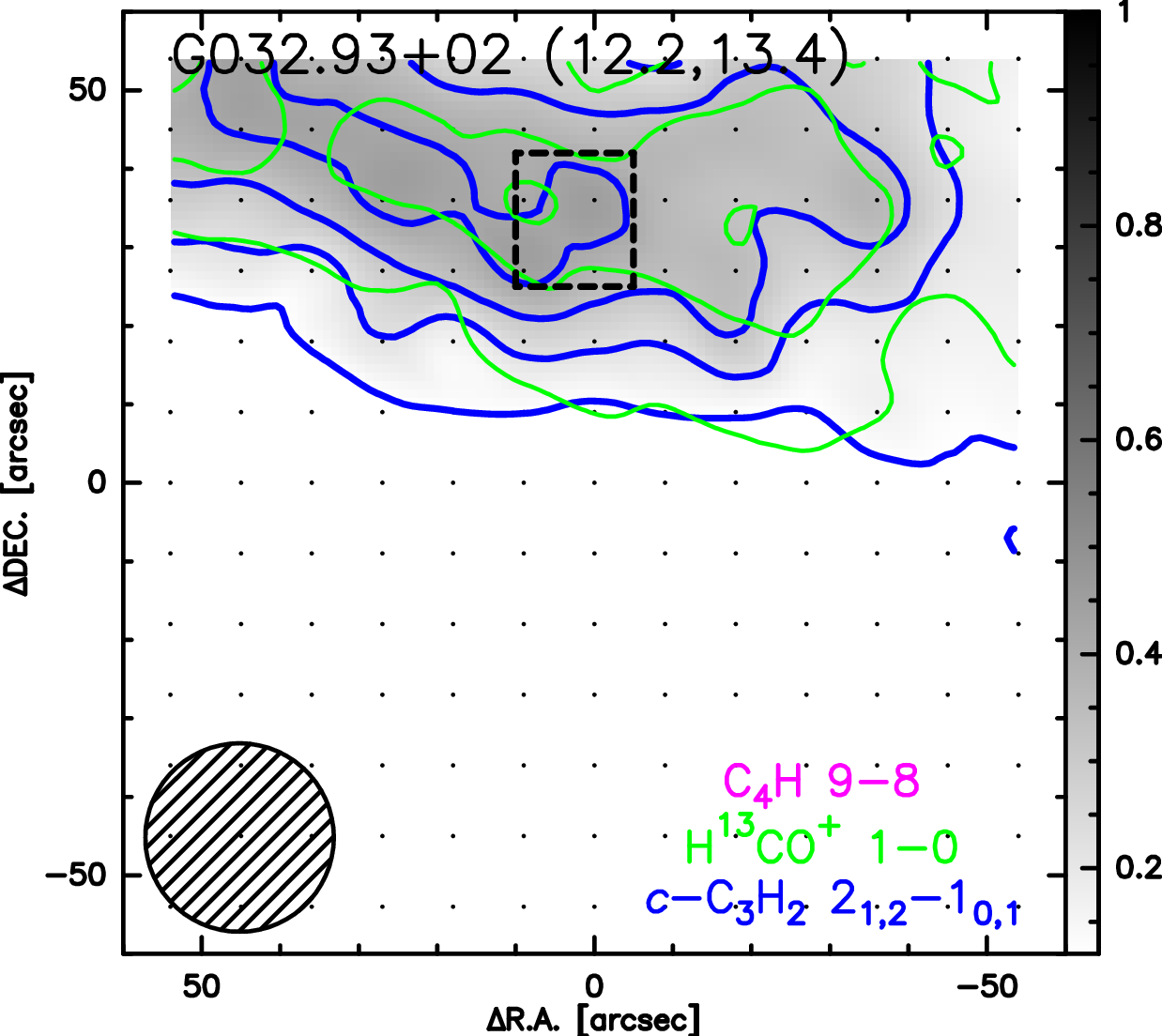} 
}

\subfigure[]{ \label{fig10:c} 
\includegraphics[height=1.8in,width=0.45\columnwidth]{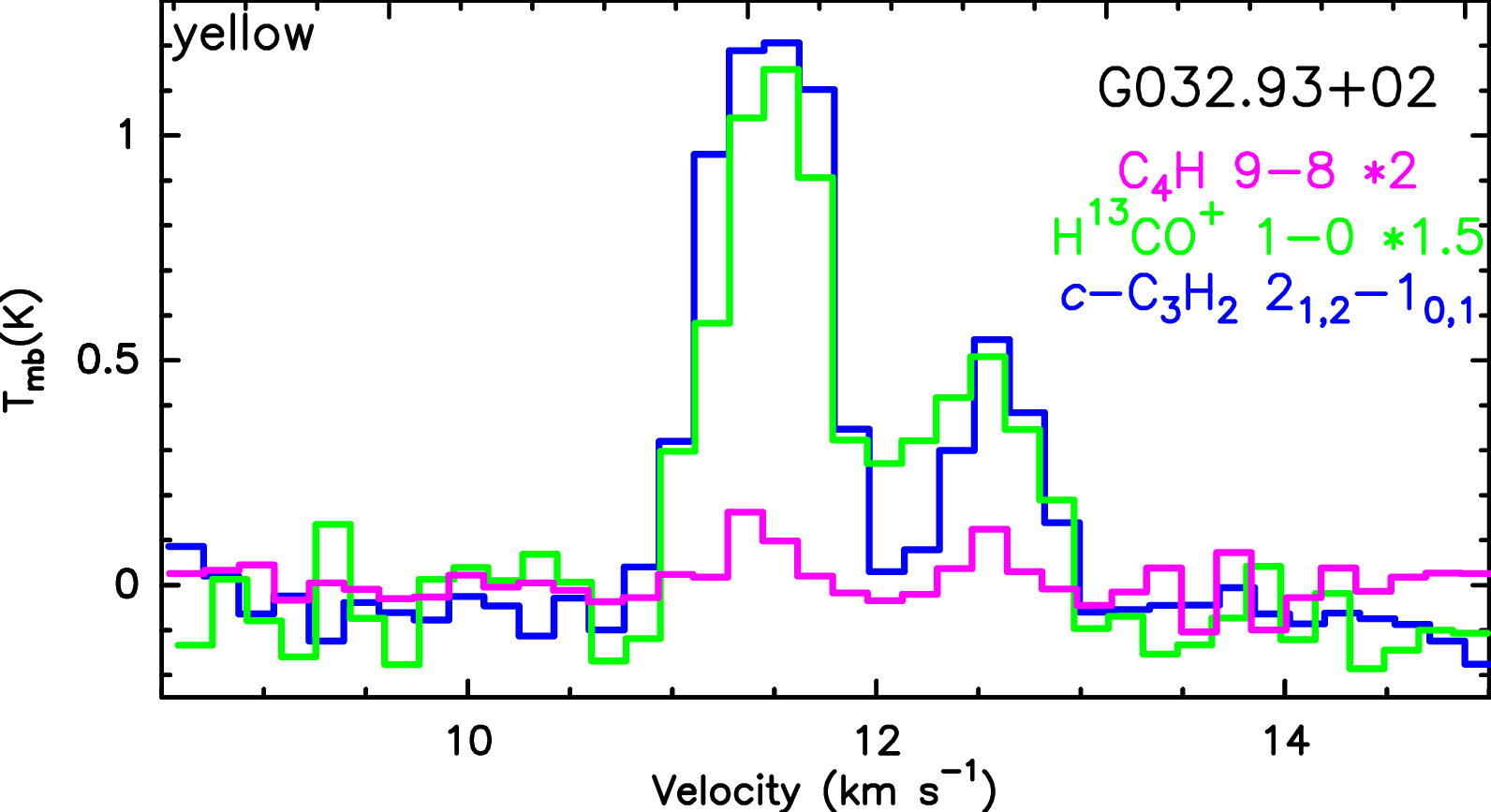} 
} 
\subfigure[]{ \label{fig10:d} 
\includegraphics[height=1.8in,width=0.45\columnwidth]{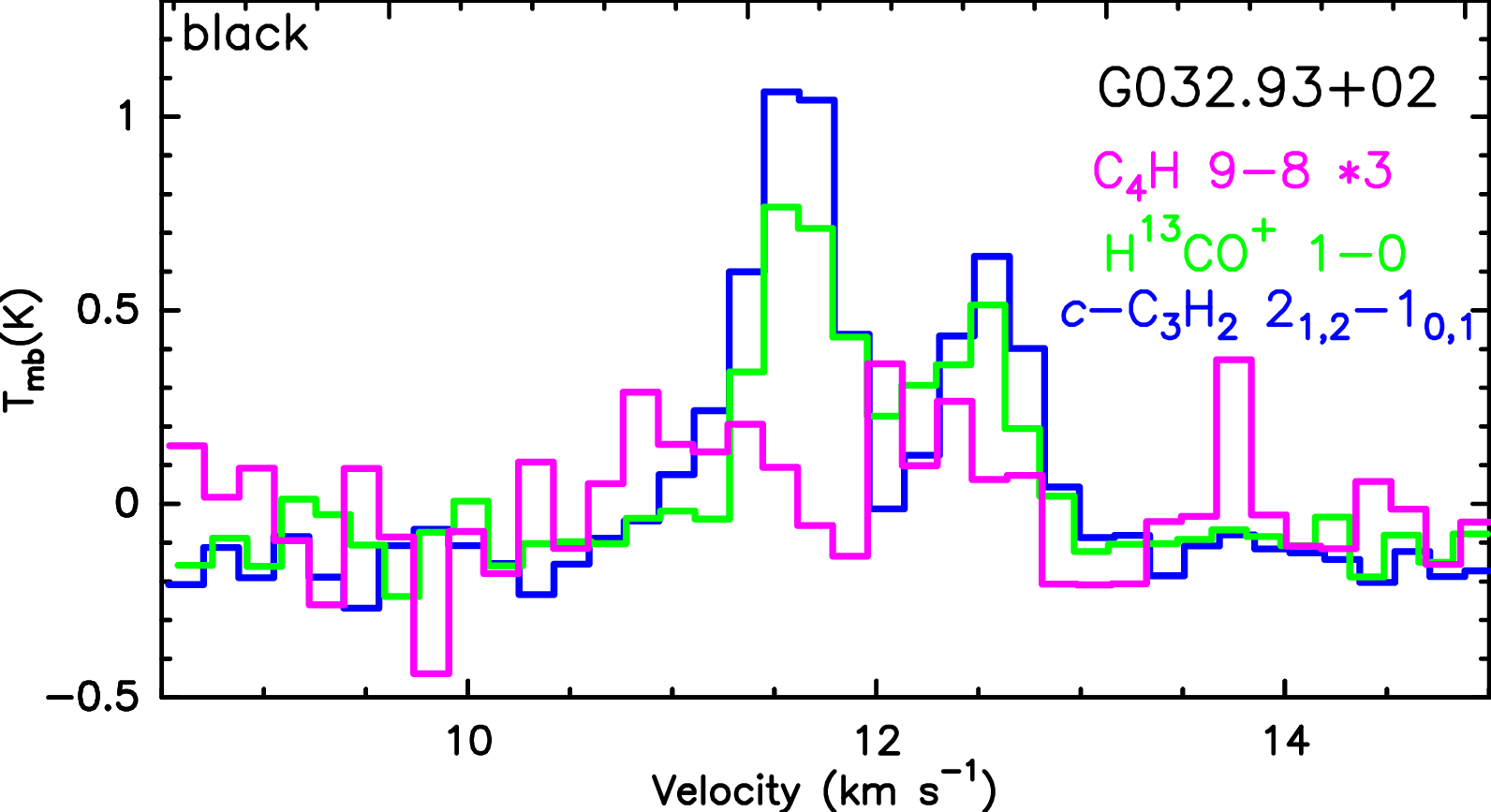} 
}

\centering 
\addtocounter{figure}{0}
\caption{(a): The velocity integrated intensity maps for  $c$-C$_3$H$_2$ 2--1 (blue contour) and H$^{13}$CO$^+$ 1--0 (green contour) of G032.93+22, velocity integral range from 10.8 to 12.2.
(b): The velocity integrated intensity maps for  $c$-C$_3$H$_2$ 2--1 and H$^{13}$CO$^+$ 1--0 of G032.93+22, velocity integral range from 12.2 to 13.4.
(c): Spectra of C$_4$H at 85672.5793 MHz, $c$-C$_3$H$_2$ at 85338.8940 MHz and H$^{13}$CO$^+$ at 86754.2884 MHz in the yellow box of G032.93+22. 
(d): Spectra of C$_4$H at 85672.5793 MHz, $c$-C$_3$H$_2$ at 85338.8940 MHz and H$^{13}$CO$^+$ at 86754.2884 MHz in the black box of G032.93+22. }	
	
\label{map-2}
\end{figure*}

\clearpage
 \begin{figure}
\centering

\includegraphics[width=0.65\textwidth]{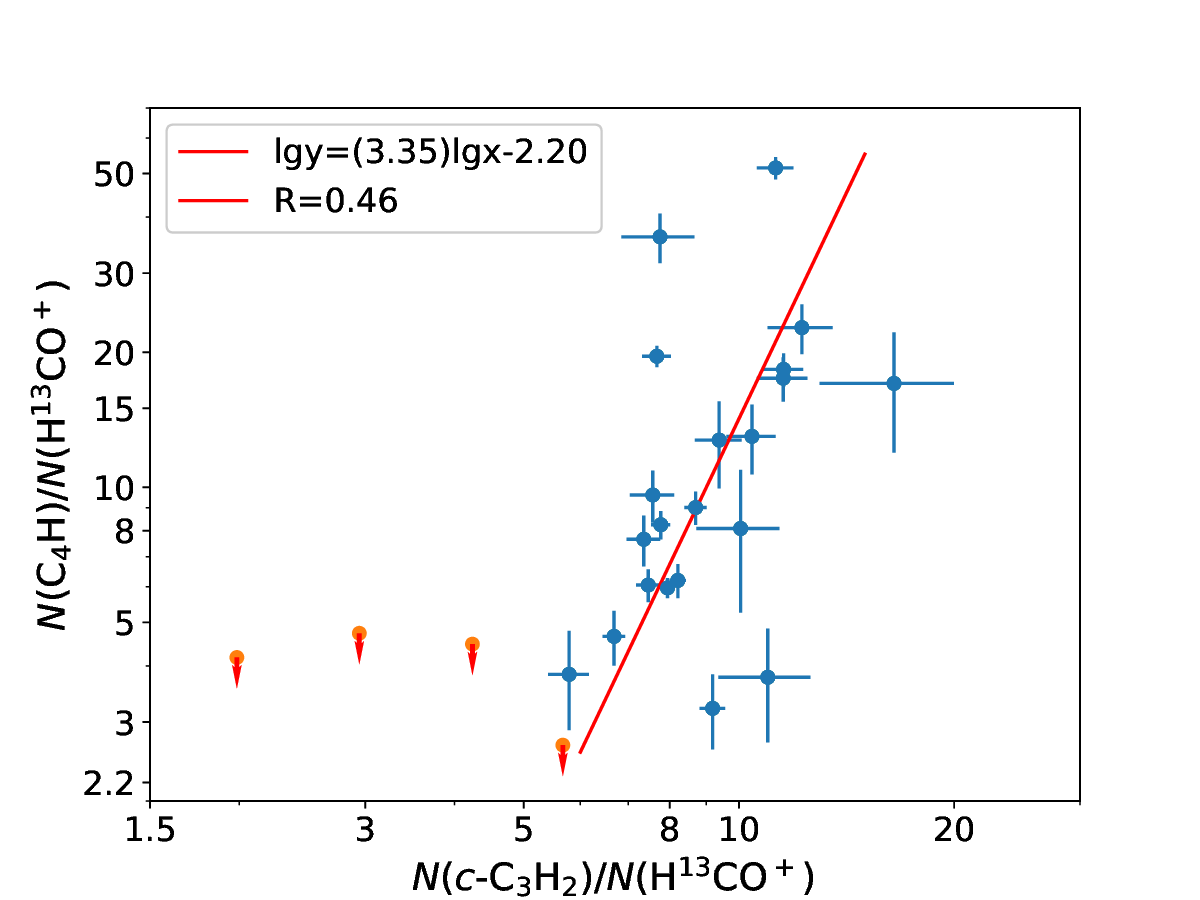}\\
\caption{The relation between $c$-C$_3$H$_2$/H$^{13}$CO$^+$ and C$_4$H/H$^{13}$CO$^+$ abundance ratio in 25 regions of the 19 sources and the red points are limits of C$_4$H.}
 
 \label{C4H_C3H2-2}	 
 \end{figure}

\begin{figure}
\centering

\includegraphics[width=0.65\textwidth]{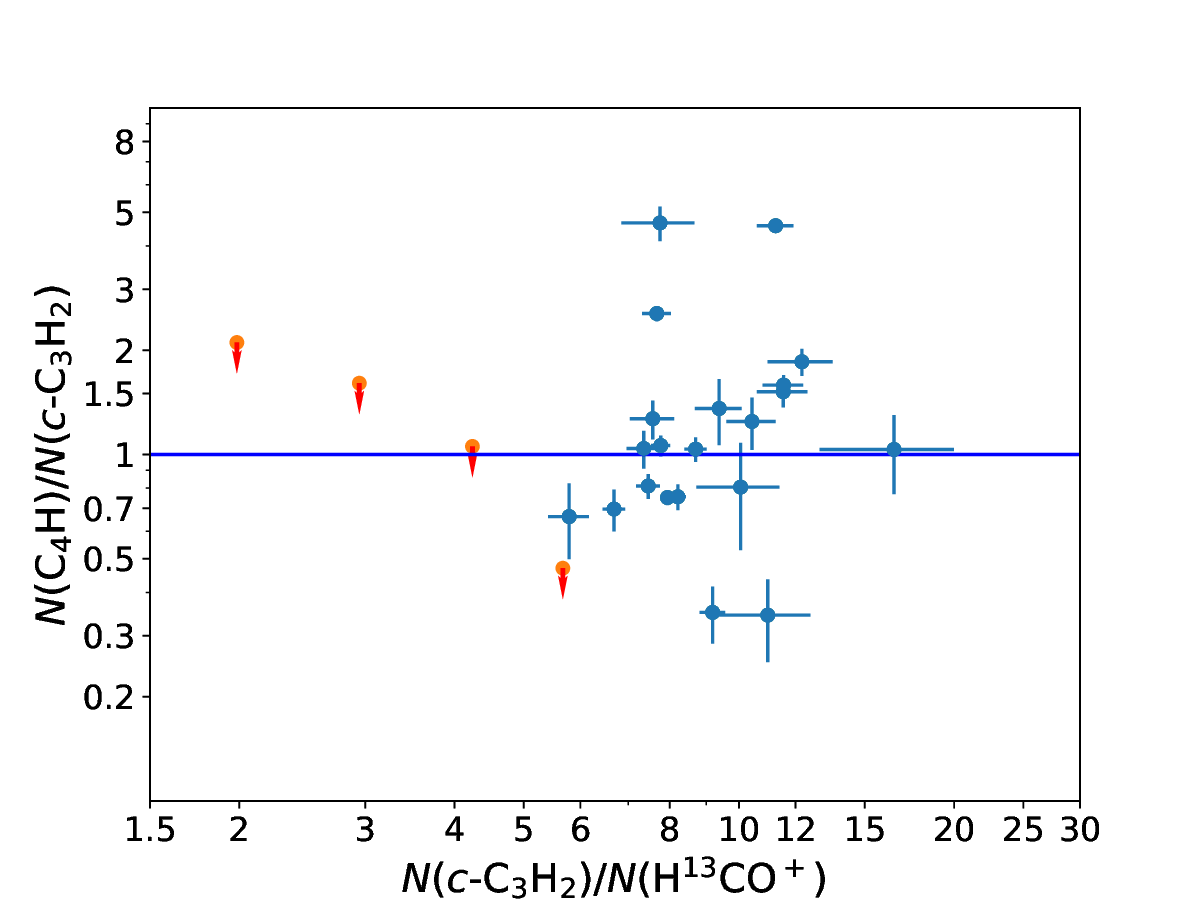}\\
\caption{The variation of C$_4$H/$c$-C$_3$H$_2$ abundance ratio along with $c$-C$_3$H$_2$/H$^{13}$CO$^+$ in 25 regions of the 19 sources. The blue line indicates where the C$_4$H/$c$-C$_3$H$_2$ abundance ratio is unity. }

 \label{C4H_C3H2-3}	 
 \end{figure}

\begin{table*}[h]
\centering
\setlength{\tabcolsep}{0.15in}
\caption{Physical parameters of C$_4$H, $c$-C$_3$H$_2$ and H$^{13}$CO$^+$.}
\label{table:Physical parameters}
\vspace{-1mm}
\begin{tabular}{cccccccccc}
\hline\hline
Molecular  &Q$_{9.375}$  & Transition & Freq  & $E_{up}$  & $g_{u}\ $   & A &\\

                 &                     &                 & (MHz)  &   (K)  &   & \colhead{($10^{-5}\ s^{-1}$)}&\\

\hline

C$_4$H	&	165.5392	&	N=9-8 J=19/2-17/2	&	85634.0044	&	20.561	&	19	&	1.5175	 	\\
 	&	165.5392	&	N=9-8 J=19/2-19/2	&	85634.0154	&	20.561 	&	21	&	1.5267		\\
 	&	165.5392	&	N=9-8 J=17/2-15/2	&	85672.5793	&	20.563 	&	17	&	1.5078		\\
 	&	165.5392	&	N=9-8 J=17/2-17/2	&	85672.5815	&	20.563 	&	19	&	1.5193		\\
$c$-C$_3$H$_2$	&	72.4028	&	J=2(1,2)-1(0,1) 	&	85338.8940	&	6.445 	&	15	&	2.3221		\\
H$^{13}$CO$^+$	&	4.8522	&	J=1-0	&	86754.2884	&	4.164	&	3	&	3.8535		\\

\hline
\end{tabular}
\end{table*}

\centering
\begin{table*} [h]
\centering
\setlength{\tabcolsep}{0.20in}
  \caption{Source information and observing parameters.  }
     \label{table:source}
\vspace{-1mm}
\begin{tabular}{cccccccl}
  \hline \hline
Source Name  & R.A.   & Decl.        &rms &Mapping Size& $v_{\rm LSR}$  &   \\
                       & (hh:mm:ss)  & (dd:mm:ss)  & (10$^{-2}$K)  & $('')$   & (km\,s$^{-1}$) &   \\

\hline

G001.38+20 	&	16:34:38.06    &        -15:46:40.7 	&	5.5 	&	220	$\times$	220		&	0.6 			&	\\
G001.84+16 	&	16:50:12.91      &      -18:04:22.3   	&	5.2 	&	120	$\times$	120		&	5.9 			&	\\
G003.73+16     	&	 16:55:21.78     &       -16:43:35.3  	&	6.8 	&	120	$\times$	120		&	6.1 			&	\\
G006.04+36     	&	15:54:10.81      &      -02:50:56.3        &	4.5 	&	240	$\times$	240		&	2.5 			&	\\
G006.32+20   	&	 16:47:40.85     &       -12:22:03.3	&	9.4 	&	120	$\times$	120		&	4.3 			&	\\
G006.41+20   	&	16:47:28.67      &      -12:13:52.5 	&	7.3 	&	120	$\times$	120		&	4.4 			&	\\
G007.14+05	&	 17:39:47.47    &        -19:45:05.1   	&	1.4 	&	100	$\times$	100		&	10.4 			&	\\
G008.52+21 	&	16:47:48.42      &      -09:53:09.8 	&	9.7 	&	110	$\times$	110		&	3.8 			&	\\
G008.67+22 	&	16:47:07.71      &      -09:35:50.0  	&	7.4 	&	120	$\times$	120		&	3.6 			&	\\
G021.20+04	&	 18:12:01.72     &       -08:05:27.7  	&	1.1 	&	220	$\times$	100		&	3.6 			&	\\
G021.66+03	&	18:17:06.23      &      -08:14:35.1       	&	4.7 	&	120	$\times$	120		&	6.7 			&	\\
G025.48+06	&	18:15:46.82       &     -03:45:19.2 	&	1.4 	&	120	$\times$	55		&	7.7 			&	\\
G026.85+06 	&	18:15:46.82        &    -03:45:19.2 	&	4.7 	&	240	$\times$	140		&	7.2 			&	\\
G028.45-06 	&	19:06:09.21       &     -06:52:51.7 	&	5.5 	&	120	$\times$	120		&	12.3 			&	\\
G028.71+03	&	18:29:55.29       &     -01:58:10.4   	&	5.2 	&	120	$\times$	120		&	7.5 			&	\\
G030.78+05	&	18:28:54.49        &     +00:28:40.0 	&	3.7 	&	120	$\times$	240		&	8.0 			&	\\
G031.44+04	&	18:28:54.49         &    +00:28:40.0 	&	4.8 	&	230	$\times$	110		&	11.1 			&	\\
G032.93+02 	&	 18:41:52.83       &      +01:13:45.3   	&	4.6 	&	120	$\times$	120		&	11.5 			&	\\
G058.16+03   	&	 18:41:52.83       &     +01:13:45.3   	&	6.3 	&	120	$\times$	120		&	9.9 			&	\\

\hline
\end{tabular}
\\
\end{table*}

\clearpage
\begin{table*}
\centering
\setlength{\tabcolsep}{0.06in}
\centering
\caption{C$_4$H 9--8, $c$-C$_3$H$_2$ 2--1 and H$^{13}$CO$^+$ 1--0 distribution information.} \label{table:distribution}

\vspace{-0.5mm}
\begin{tabular}{cccccccccl}
  \hline
    \hline
             
  \multirow{2}{*}{Source Name}          &   \multicolumn{2}{c}{\cfh(9-8)}   &   \multicolumn{2}{c}{\cctht(2-1) }       &\multicolumn{2}{c}{\hcfn(1-0)}   & \multicolumn{1}{c}{Difference between}   \\
	& Detection   &Clear Feature  &  Detection  & Clear  Feature      &  Detection     &Clear Feature  &  \cfh (9-8) and  \cctht (2-1)    \\
		 	 
\hline
 
G001.38+20	&	$\surd$	&	$\surd$	&	$\surd$	&	$\surd$	&	$\surd$	&	$\surd$	&	...	&	\\
G001.84+16	&	$\surd$	&	...	&	$\surd$	&	$\surd$	&	$\surd$	&	$\surd$	&	$\surd$	&	\\
G003.73+16	&	...	&	...	&	$\surd$	&	$\surd$	&	$\surd$	&	$\surd$	&	...	&	\\
G006.04+36	&	$\surd$	&	$\surd$	&	$\surd$	&	$\surd$	&	$\surd$	&	$\surd$	&	...	&	\\
G006.32+20	&	...	&	...	&	$\surd$	&	...	&	$\surd$	&	$\surd$	&	...	&	\\
G006.41+20	&	...	&	...	&	$\surd$	&	...	&	$\surd$	&	$\surd$	&	...	&	\\
G007.14+05	&	$\surd$	&	...	&	$\surd$	&	$\surd$	&	$\surd$	&	$\surd$	&	$\surd$	&	\\
G008.52+21	&	...	&	...	&	...	&	...	&	$\surd$	&	$\surd$	&	...	&	\\
G008.67+22	&	$\surd$	&	$\surd$	&	$\surd$	&	$\surd$	&	$\surd$	&	$\surd$	&	...	&	\\
G021.20+04	&	$\surd$	&	...	&	$\surd$	&	$\surd$	&	$\surd$	&	$\surd$	&	$\surd$	&	\\
G021.66+03	&	$\surd$	&	$\surd$	&	$\surd$	&	$\surd$	&	$\surd$	&	$\surd$	&	$\surd$	&	\\
G025.48+06	&	$\surd$	&	$\surd$	&	$\surd$	&	$\surd$	&	$\surd$	&	$\surd$	&	...	&	\\
G026.85+06	&	$\surd$	&	$\surd$	&	$\surd$	&	$\surd$	&	$\surd$	&	$\surd$	&	...	&	\\
 G028.45-06	&	$\surd$	&	$\surd$	&	$\surd$	&	$\surd$	&	$\surd$	&	$\surd$	&	...	&	\\
G028.71+03	&	$\surd$	&	$\surd$	&	$\surd$	&	$\surd$	&	$\surd$	&	$\surd$	&	...	&	\\
G030.78+05	&	$\surd$	&	$\surd$	&	$\surd$	&	$\surd$	&	$\surd$	&	$\surd$	&	...	&	\\
G031.44+04	&	...	&	...	&	$\surd$	&	$\surd$	&	$\surd$	&	$\surd$	&	...	&	\\
G032.93+02	&	...	&	...	&	$\surd$	&	$\surd$	&	$\surd$	&	$\surd$	&	...	&	\\
G058.16+03	&	$\surd$	&	$\surd$	&	$\surd$	&	$\surd$	&	$\surd$	&	$\surd$	&	...	&	\\\hline
\end{tabular}
\\

\end{table*}

\clearpage
\begin{table*}
\centering
\setlength{\tabcolsep}{0.06in}
\centering
\caption{Information parameters for map setting of  C$_4$H 9--8, $c$-C$_3$H$_2$ 2--1 and H$^{13}$CO$^+$ 1--0. \label{table:step}}

\vspace{-0.5mm}
\begin{tabular}{cccccccccl}
  \hline
    \hline
             
  \multirow{2}{*}{Source name}          &   \multicolumn{3}{c}{\cfh(9-8)}   &   \multicolumn{3}{c}{\cctht(2-1) }       &\multicolumn{3}{c}{\hcfn(1-0)}    \\
	&  1 $\sigma$    & Starting  & Step &  1 $\sigma$      &  Starting    & Step  &  1 $\sigma$    & Starting & Step  \\
	&  K km\,s$^{-1}$ &  K km\,s$^{-1}$ &  K km\,s$^{-1}$  &  K km\,s$^{-1}$ &  K km\,s$^{-1}$   &  K km\,s$^{-1}$    &  K km\,s$^{-1}$  & K km\,s$^{-1}$      & K km\,s$^{-1}$    \\
	 	 
\hline
 
G001.38+20	&	0.07	&	0.37	&	0.1	&	0.05	&	0.8	&	0.3	&	0.05	&	0.8	&	0.25	\\
G001.84+16	&	0.12	&	0.35	&	0.13	&	0.08	&	0.6	&	0.11	&	0.08	&	0.7	&	0.1	\\
G003.73+16	&	0.09	&	 ...	&	 ...	&	0.07	&	0.16	&	0.08	&	0.07	&	0.25	&	0.06	\\
G006.04+36	&	0.06	&	0.2	&	0.08	&	0.05	&	0.52	&	0.1	&	0.05	&	0.53	&	0.1	\\
G006.32+20	&	0.09	&	 ...	&	...	&	0.07	&	0.26	&	0.08	&	0.07	&	0.3	&	0.08	\\
G006.41+20	&	0.08	&	 ...	&	 ...	&	0.07	&	0.25	&	0.06	&	0.07	&	0.2	&	0.08	\\
G007.14+05	&	0.07	&	0.25	&	0.08	&	0.09	&	0.45	&	0.12	&	0.09	&	0.3	&	0.1	\\
G008.52+21	&	0.06	&	 ...	&	...	&	0.05	&	...	&	...	&	0.05	&	0.25	&	0.1	\\
G008.67+22	&	0.05	&	0.2	&	0.08	&	0.06	&	0.3	&	0.08	&	0.06	&	0.2	&	0.06	\\
G021.20+04	&	0.08	&	0.35	&	0.1	&	0.07	&	0.55	&	0.2	&	0.07	&	0.5	&	0.13	\\
G021.66+03	&	0.06	&	0.24	&	0.07	&	0.05	&	0.6	&	0.15	&	0.05	&	0.55	&	0.1	\\
G025.48+06	&	0.08	&	0.65	&	0.12	&	0.09	&	0.7	&	0.15	&	0.09	&	0.5	&	0.12	\\
G026.85+06	&	0.07	&	0.36	&	0.1	&	0.06	&	0.4	&	0.15	&	0.07	&	0.5	&	0.1	\\
 G028.45-06	&	0.06	&	0.35	&	0.1	&	0.06	&	0.65	&	0.2	&	0.07	&	0.6	&	0.15	\\
G028.71+03	&	0.08	&	0.8	&	0.18	&	0.07	&	0.8	&	0.12	&	0.07	&	0.7	&	0.08	\\
G030.78+05	&	0.07	&	0.45	&	0.12	&	0.07	&	1.7	&	0.4	&	0.07	&	1.5	&	0.3	\\
G031.44+04	&	0.08	&	...	&	 ...	&	0.06	&	0.6	&	0.15	&	0.07	&	0.6	&	0.1	\\
G032.93+02(yellow)	&	0.1	&	...	&	...	&	0.04	&	0.6	&	0.12	&	0.08	&	0.5	&	0.08	\\
G032.93+02(black)	&	0.1	&	...	&	...	&	0.04	&	0.12	&	0.1	&	0.08	&	0.12	&	0.08	\\
G058.16+03	&	0.08	&	0.3	&	0.1	&	0.07	&	0.45	&	0.15	&	0.07	&	0.4	&	0.1	\\
\hline
\end{tabular}
\\

\end{table*}

\clearpage
\small

\begin{longtable}{ccccccl}

\caption{\label{table:Observed date} Observed data of C$_4$H, $c$-C$_3$H$_2$ and H$^{13}$CO$^+$.}\\
\hline
 \hline
\multicolumn{1}{l}{Source name} & \multicolumn{2}{c}{Molecular Line} & \multicolumn{1}{c}{$\int T_{\rm mb}\rm  d\rm{v}$} & \multicolumn{1}{c}{FWHM} & $T_{\rm peak}$  & box color \\

\multicolumn{1}{l}{  } & \multicolumn{2}{c}{ } & \multicolumn{1}{c}{(K·km\,s$^{-1}$)} & \multicolumn{1}{c}{ (km\,s$^{-1}$) } &  (K)   &  \\

\hline
\endfirsthead
\caption{continued.}\\
\hline
\hline
\multicolumn{1}{l}{Source Name} & \multicolumn{2}{c}{Molecular Line} & \multicolumn{1}{c}{$\int T_{\rm mb}\rm  d\rm{v}$} & \multicolumn{1}{c}{FWHM} & $T_{\rm peak}$  & Box Color \\
\multicolumn{1}{l}{  } & \multicolumn{2}{c}{ } & \multicolumn{1}{c}{(K km\,s$^{-1}$)} & \multicolumn{1}{c}{ (km\,s$^{-1}$) } &  (K)   &  \\
\hline
\endhead
\hline
\endfoot
G001.38+20	&	C$_4$H 	&	N=9-8   J=19/2-17/2	&		0.21 	$\pm$	0.02 	&	0.46 	$\pm$	0.06 	&	0.43 	&	yellow\\
	&	C$_4$H 	&	N=9-8   J=17/2-15/2	&		0.17 	$\pm$	0.02 	&	0.42 	$\pm$	0.08 	&	0.40 	&		\\
	&	  $c$-C$_3$H$_2$	&	J=2(1,2)-1(0,1)	&		1.59 	$\pm$	0.02 	&	0.55 	$\pm$	0.01 	&	2.73 	&		\\
	&	H$^{13}$CO$^+$  	&	J=1-0	&		1.19 	$\pm$	0.03 	&	0.53 	$\pm$	0.02 	&	2.01 	&		\\
G001.84+16	&	C$_4$H 	&	N=9-8   J=19/2-17/2	&		0.17 	$\pm$	0.03 	&	0.48 	$\pm$	0.09 	&	0.34 	&	yellow	\\
	&	C$_4$H &	N=9-8   J=17/2-15/2	&		0.13 	$\pm$	0.02 	&	0.23 	$\pm$	0.06 	&	0.55 	&		\\
	&	 $c$-C$_3$H$_2$ 	&	J=2(1,2)-1(0,1)	&		0.75 	$\pm$	0.03 	&	0.33 	$\pm$	0.02 	&	2.11 	&		\\
	&	H$^{13}$CO$^+$	&	J=1-0	&		0.60 	$\pm$	0.03 	&	0.43 	$\pm$	0.03 	&	1.32 	&		\\
G003.73+16	&	C$_4$H 	&	N=9-8   J=19/2-17/2	&		0.05 	$\pm$	0.01 	&	0.29 	$\pm$	0.09 	&	0.16 	&	yellow\\
	&	C$_4$H &	N=9-8   J=17/2-15/2	&		0.03 	$\pm$	0.01 	&	0.18 	$\pm$	0.22 	&	0.17 	&		\\
	&	 $c$-C$_3$H$_2$ 	&	J=2(1,2)-1(0,1)	&		0.37 	$\pm$	0.01 	&	0.24 	$\pm$	0.01 	&	1.42 	&		\\
	&	H$^{13}$CO$^+$	&	J=1-0	&		0.39 	$\pm$	0.02 	&	0.39 	$\pm$	0.02 	&	0.94 	&		\\
G006.04+36	&	C$_4$H 	&	N=9-8   J=19/2-17/2	&		0.05 	$\pm$	0.01 	&	0.20 	$\pm$	0.18 	&	0.23 	&	yellow\\
	&	C$_4$H 	&	N=9-8   J=17/2-15/2	&		0.08 	$\pm$	0.01 	&	0.29 	$\pm$	0.04 	&	0.26 	&		\\
	&	$c$-C$_3$H$_2$	&	J=2(1,2)-1(0,1)	&		0.58 	$\pm$	0.01 	&	0.32 	$\pm$	0.01 	&	1.69 	&		\\
	&	H$^{13}$CO$^+$  	&	J=1-0	&		0.53 	$\pm$	0.02 	&	0.48 	$\pm$	0.02 	&	1.05 	&		\\
G006.04+36	&	 C$_4$H	&	N=9-8   J=19/2-17/2	&		0.12 	$\pm$	0.01 	&	0.33 	$\pm$	0.03 	&	0.35 	&	black	\\
	&	 C$_4$H 	&	N=9-8   J=17/2-15/2	&		0.09 	$\pm$	0.01 	&	0.28 	$\pm$	0.03 	&	0.29 	&		\\
	&	  $c$-C$_3$H$_2$	&	J=2(1,2)-1(0,1)	&		0.61 	$\pm$	0.01 	&	0.32 	$\pm$	0.01 	&	1.81 	&		\\
	&	H$^{13}$CO$^+$ 	&	J=1-0	&		0.48 	$\pm$	0.01 	&	0.46 	$\pm$	0.01 	&	1.00 	&		\\
G006.32+20	&	C$_4$H 	&	N=9-8   J=19/2-17/2	&	 $\le$	0.052 	 		&	...	 		&	...	&	yellow	\\
	&	C$_4$H 	&	N=9-8   J=17/2-15/2	&	 $\le$	0.053 	 		&	...	 		&	...	&		\\
	&	  $c$-C$_3$H$_2$	&	J=2(1,2)-1(0,1)	&		0.31 	$\pm$	0.03 	&	0.37 	$\pm$	0.05 	&	0.79 	&		\\
	&	H$^{13}$CO$^+$ 	&	J=1-0	&		0.45 	$\pm$	0.05 	&	0.43 	$\pm$	0.06 	&	0.98 	&		\\
G006.41+20	&	C$_4$H	&	N=9-8   J=19/2-17/2	&	 $\le$	0.045 	 		&	...	 		&	...	&	yellow	\\
	&	C$_4$H	&	N=9-8   J=17/2-15/2	&	 $\le$	0.049 	 		&		 ...		&	...	&		\\
	&	 $c$-C$_3$H$_2$	&	J=2(1,2)-1(0,1)	&		0.18 	$\pm$	0.04 	&	0.61 	$\pm$	0.15 	&	0.28 	&		\\
	&	H$^{13}$CO$^+$	&	J=1-0	&		0.38 	$\pm$	0.03 	&	0.50 	$\pm$	0.06 	&	0.71 	&		\\

G007.14+05	&	C$_4$H 	&	N=9-8   J=19/2-17/2	&		0.11 	$\pm$	0.03 	&	0.20 	$\pm$	1.88 	&	0.52 	&	yellow	\\
	&	 C$_4$H 	&	N=9-8   J=17/2-15/2	&		0.15 	$\pm$	0.05 	&	0.67 	$\pm$	0.20 	&	0.22 	&		\\
	&	  $c$-C$_3$H$_2$	&	J=2(1,2)-1(0,1)	&		0.81 	$\pm$	0.09 	&	0.98 	$\pm$	0.11 	&	0.77 	&		\\
	&	H$^{13}$CO$^+$ 	&	J=1-0	&		0.30 	$\pm$	0.06 	&	0.46 	$\pm$	0.08 	&	0.61 	&		\\
G007.14+05	&	C$_4$H 	&	N=9-8   J=19/2-17/2	&		0.08 	$\pm$	0.04 	&	0.53 	$\pm$	0.24 	&	0.15 	&	black	\\
	&	C$_4$H 	&	N=9-8   J=17/2-15/2	&		0.13 	$\pm$	0.06 	&	0.55 	$\pm$	0.31 	&	0.23 	&		\\
	&	 $c$-C$_3$H$_2$	&	J=2(1,2)-1(0,1)	&		0.86 	$\pm$	0.07 	&	0.70 	$\pm$	0.06 	&	1.16 	&		\\
	&	H$^{13}$CO$^+$	&	J=1-0	&		0.52 	$\pm$	0.06 	&	0.69 	$\pm$	0.09 	&	0.71 	&		\\
G008.52+21	&	 C$_4$H 	&	N=9-8   J=19/2-17/2	&	 $\le$	0.051 	 		&	...	 		&	...	&	yellow	\\
	&	C$_4$H 	&	N=9-8   J=17/2-15/2	&	 $\le$	0.053 	 		&	...	 		&	...	&		\\
	&	  $c$-C$_3$H$_2$	&	J=2(1,2)-1(0,1)	&		0.16 	$\pm$	0.05 	&	0.50 	$\pm$	0.26 	&	0.29 	&		\\
	&	H$^{13}$CO$^+$ 	&	J=1-0	&		0.48 	$\pm$	0.03 	&	0.35 	$\pm$	0.03 	&	1.29 	&		\\
G008.67+22	&	 C$_4$H 	&	N=9-8   J=19/2-17/2	&		0.23 	$\pm$	0.02 	&	0.26 	$\pm$	0.02 	&	0.84 	&	yellow	\\
	&	C$_4$H 	&	N=9-8   J=17/2-15/2	&		0.23 	$\pm$	0.03 	&	0.31 	$\pm$	0.05 	&	0.71 	&		\\
	&	 $c$-C$_3$H$_2$ 	&	J=2(1,2)-1(0,1)	&		0.31 	$\pm$	0.02 	&	0.19 	$\pm$	0.04 	&	1.52 	&		\\
	&	H$^{13}$CO$^+$ 	&	J=1-0	&		0.24 	$\pm$	0.02 	&	0.33 	$\pm$	0.03 	&	0.69 	&		\\
G021.20+04	&	 C$_4$H 	&	N=9-8   J=19/2-17/2	&		0.20 	$\pm$	0.06 	&	0.31 	$\pm$	0.13 	&	0.62 	&	yellow\\
	&	C$_4$H 	&	N=9-8   J=17/2-15/2	&		0.17 	$\pm$	0.05 	&	0.39 	$\pm$	0.14 	&	0.41 	&		\\
	&	 $c$-C$_3$H$_2$	&	J=2(1,2)-1(0,1)	&		0.86 	$\pm$	0.04 	&	0.34 	$\pm$	0.02 	&	2.36 	&		\\
	&	H$^{13}$CO$^+$	&	J=1-0	&		0.56 	$\pm$	0.03 	&	0.35 	$\pm$	0.03 	&	1.50 	&		\\
G021.20+04	&	C$_4$H 	&	N=9-8   J=19/2-17/2	&		0.18 	$\pm$	0.04 	&	0.37 	$\pm$	0.13 	&	0.45 	&	black	\\
	&	C$_4$H	&	N=9-8   J=17/2-15/2	&		0.17 	$\pm$	0.04 	&	0.39 	$\pm$	0.09 	&	0.41 	&		\\
	&	  $c$-C$_3$H$_2$ 	&	J=2(1,2)-1(0,1)	&		0.88 	$\pm$	0.04 	&	0.31 	$\pm$	0.02 	&	2.64 	&		\\
	&	H$^{13}$CO$^+$	&	J=1-0	&		0.52 	$\pm$	0.03 	&	0.37 	$\pm$	0 .03    	&	1.32 	&		\\
G021.66+03	&	C$_4$H 	&	N=9-8   J=19/2-17/2	&		0.12 	$\pm$	0.02 	&	0.26 	$\pm$	0.04 	&	0.43 	&	yellow	\\
	&	 C$_4$H 	&	N=9-8   J=17/2-15/2	&		0.17 	$\pm$	0.02 	&	0.51 	$\pm$	0.05 	&	0.31 	&		\\
	&	 $c$-C$_3$H$_2$ 	&	J=2(1,2)-1(0,1)	&		1.13 	$\pm$	0.03 	&	0.59 	$\pm$	0.02 	&	1.80 	&		\\
	&	H$^{13}$CO$^+$ 	&	 N=1-0   	&		0.93 	$\pm$	0.03 	&	0.67 	$\pm$	0.02 	&	1.30 	&		\\
G025.48+06	&	 C$_4$H 	&	N=9-8   J=19/2-17/2	&		0.42 	$\pm$	0.03 	&	0.37 	$\pm$	0.03 	&	1.07 	&	yellow	\\
	&	 C$_4$H 	&	N=9-8   J=17/2-15/2	&		0.35 	$\pm$	0.04 	&	0.35 	$\pm$	0.04 	&	0.94 	&		\\
	&	  $c$-C$_3$H$_2$	&	J=2(1,2)-1(0,1)	&		1.52 	$\pm$	0.04 	&	0.43 	$\pm$	0.02 	&	3.28 	&		\\
	&	H$^{13}$CO$^+$	&	J=1-0	&		0.80 	$\pm$	0.05 	&	0.45 	$\pm$	0.03 	&	1.67 	&		\\
G025.48+06	&	C$_4$H 	&	N=9-8   J=19/2-17/2	&		0.44 	$\pm$	0.04 	&	0.37 	$\pm$	0.04 	&	1.12 	&	black	\\
	&	C$_4$H 	&	N=9-8   J=17/2-15/2	&		0.33 	$\pm$	0.05 	&	0.30 	$\pm$	0.04 	&	1.03 	&		\\
	&	 $c$-C$_3$H$_2$	&	J=2(1,2)-1(0,1)	&		1.29 	$\pm$	0.05 	&	0.41 	$\pm$	0.02 	&	2.99 	&		\\
	&	H$^{13}$CO$^+$	&	J=1-0	&		0.65 	$\pm$	0.06 	&	0.38 	$\pm$	0.05 	&	1.58 	&		\\
G026.85+06	&	C$_4$H	&	N=9-8   J=19/2-17/2	&		0.28 	$\pm$	0.02 	&	0.28 	$\pm$	0.02 	&	0.94 	&	yellow	\\
	&	 C$_4$H	&	N=9-8   J=17/2-15/2	&		0.24 	$\pm$	0.01 	&	0.29 	$\pm$	0.02 	&	0.79 	&		\\
	&	  $c$-C$_3$H$_2$ &	J=2(1,2)-1(0,1)	&		0.64 	$\pm$	0.02 	&	0.44 	$\pm$	0.01 	&	1.39 	&		\\
	&	H$^{13}$CO$^+$	&	J=1-0	&		0.51 	$\pm$	0.02 	&	0.54 	$\pm$	0.02 	&	0.90 	&		\\
 G028.45-06	&	C$_4$H	&	N=9-8   J=19/2-17/2	&		0.23 	$\pm$	0.02 	&	0.53 	$\pm$	0.06 	&	0.40 	&	yellow	\\
	&	 C$_4$H 	&	N=9-8   J=17/2-15/2	&		0.18 	$\pm$	0.02 	&	0.46 	$\pm$	0.06 	&	0.36 	&		\\
	&	  $c$-C$_3$H$_2$ 	&	J=2(1,2)-1(0,1)	&		1.21 	$\pm$	0.02 	&	0.56 	$\pm$	0.01 	&	2.04 	&		\\
	&	H$^{13}$CO$^+$  	&	J=1-0	&		0.85 	$\pm$	0.03 	&	0.65 	$\pm$	0.02 	&	1.22 	&		\\
G028.71+03	&	 C$_4$H 	&	N=9-8   J=19/2-17/2	&		0.93 	$\pm$	0.03 	&	0.52 	$\pm$	0.02 	&	1.68 	&	yellow	\\
	&	C$_4$H 	&	N=9-8   J=17/2-15/2	&		0.88 	$\pm$	0.02 	&	0.52 	$\pm$	0.02 	&	1.60 	&		\\
	&	   $c$-C$_3$H$_2$	&	J=2(1,2)-1(0,1)	&		1.24 	$\pm$	0.03 	&	0.99 	$\pm$	0.03 	&	1.18 	&		\\
	&	H$^{13}$CO$^+$	&	J=1-0	&		0.67 	$\pm$	0.04 	&	0.89 	$\pm$	0.05 	&	0.72 	&		\\
G030.78+05	&	C$_4$H	&	N=9-8   J=19/2-17/2	&		0.33 	$\pm$	0.02 	&	0.89 	$\pm$	0.07 	&	0.35 	&	yellow\\
	&	C$_4$H 	&	N=9-8   J=17/2-15/2	&		0.33 	$\pm$	0.02 	&	0.97 	$\pm$	0.08 	&	0.32 	&		\\
	&	  $c$-C$_3$H$_2$	&	J=2(1,2)-1(0,1)	&		2.75 	$\pm$	0.02 	&	1.04 	$\pm$	0.01 	&	2.49 	&		\\
	&	H$^{13}$CO$^+$	&	 N=1-0   	&		2.12 	$\pm$	0.03 	&	1.22 	$\pm$	0.02 	&	1.63 	&		\\
G031.44+04	&	C$_4$H 	&	N=9-8   J=19/2-17/2	&		0.07 	$\pm$	0.02 	&	0.55 	$\pm$	0.15 	&	0.12 	&	yellow	\\
	&	C$_4$H	&	N=9-8   J=17/2-15/2	&		0.04 	$\pm$	0.01 	&	0.19 	$\pm$	0.53 	&	0.20 	&		\\
	&	  $c$-C$_3$H$_2$	&	J=2(1,2)-1(0,1)	&		0.99 	$\pm$	0.02 	&	0.60 	$\pm$	0.01 	&	1.56 	&		\\
	&	H$^{13}$CO$^+$ 	&	J=1-0	&		0.66 	$\pm$	0.02 	&	0.61 	$\pm$	0.02 	&	1.01 	&		\\
G032.93+02	&	C$_4$H 	&	N=9-8   J=19/2-17/2	&		0.04 	$\pm$	0.02 	&	0.35 	$\pm$	0.17 	&	0.12 	&	yellow	\\
	&	C$_4$H 	&	N=9-8   J=17/2-15/2	&		0.06 	$\pm$	0.02 	&	0.29 	$\pm$	0.11 	&	0.18 	&		\\
	&	 $c$-C$_3$H$_2$	&	J=2(1,2)-1(0,1)	&		0.93 	$\pm$	0.07 	&	0.65 	$\pm$	0.06 	&	1.35 	&		\\
	&	H$^{13}$CO$^+$	&	J=1-0	&		0.52 	$\pm$	0.06 	&	0.63 	$\pm$	0.10 	&	0.76 	&		\\
G032.93+02	&	C$_4$H 	&	N=9-8   J=19/2-17/2	&	 $\le$	0.046 	 		&	...	 		&	...	&	black	\\
	&	C$_4$H 	&	N=9-8   J=17/2-15/2	&	 $\le$	0.043 	 		&		... 		&	...	&		\\
	&	 $c$-C$_3$H$_2$	&	J=2(1,2)-1(0,1)	&		0.60 	$\pm$	0.09 	&	0.49 	$\pm$	0.09 	&	1.15 	&		\\
	&	H$^{13}$CO$^+$	&	J=1-0	&		0.64 	$\pm$	0.07 	&	1.08 	$\pm$	0.13 	&	0.56 	&		\\
G058.16+03	&	 C$_4$H 	&	N=9-8   J=19/2-17/2	&		0.20 	$\pm$	0.03 	&	0.38 	$\pm$	0.08 	&	0.48 	&	yellow	\\
	&	 C$_4$H 	&	N=9-8   J=17/2-15/2	&		0.20 	$\pm$	0.02 	&	0.44 	$\pm$	0.05 	&	0.43 	&		\\
	&	$c$-C$_3$H$_2$	&	J=2(1,2)-1(0,1)	&		0.82 	$\pm$	0.03 	&	0.37 	$\pm$	0.02 	&	2.07 	&		\\
	&	H$^{13}$CO$^+$  	&	J=1-0	&		0.44 	$\pm$	0.03 	&	0.41 	$\pm$	0.04 	&	0.99 	&		\\
G058.16+03	&	 C$_4$H &	N=9-8   J=19/2-17/2	&		0.16 	$\pm$	0.03 	&	0.55 	$\pm$	0.14 	&	0.27 	&	black	\\
	&	C$_4$H 	&	N=9-8   J=17/2-15/2	&		0.16 	$\pm$	0.03 	&	0.61 	$\pm$	0.10 	&	0.25 	&		\\
	&	 $c$-C$_3$H$_2$	&	J=2(1,2)-1(0,1)	&		0.97 	$\pm$	0.03 	&	0.64 	$\pm$	0.02 	&	1.42 	&		\\
	&	H$^{13}$CO$^+$	&	J=1-0	&		0.81 	$\pm$	0.04 	&	0.71 	$\pm$	0.04 	&	1.06 	&		\\

\end{longtable}

\clearpage
\begin{table*}
\centering
\setlength{\tabcolsep}{0.06in}
\centering
\caption{Colunm Density and Relative Abundance of \cfh,  \cctht and \hcfn.  \label{table_Colunm density}}
\vspace{-0.5mm}
\begin{tabular}{cccccccccl}
  \hline
    \hline
             
  \multirow{2}{*}{Source Name}          &       $N$(\cfh)       &    $N$(\cctht)    &        $N$(\hcfn)        & $\dfrac{N_{C_4H}}{N_{H^{13}CO^+}}$  &    $\dfrac{N_{c-C_3H_2}}{N_{H^{13}CO^+}}$   &   $\dfrac{N_{C_4H}}{N_{c-C_3H_2}}$         & Box color   & \\
	& $10^{12}$cm$^{-2}$ & $10^{12}$cm$^{-2}$  & $10^{12}$cm$^{-2}$ &     &      &   &      &   \\
\hline

G001.38+20	&		7.04 	 $\pm$	0.60 	&	9.32 	 $\pm$	0.13 	&	1.14 	 $\pm$	0.025 	&		6.20 	 $\pm$	0.54 	&	8.21 	 $\pm$	0.21 	&		0.76 	 $\pm$	0.06 	&	yellow	\\
G001.84+16	&		5.56 	 $\pm$	0.67 	&	4.38 	 $\pm$	0.19 	&	0.58 	 $\pm$	0.032 	&		9.61 	 $\pm$	1.27 	&	7.58 	 $\pm$	0.54 	&		1.27 	 $\pm$	0.16 	&	yellow	\\
G003.73+16	&		1.51	 $\pm$	0.35 	&	2.16 	 $\pm$	0.09 	&	0.37 	 $\pm$	0.020 	&		3.83 	 $\pm$	0.95 	&	5.78 	 $\pm$	0.38 	&		0.66 	 $\pm$	0.16 	&	yellow 	\\
G006.04+36	&		2.37	 $\pm$	0.32 	&	3.40 	 $\pm$	0.07 	&	0.51 	 $\pm$	0.015 	&		4.65 	 $\pm$	0.65 	&	6.69 	 $\pm$	0.24 	&		0.70 	 $\pm$	0.10 	&	yellow	\\
G006.04+36	&		3.82 	 $\pm$	0.26 	&	3.60 	 $\pm$	0.05 	&	0.46 	 $\pm$	0.012 	&		8.25 	 $\pm$	0.60 	&	7.77 	 $\pm$	0.24 	&		1.06 	 $\pm$	0.07 	&	black	\\
G006.32+20	&	 $\le$	1.94	  	 	&	1.83 	 $\pm$	0.20 	&	0.43 	 $\pm$	0.049 	&	 $\le$	4.47 	  	 	&	4.24 	 $\pm$	0.67 	&	 $\le$	1.06 	 	 	&	yellow	\\
G006.41+20	&	 $\le$	1.72	 	 	&	1.07 	 $\pm$	0.22 	&	0.37 	 $\pm$	0.030 	&	 $\le$	4.73 		 	&	2.94 	 $\pm$	0.66 	&	 $\le$	1.61 		 	&	yellow	\\
G007.14+05	&		4.90 	 $\pm$	1.15 	&	4.74 	 $\pm$	0.50 	&	0.29 	 $\pm$	0.053 	&		17.04 	 $\pm$	5.09 	&	16.48 	 $\pm$	3.5 	&		1.03 	 $\pm$	0.27 	&	yellow 	\\
G007.14+05	&		4.03 	 $\pm$	1.34 	&	5.01 	 $\pm$	0.39 	&	0.50 	 $\pm$	0.054 	&		8.10 	 $\pm$	2.84 	&	10.06 	 $\pm$	1.34 	&		0.80 	 $\pm$	0.28 	&	black	\\
G008.52+21	&	 $\le$	1.91	 	 	&	0.91 	 $\pm$	0.32 	&	0.46 	 $\pm$	0.029 	&	 $\le$	4.18 		 	&	1.98 	 $\pm$	0.70 	&	 $\le$	2.10 		 	&	yellow	\\
G008.67+22	&		8.43 	 $\pm$	0.73 	&	1.81 	 $\pm$	0.13 	&	0.23 	 $\pm$	0.021 	&		36.14 	 $\pm$	4.57 	&	7.77 	 $\pm$	0.91 	&		4.65 	 $\pm$	0.50 	&	yellow	\\
G021.20+04	&		6.87	 $\pm$	1.45 	&	5.06 	 $\pm$	0.23 	&	0.54 	 $\pm$	0.033 	&		12.74 	 $\pm$	2.79 	&	9.37 	 $\pm$	0.71 	&		1.36 	 $\pm$	0.29 	&	yellow	\\
G021.20+04	&		6.43	 $\pm$	1.06 	&	5.16 	 $\pm$	0.24 	&	0.50 	 $\pm$	0.032 	&		12.99 	 $\pm$	2.30 	&	10.44 	 $\pm$	0.82 	&		1.24 	 $\pm$	0.21 	&	black	\\
G021.66+03	&		5.38	 $\pm$	0.42 	&	6.63 	 $\pm$	0.17 	&	0.89 	 $\pm$	0.026 	&		6.06 	 $\pm$	0.51 	&	7.46 	 $\pm$	0.29 	&		0.81 	 $\pm$	0.07 	&	yellow 	\\
G025.48+06	&		14.08 	 $\pm$	0.89 	&	8.88 	 $\pm$	0.26 	&	0.77 	 $\pm$	0.045 	&		18.31 	 $\pm$	1.57 	&	11.55 	 $\pm$	0.75 	&		1.59 	 $\pm$	0.11 	&	yellow	\\
G025.48+06	&		14.02	 $\pm$	1.14 	&	7.57 	 $\pm$	0.29 	&	0.62 	 $\pm$	0.060 	&		22.68 	 $\pm$	2.87 	&	12.24 	 $\pm$	1.28 	&		1.85 	 $\pm$	0.17 	&	black	\\
G026.85+06	&		9.62 	 $\pm$	0.40 	&	3.77 	 $\pm$	0.11 	&	0.49 	 $\pm$	0.017 	&		19.59 	 $\pm$	1.06 	&	7.68 	 $\pm$	0.36 	&		2.55 	 $\pm$	0.13 	&	yellow	\\
 G028.45-06	&		7.34	 $\pm$	0.58 	&	7.09 	 $\pm$	0.12 	&	0.81 	 $\pm$	0.026 	&		9.01 	 $\pm$	0.77 	&	8.70 	 $\pm$	0.31 	&		1.04 	 $\pm$	0.08 	&	yellow	\\
G028.71+03	&		33.25	 $\pm$	0.65 	&	7.27 	 $\pm$	0.18 	&	0.65 	 $\pm$	0.035 	&		51.48 	 $\pm$	2.90 	&	11.25 	 $\pm$	0.66 	&		4.58 	 $\pm$	0.14 	&	yellow 	\\
G030.78+05	&		12.10	 $\pm$	0.59 	&	16.10 	 $\pm$	0.13 	&	2.03 	 $\pm$	0.027 	&		5.97 	 $\pm$	0.30 	&	7.95 	 $\pm$	0.13 	&		0.75 	 $\pm$	0.04 	&	yellow	\\
G031.44+04	&		2.03 	 $\pm$	0.38 	&	5.80 	 $\pm$	0.12 	&	0.63 	 $\pm$	0.022 	&		3.22 	 $\pm$	0.61 	&	9.19 	 $\pm$	0.38 	&		0.35 	 $\pm$	0.07 	&	yellow\\
G032.93+02 	&		1.86 	 $\pm$	0.48 	&	5.42 	 $\pm$	0.44 	&	0.49 	 $\pm$	0.060 	&		3.77 	 $\pm$	1.07 	&	10.97 	 $\pm$	1.61 	&		0.34 	 $\pm$	0.09 	&	yellow	\\
G032.93+02 	&	 $\le$	1.64	 	 	&	3.49 	 $\pm$	0.53 	&	0.62 	 $\pm$	0.068 	&	 $\le$	2.66 		 	&	5.67 	 $\pm$	1.07 	&	 $\le$	0.47 		 	&	black	\\
G058.16+03	&		7.31 	 $\pm$	0.65 	&	4.82 	 $\pm$	0.20 	&	0.42 	 $\pm$	0.029 	&		17.50 	 $\pm$	1.98 	&	11.54 	 $\pm$	0.94 	&		1.52 	 $\pm$	0.15 	&	yellow	\\
G058.16+03	&		5.91 	 $\pm$	0.72 	&	5.68 	 $\pm$	0.17 	&	0.77 	 $\pm$	0.035 	&		7.66 	 $\pm$	0.99 	&	7.36 	 $\pm$	0.40 	&		1.04 	 $\pm$	0.13 	&	black	\\

\hline 

\end{tabular}
\\

\end{table*}

\clearpage

\appendix
\setcounter{figure}{0}
\renewcommand{\thefigure}{A\arabic{figure}}
\section{The spatial distribution maps and spectral lines of all sources.}

\begin{figure*}

\centering 

\includegraphics[width=0.38\columnwidth]{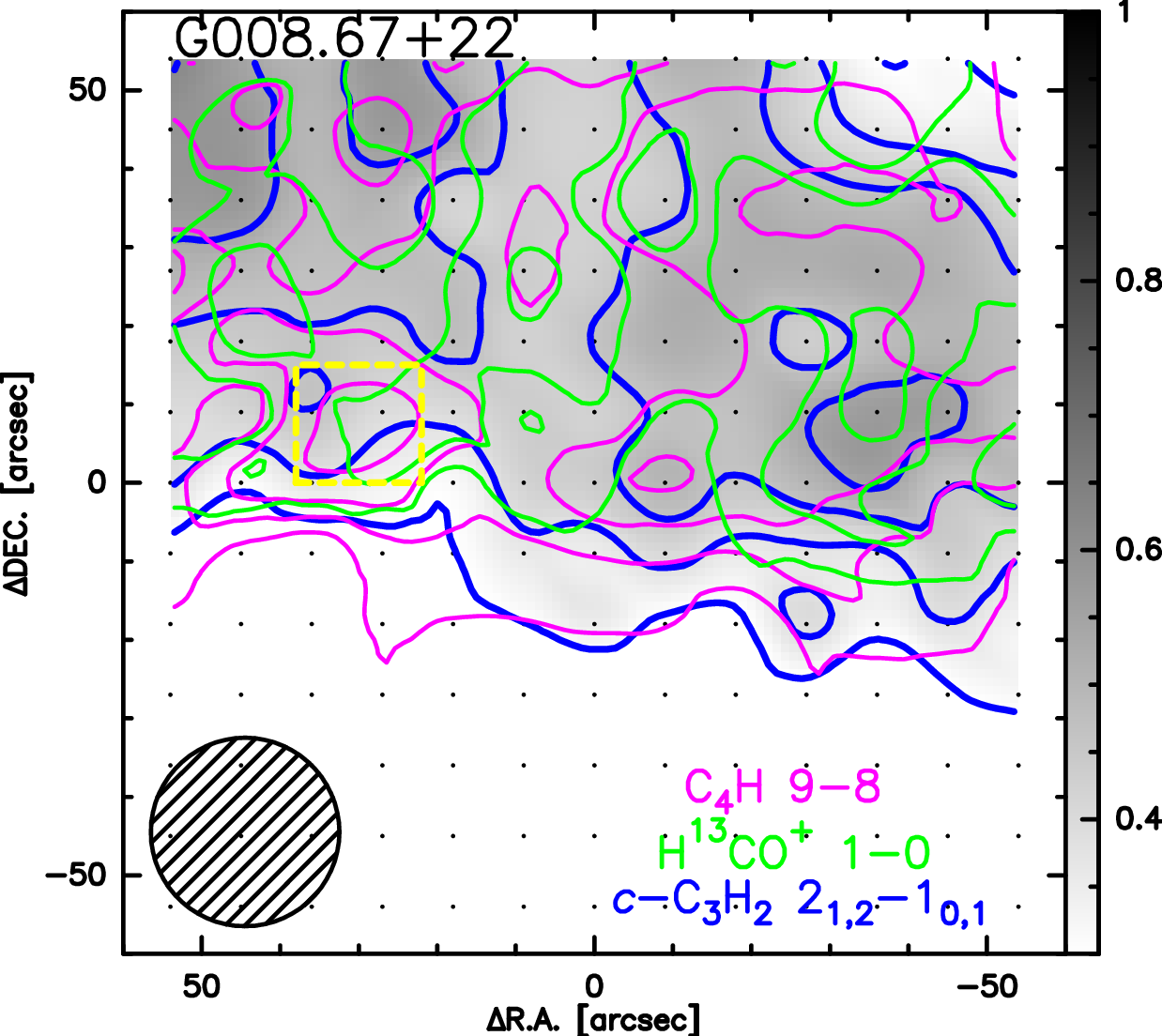}\includegraphics[width=0.38\columnwidth]{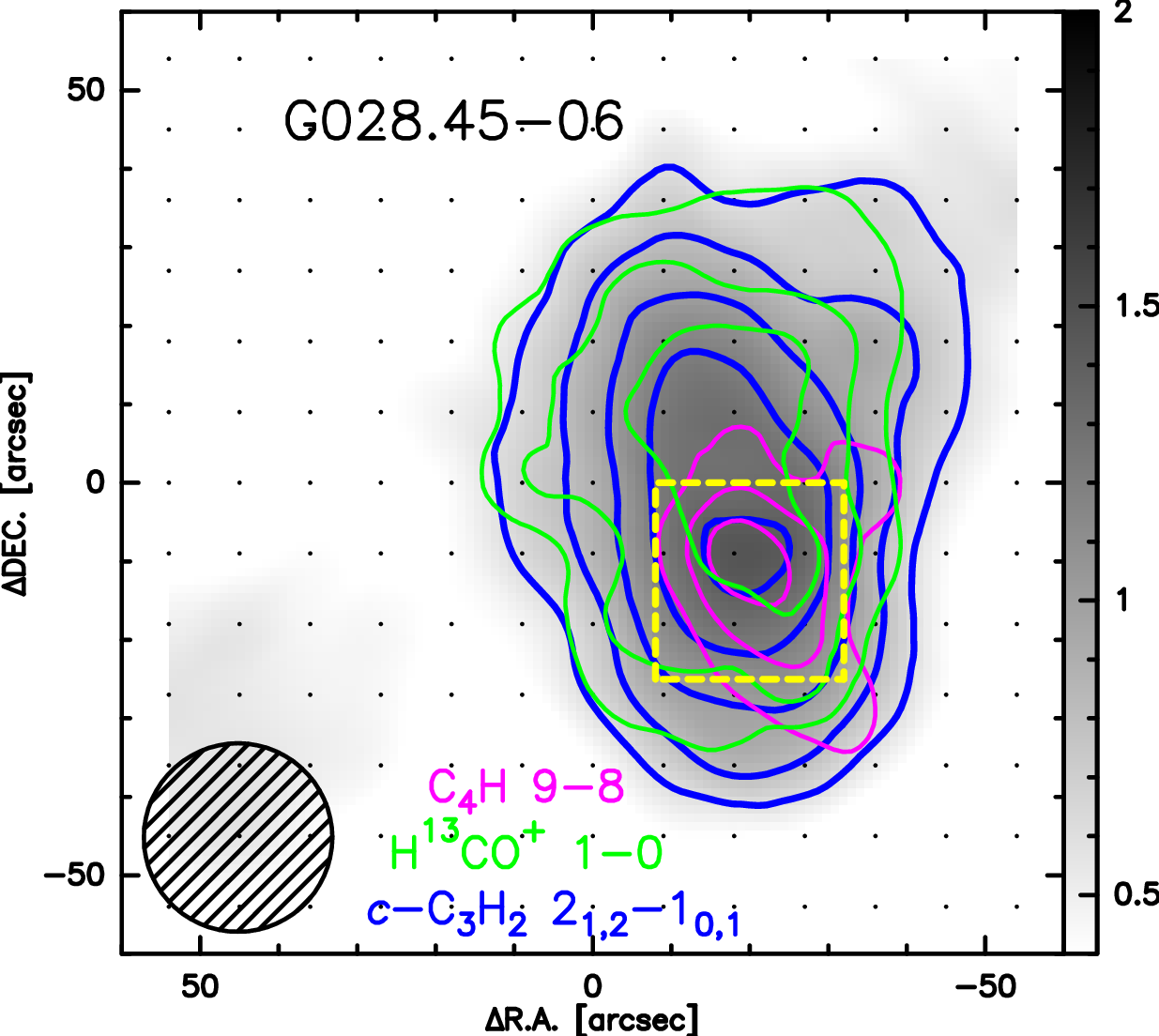}\\ 
\includegraphics[width=0.38\columnwidth]{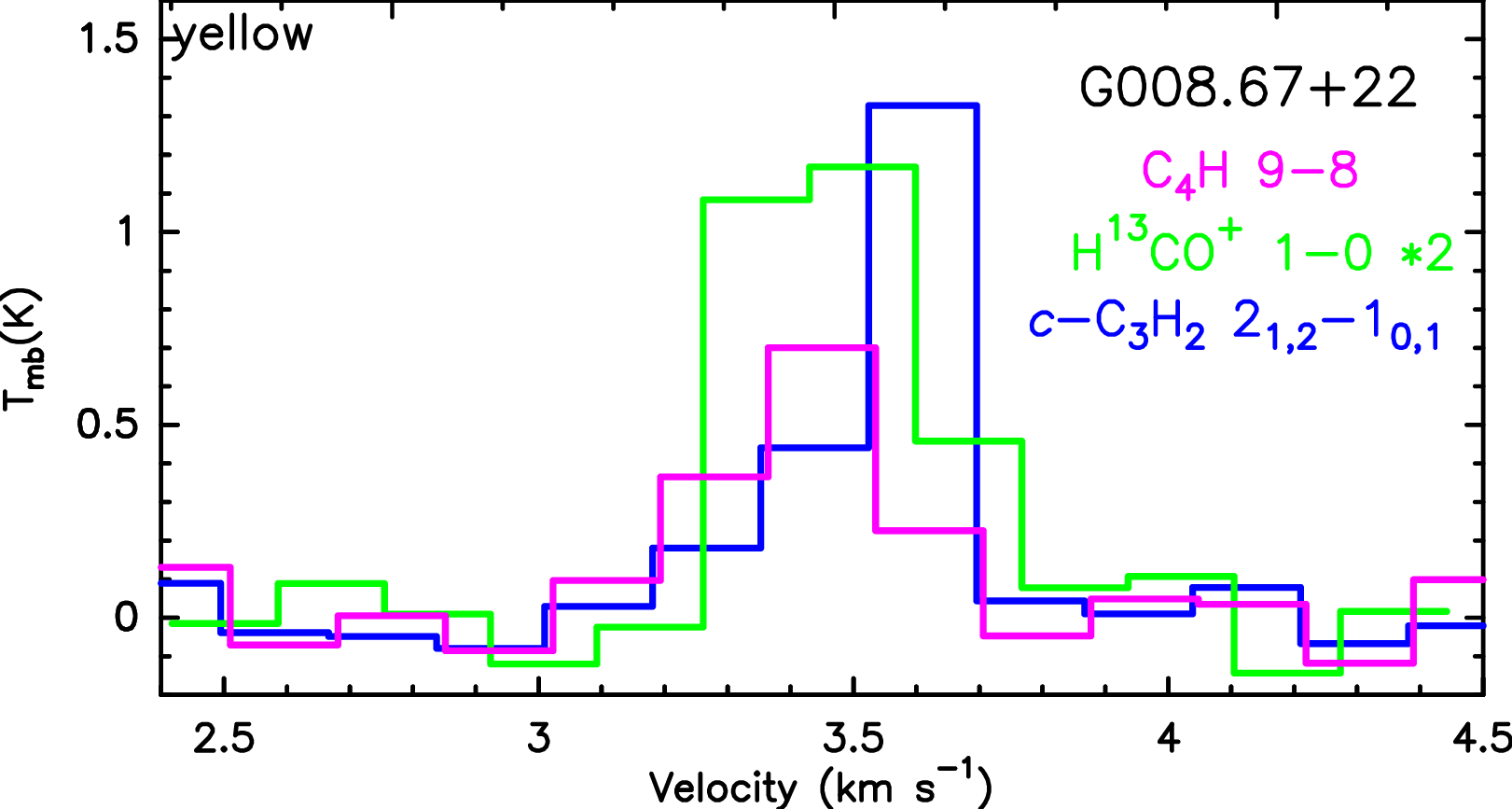}\includegraphics[width=0.38\columnwidth]{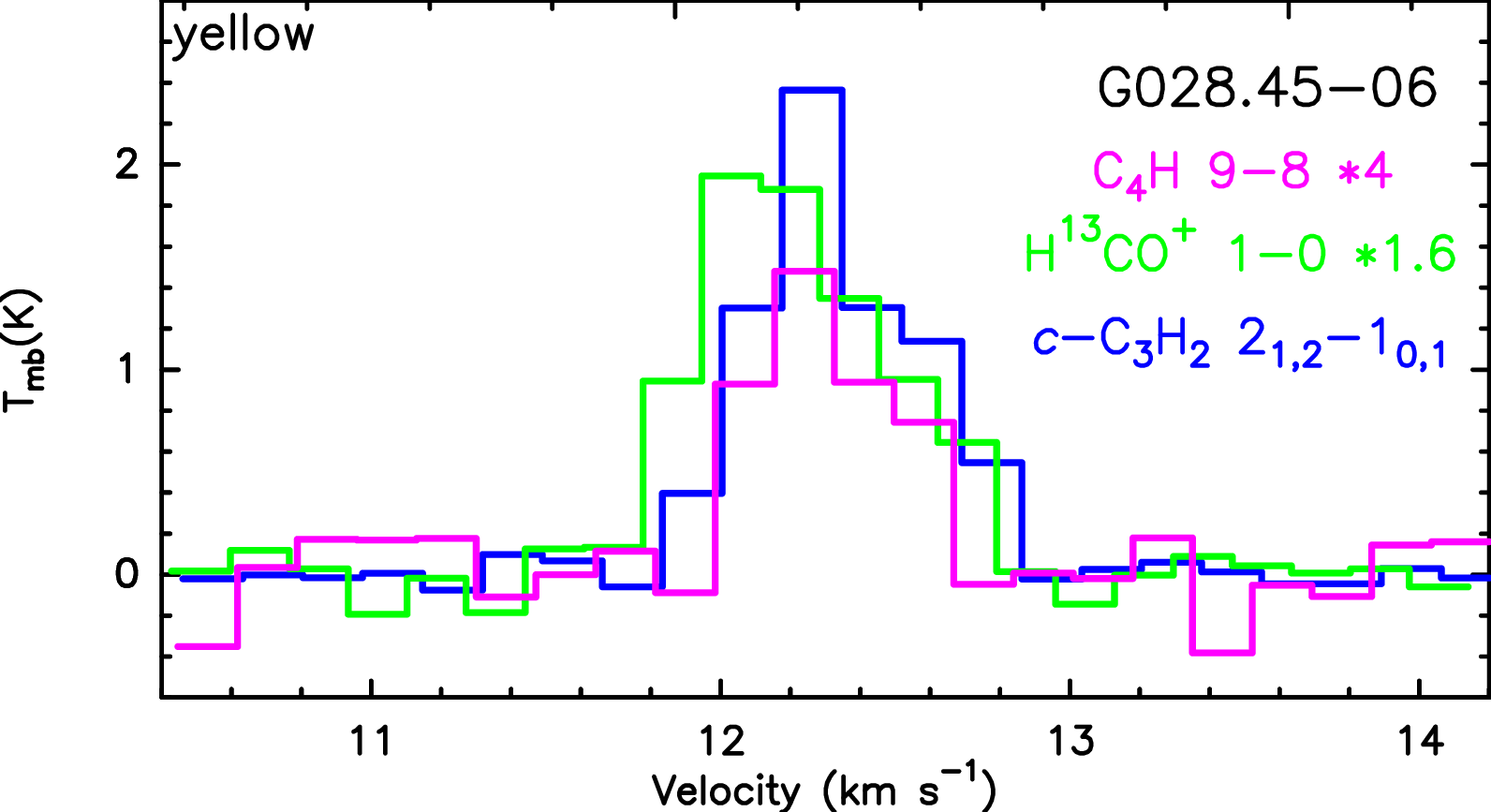}\\
 \includegraphics[width=0.38\columnwidth]{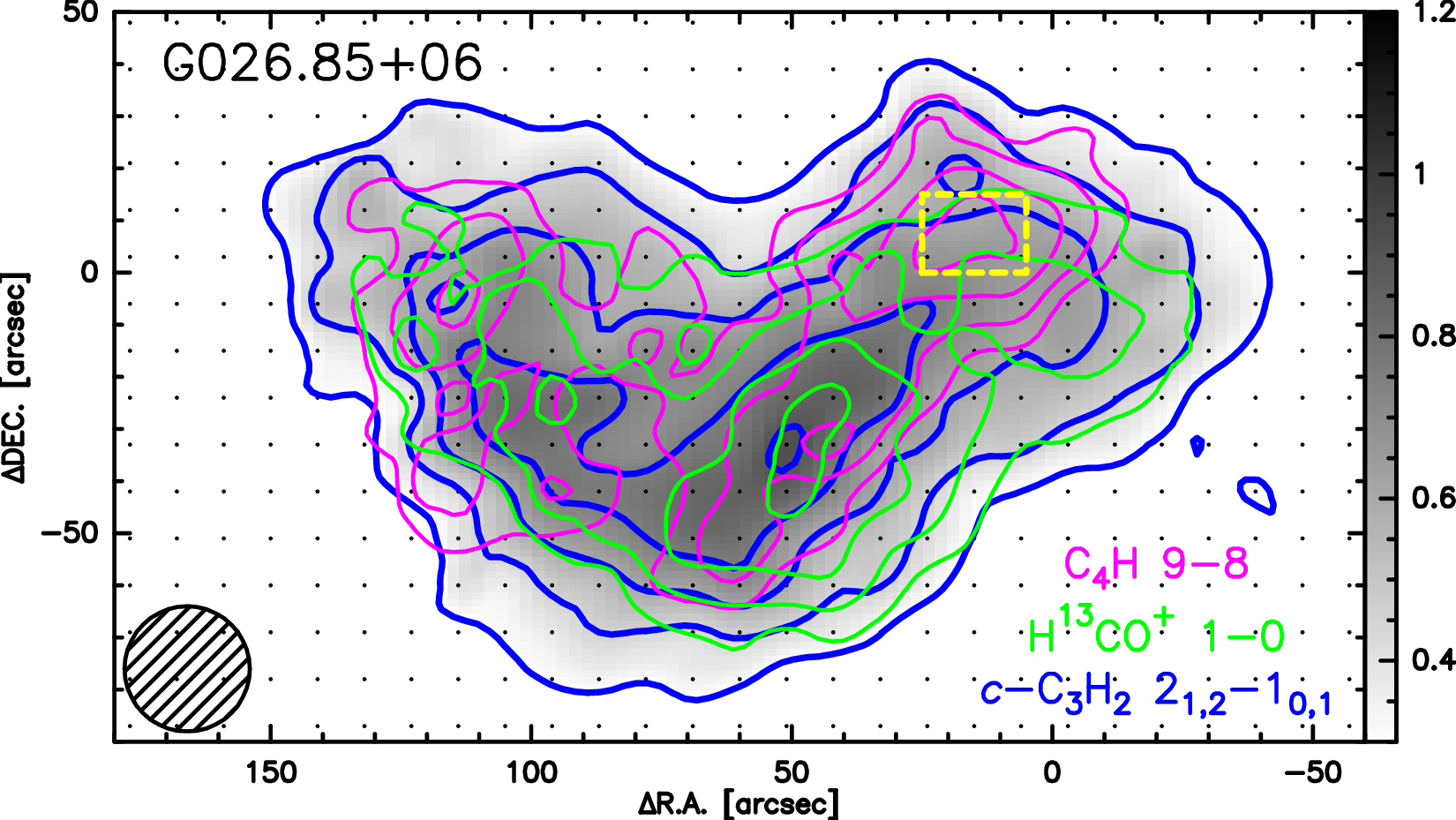}\includegraphics[width=0.38\columnwidth]{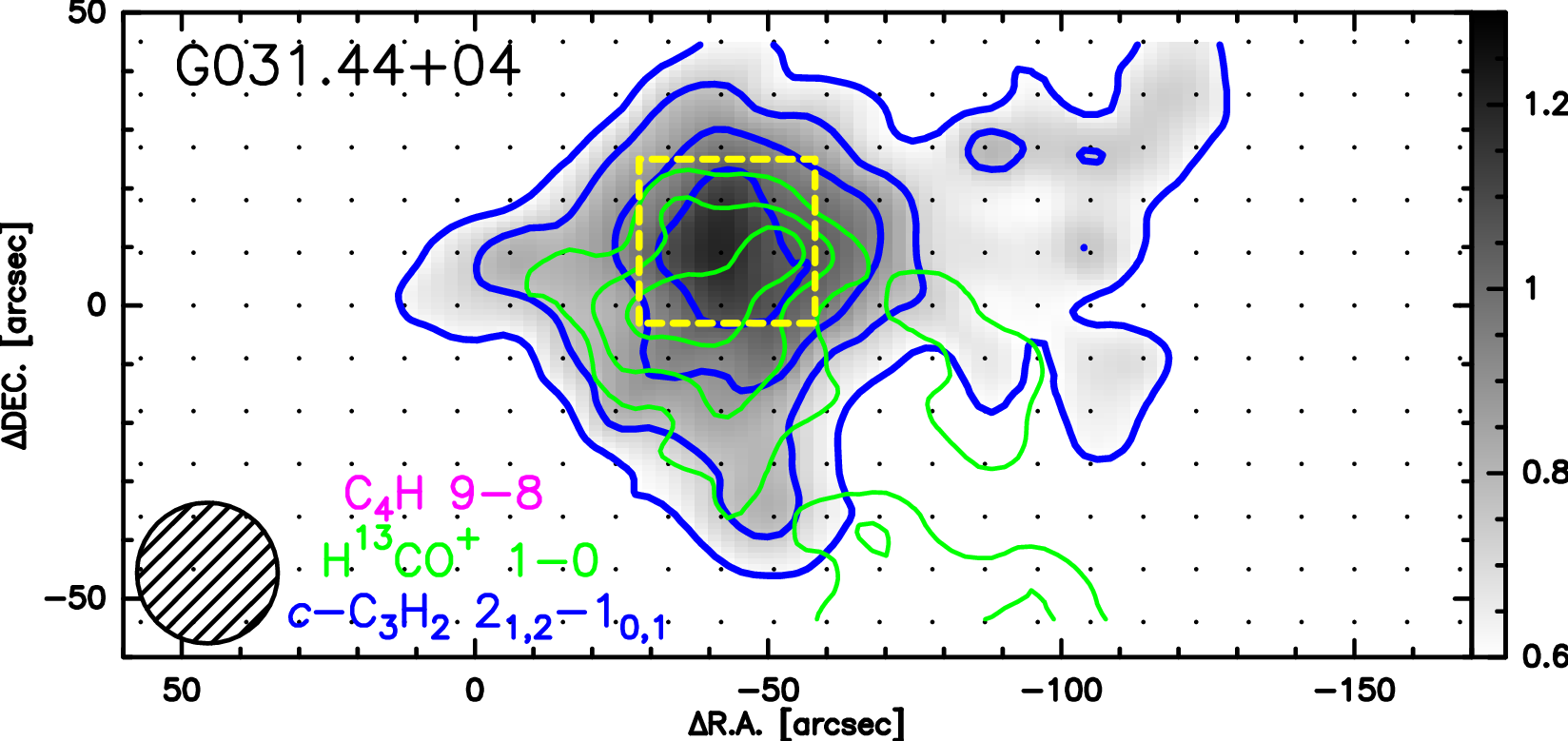}\\ 
\includegraphics[width=0.38\columnwidth]{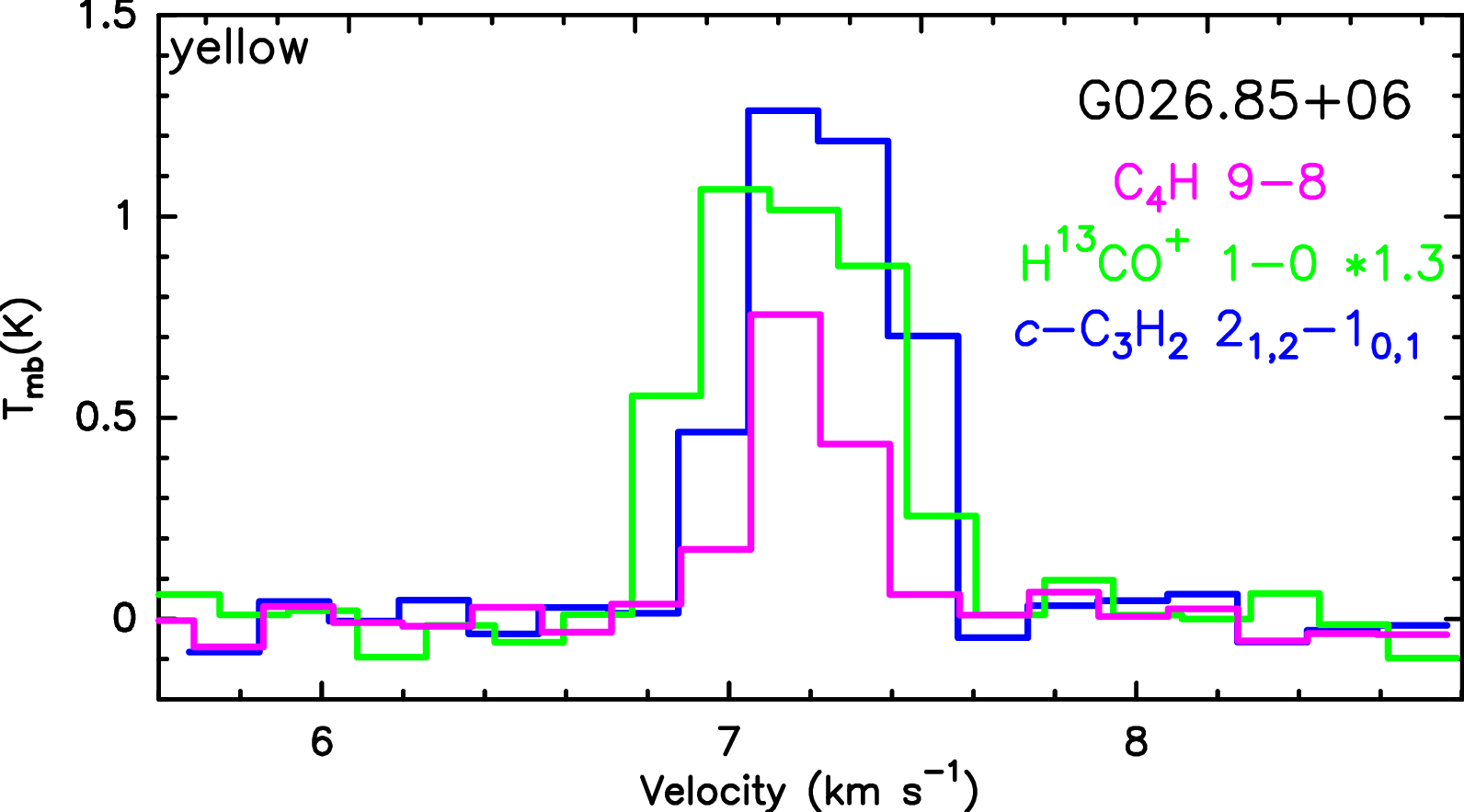}\includegraphics[width=0.38\columnwidth]{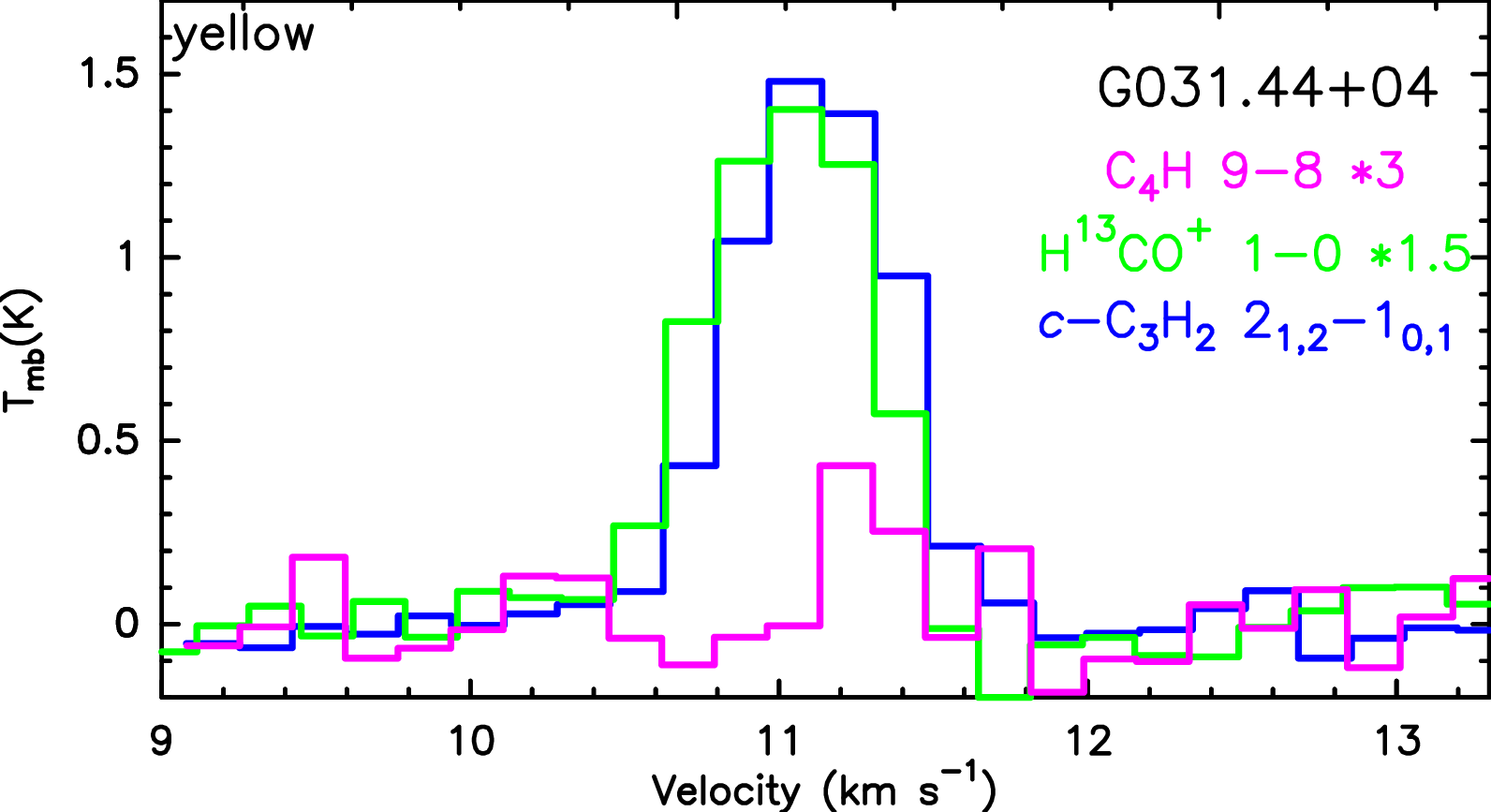} 
\caption{The same as Figure 1 for more sources.}
\label{appendix}
\end{figure*}

\clearpage
\addtocounter{figure}{-1}
\centering 

\begin{figure*}

\centering 
\includegraphics[width=0.4\columnwidth]{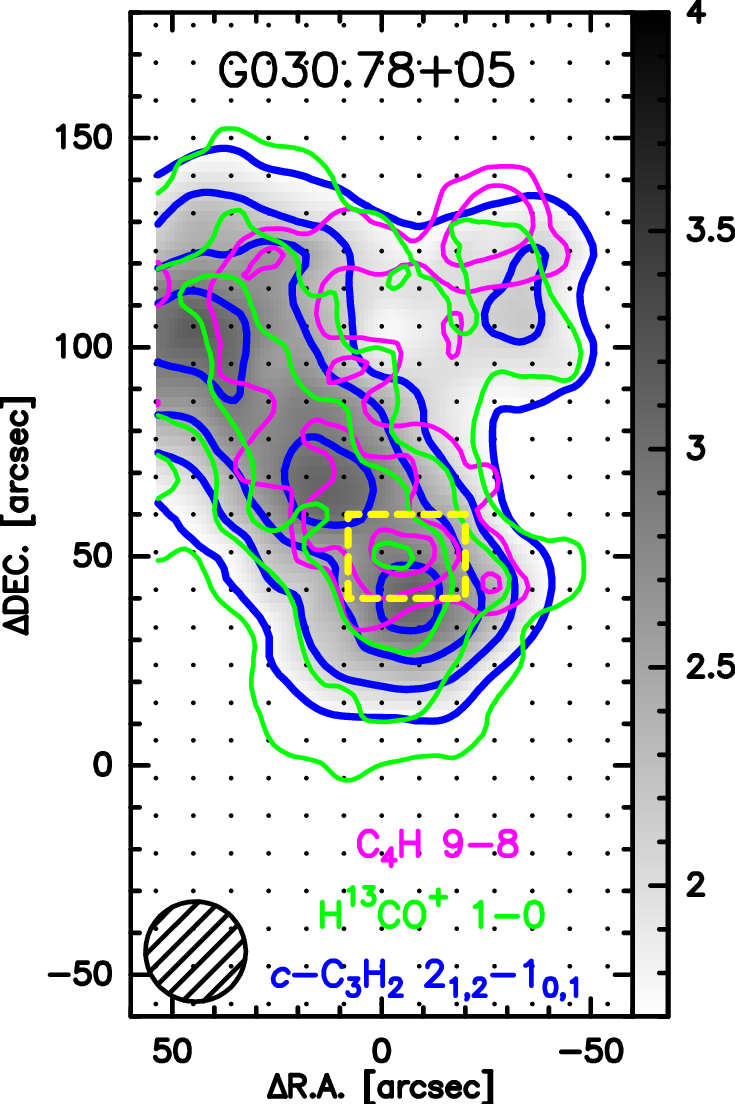} 
\includegraphics[width=0.4\columnwidth]{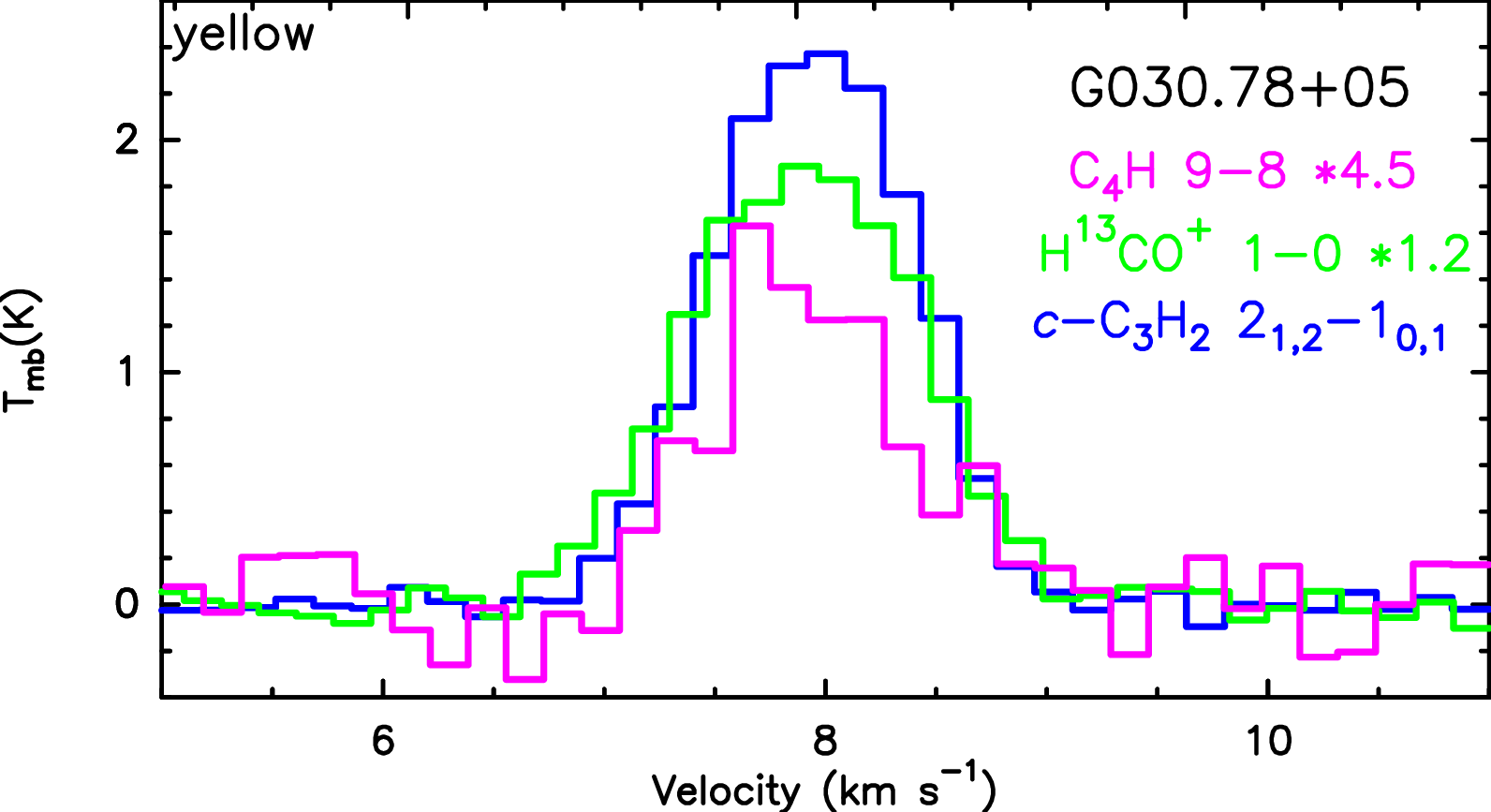} 

	



\centering 
\includegraphics[width=0.4\columnwidth]{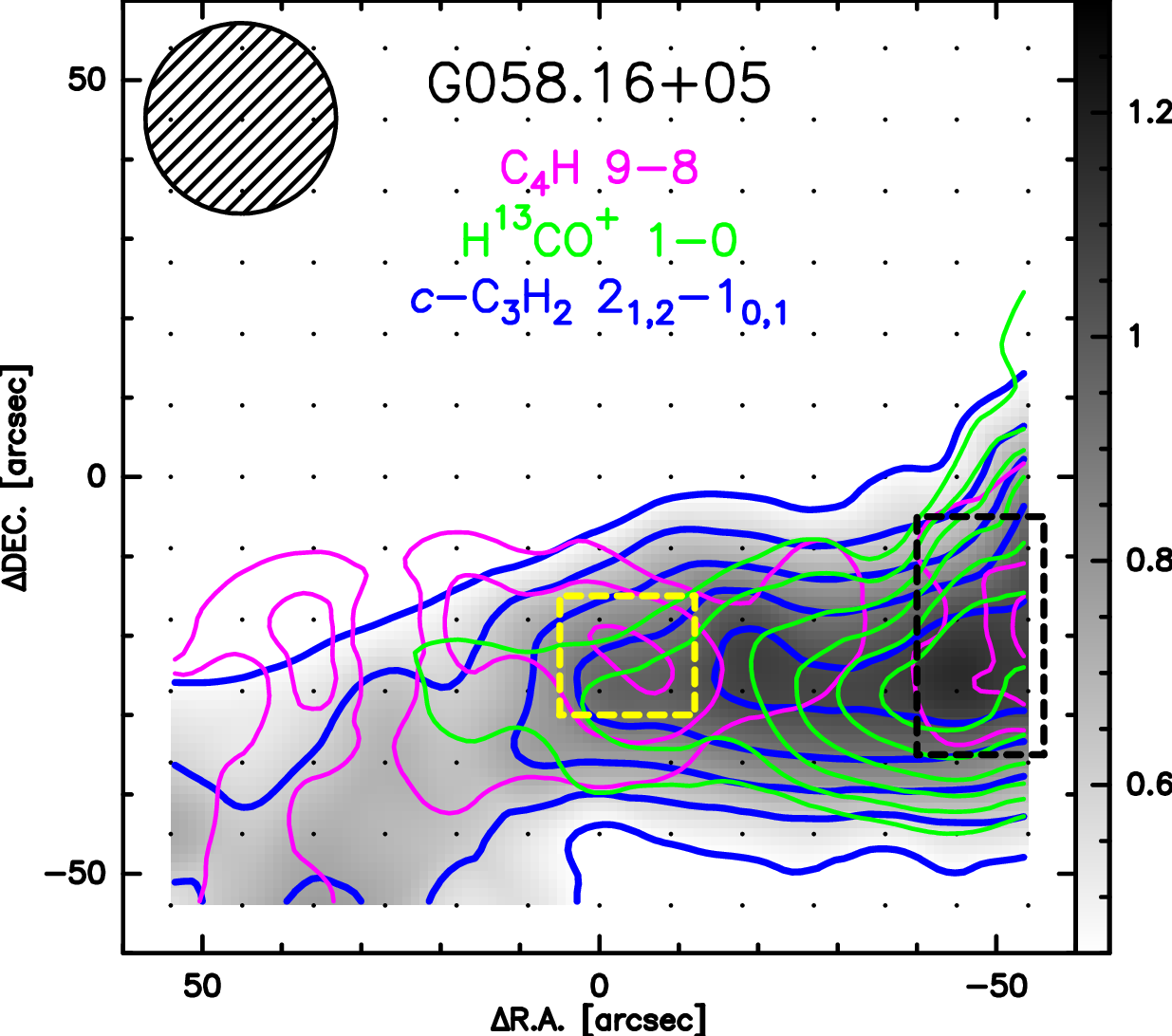} 

\centering 
\includegraphics[width=0.4\columnwidth]{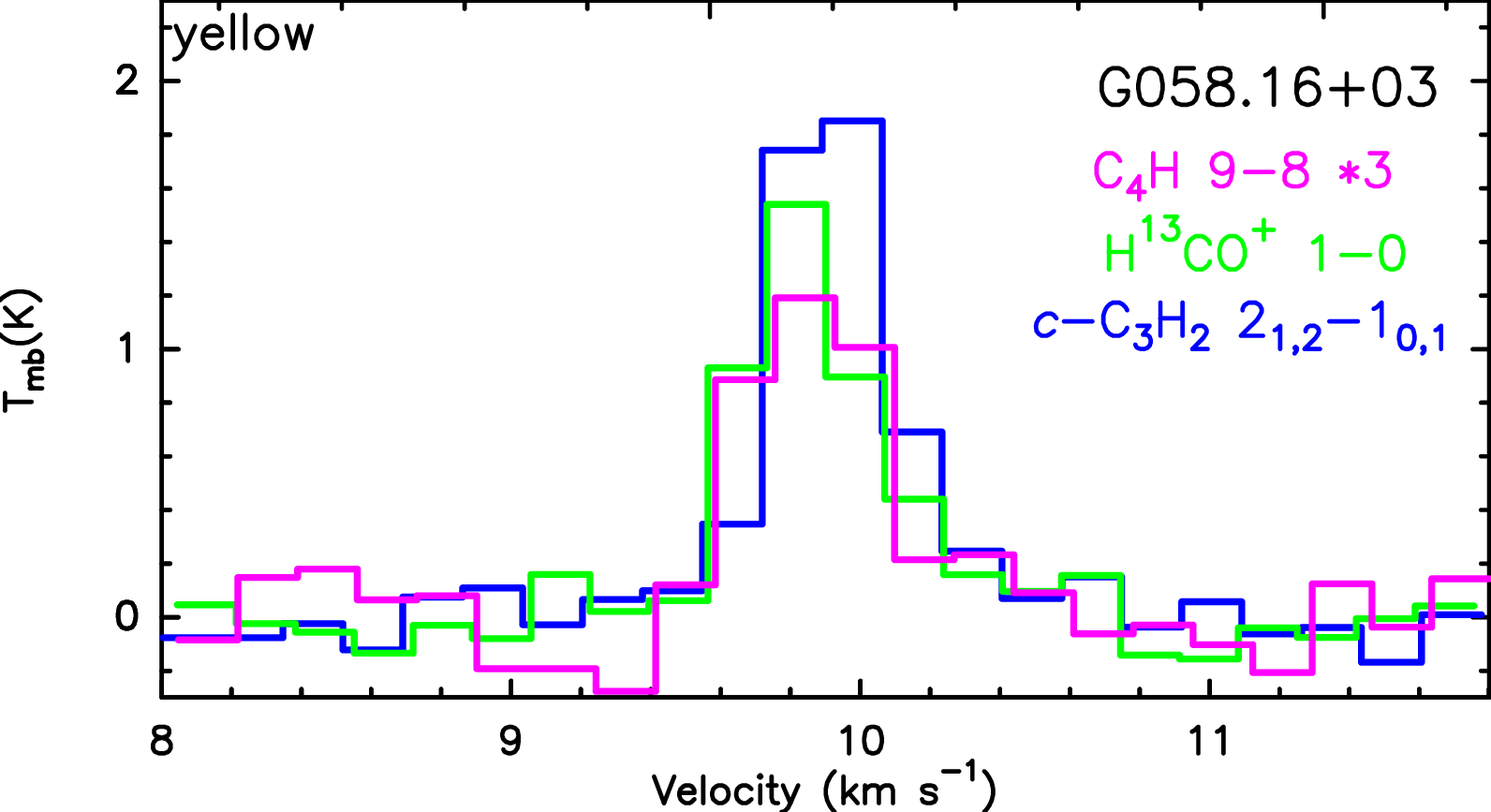} 
\includegraphics[width=0.4\columnwidth]{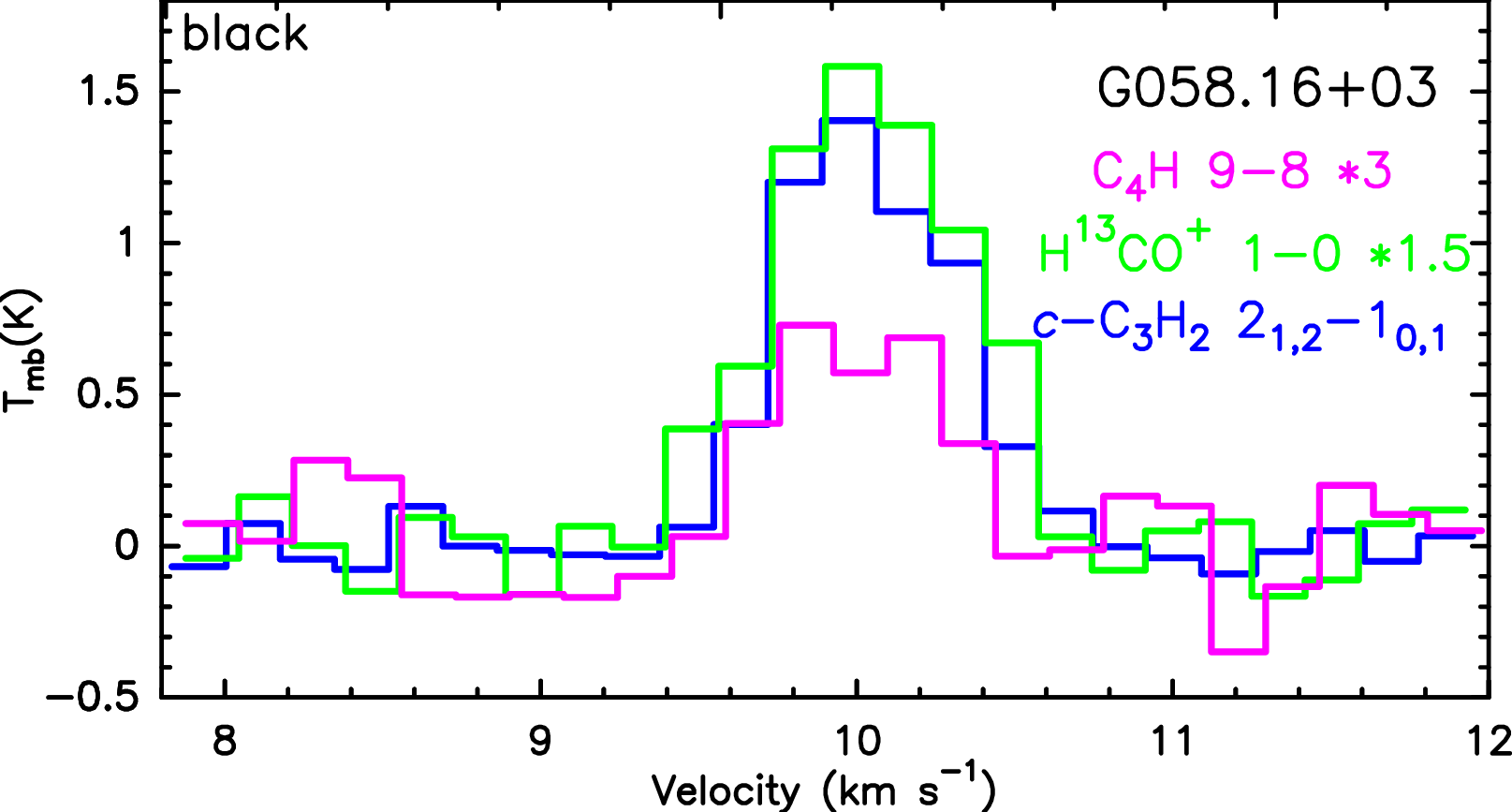} 

\centering 
\caption{Continued.}	
 \label{appendix}
\end{figure*}

\clearpage

\begin{figure*}
\addtocounter{figure}{-1}
\centering

\centering 
\includegraphics[width=0.4\columnwidth]{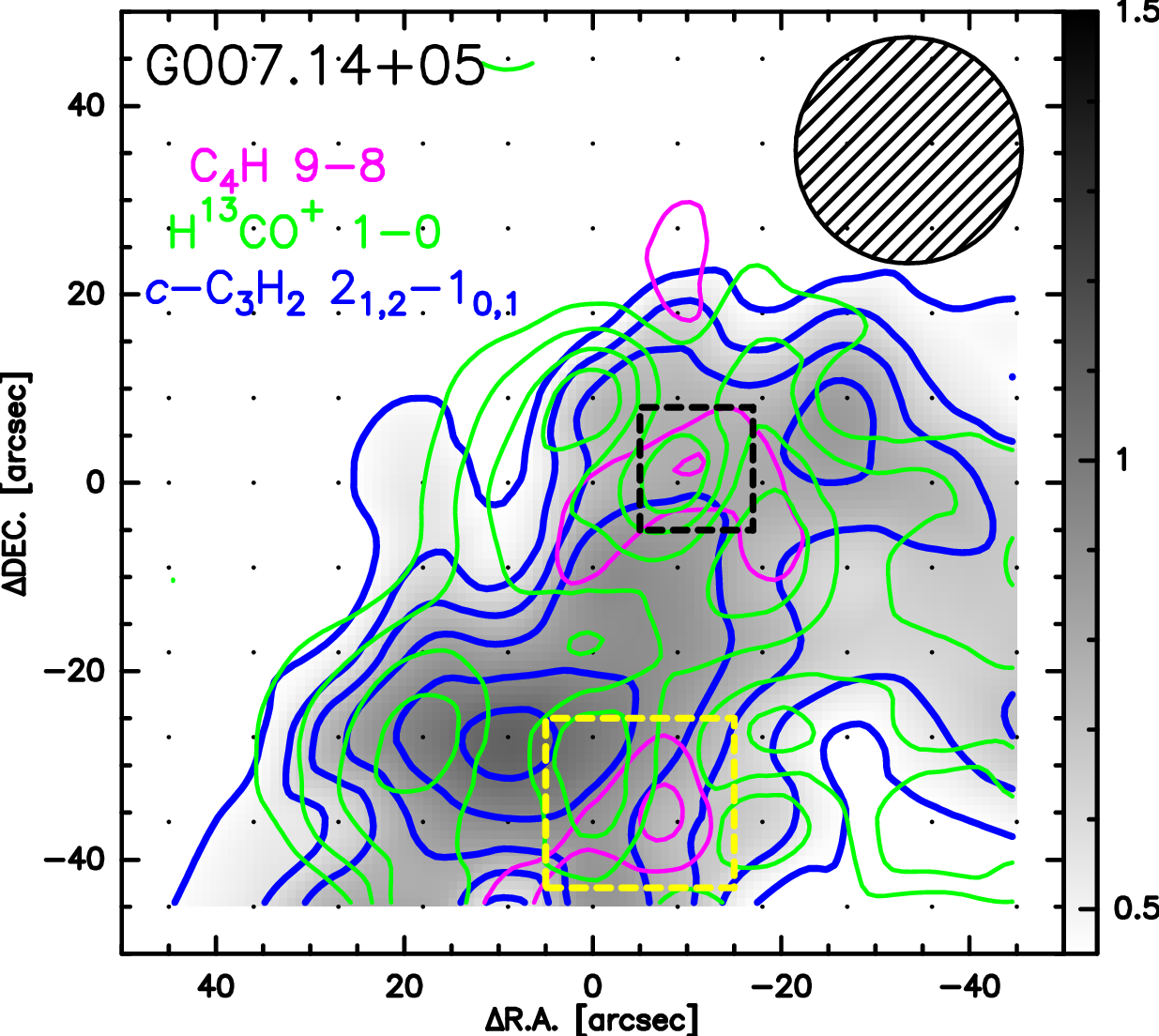} 

\centering 
\includegraphics[width=0.4\columnwidth]{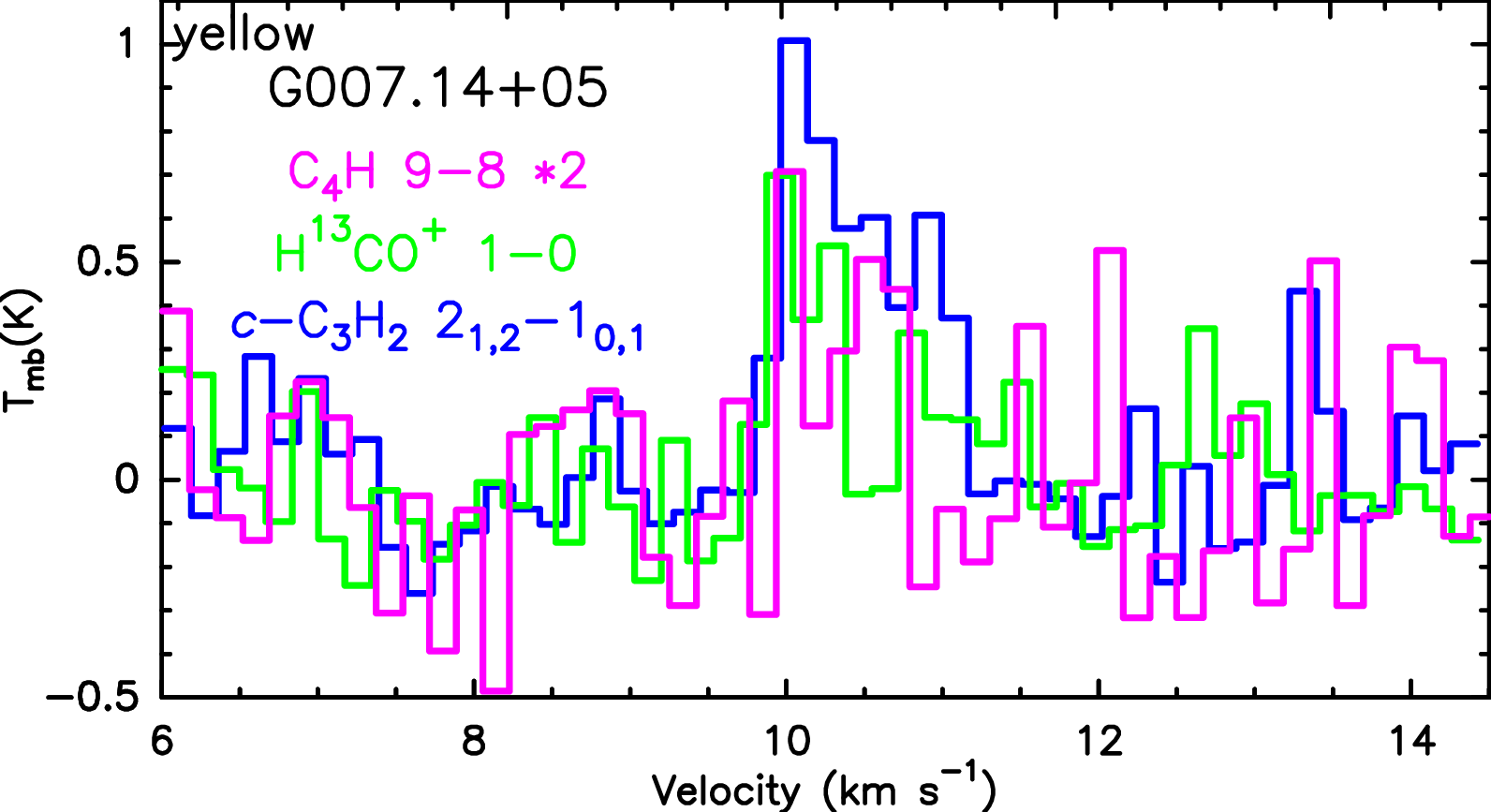} 
\includegraphics[width=0.4\columnwidth]{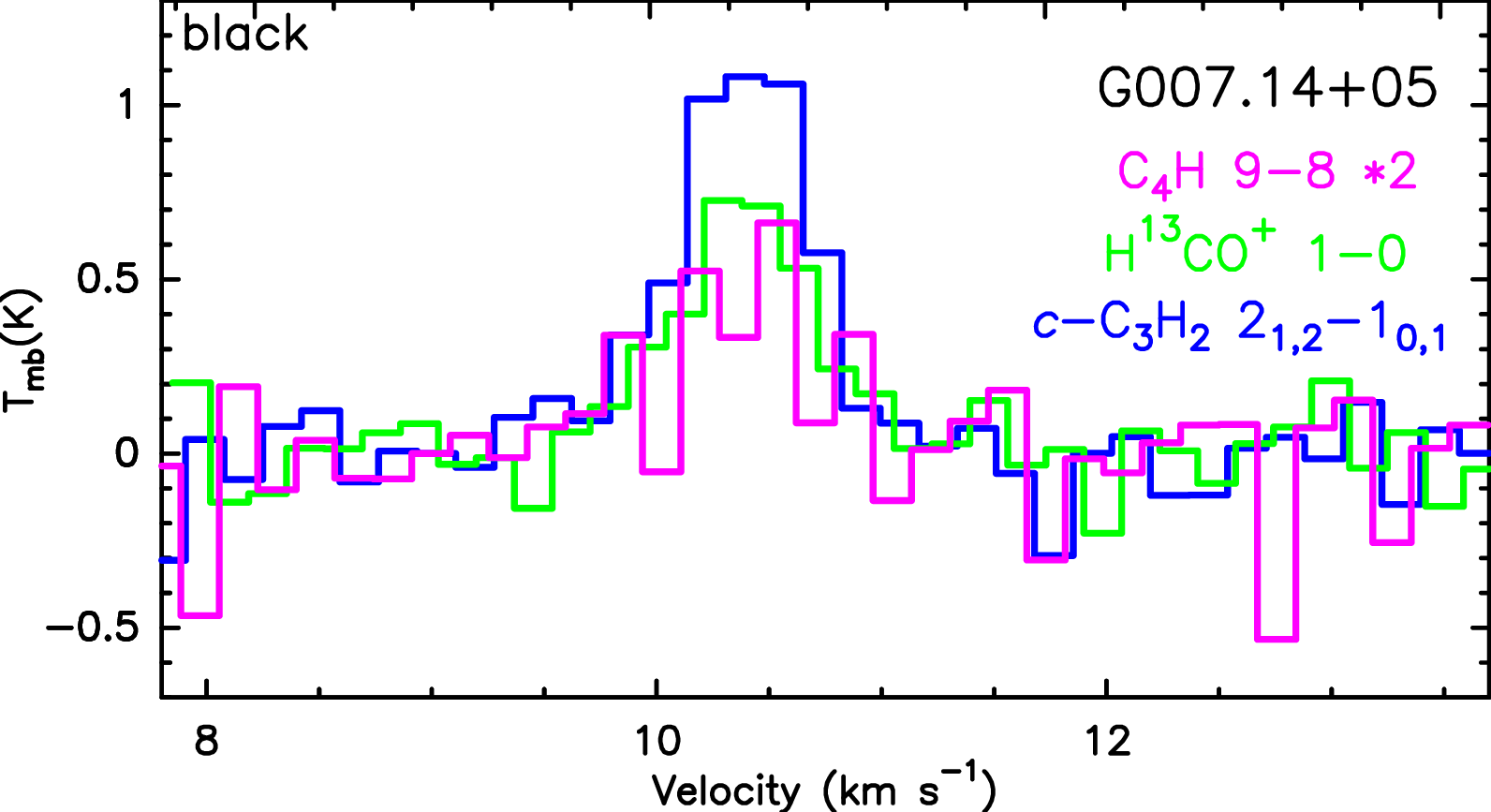} 

 
	



\centering 
\includegraphics[width=0.4\columnwidth]{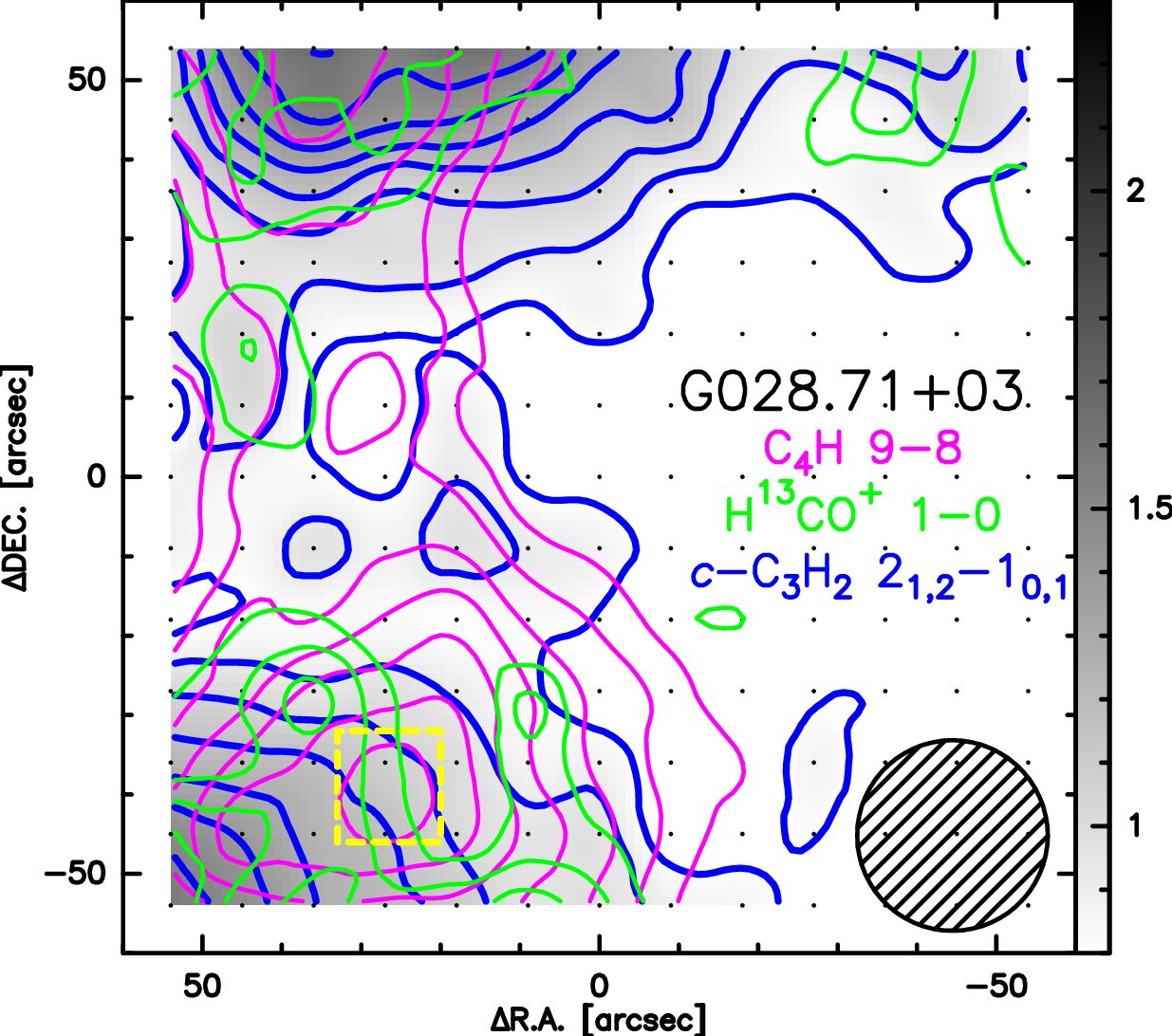} 
\includegraphics[width=0.4\columnwidth]{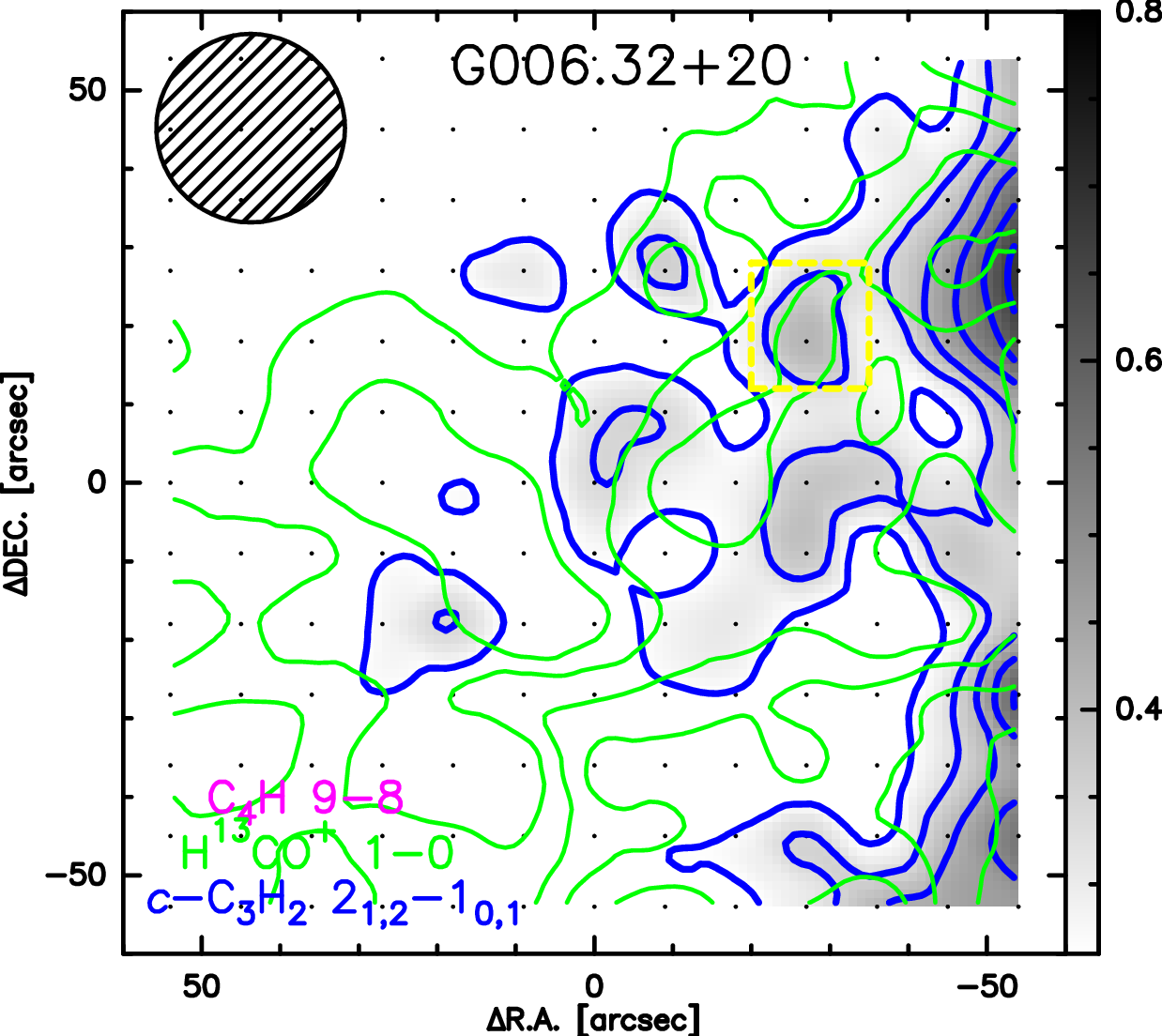} 
\centering 
\includegraphics[width=0.4\columnwidth]{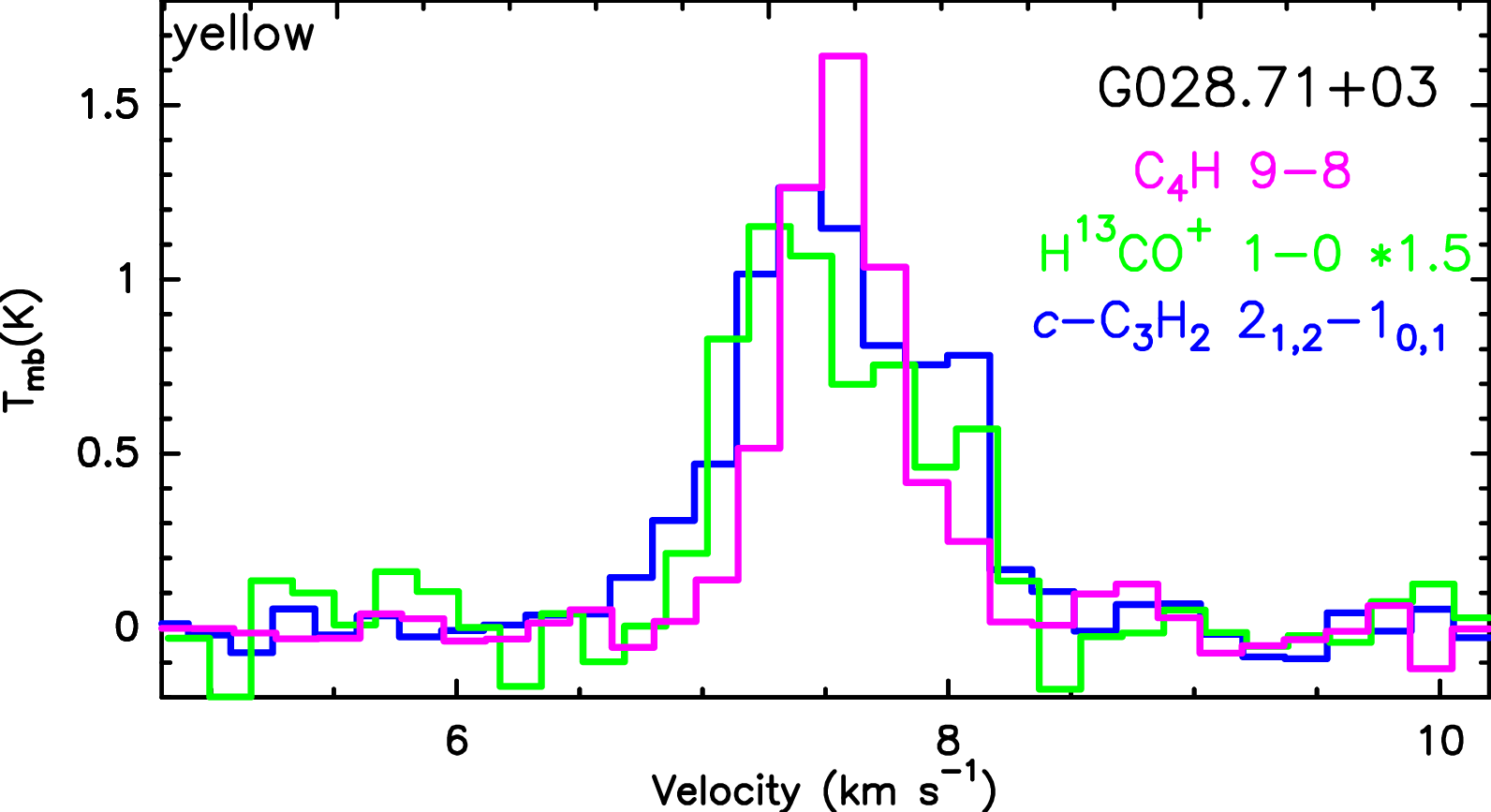} 
\includegraphics[width=0.4\columnwidth]{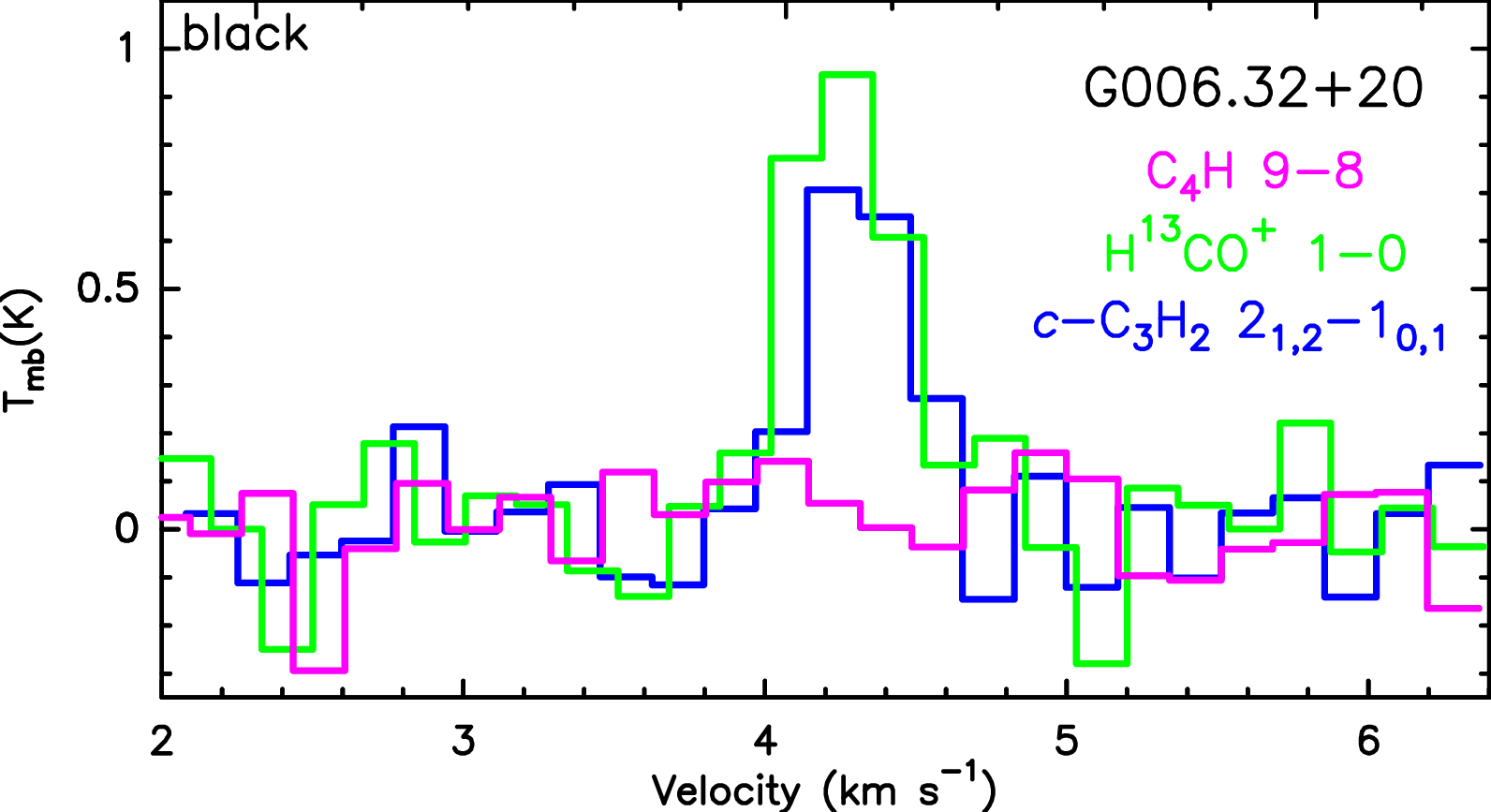} 

\centering 
\caption{Continued.}	
	
\label{appendix}
\end{figure*}

\clearpage
\addtocounter{figure}{-1}
\centering

\begin{figure*}

\centering 
\includegraphics[width=0.4\columnwidth]{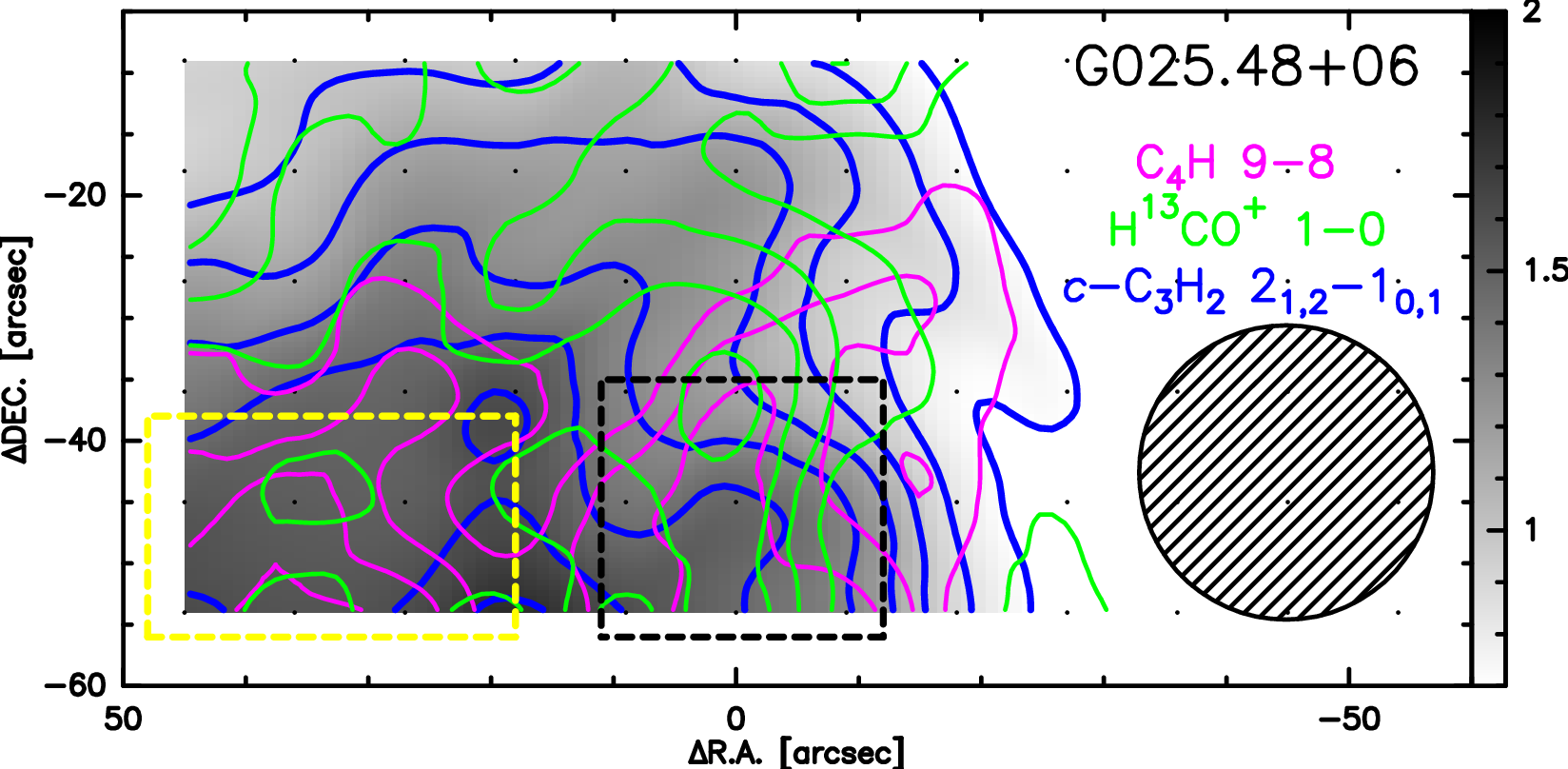} 

\centering 
\includegraphics[width=0.4\columnwidth]{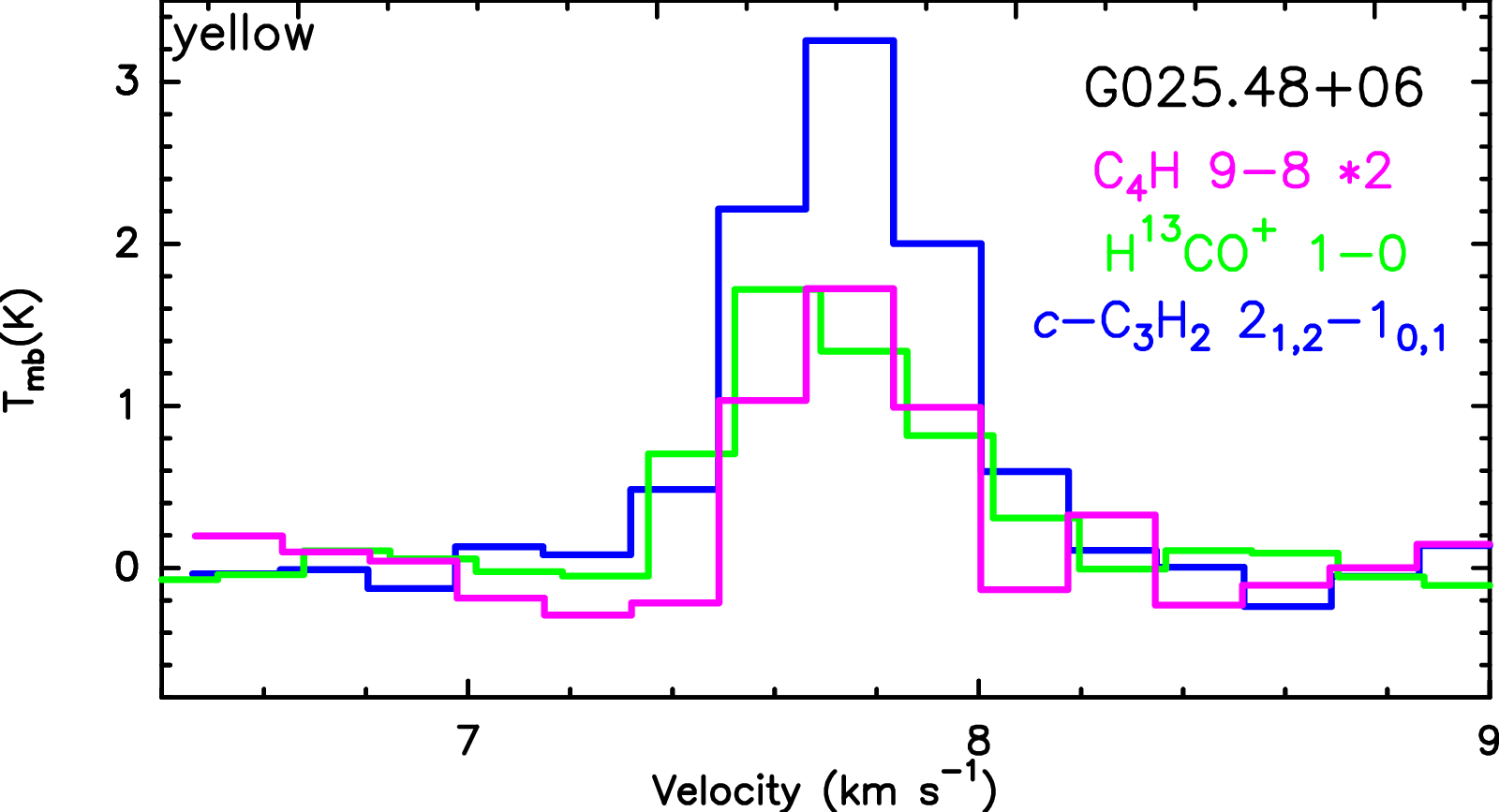} 
\includegraphics[width=0.4\columnwidth]{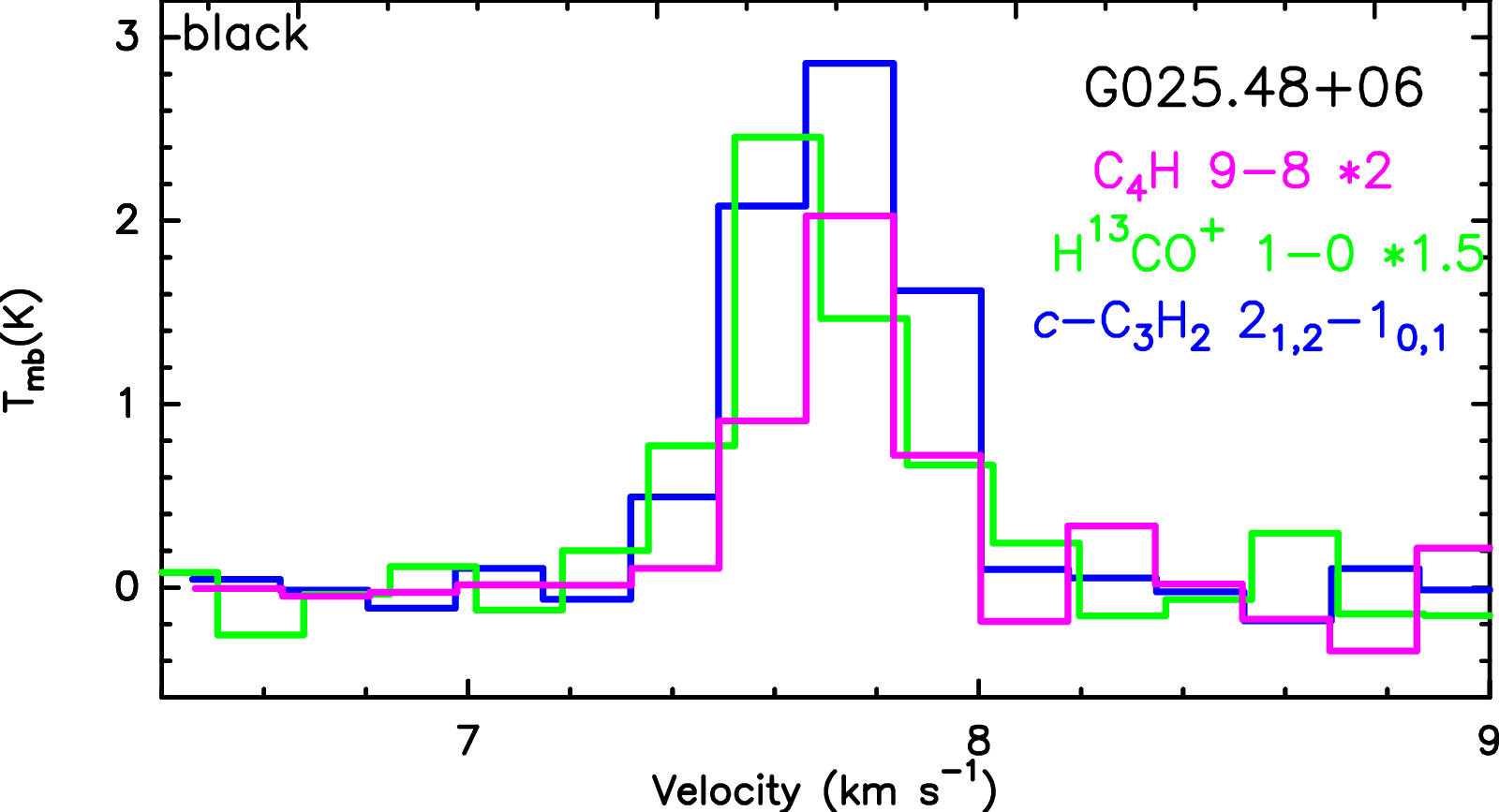} 

	



\centering 
\includegraphics[width=0.4\columnwidth]{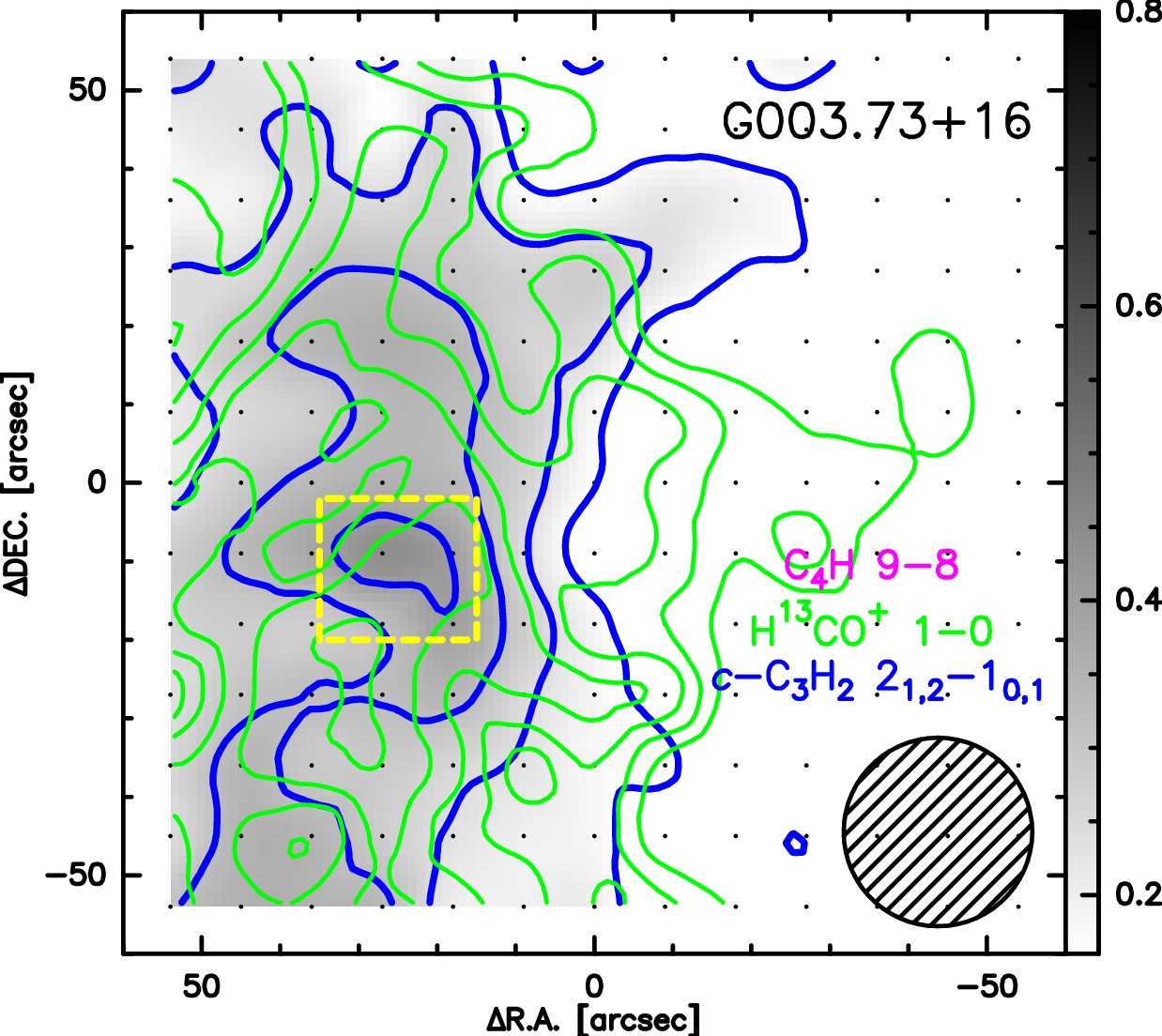} 
\includegraphics[width=0.4\columnwidth]{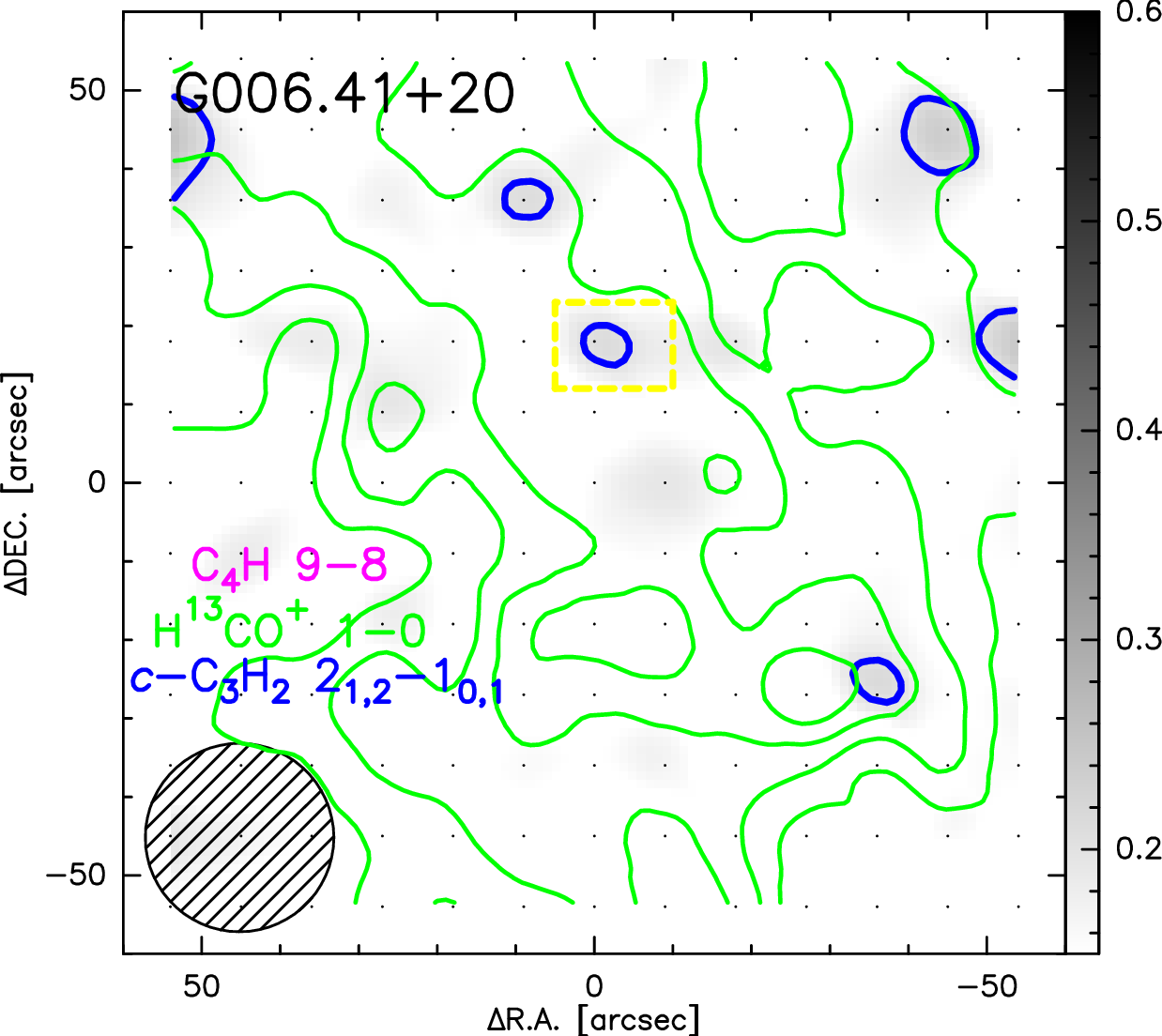} 
\centering 
\includegraphics[width=0.4\columnwidth]{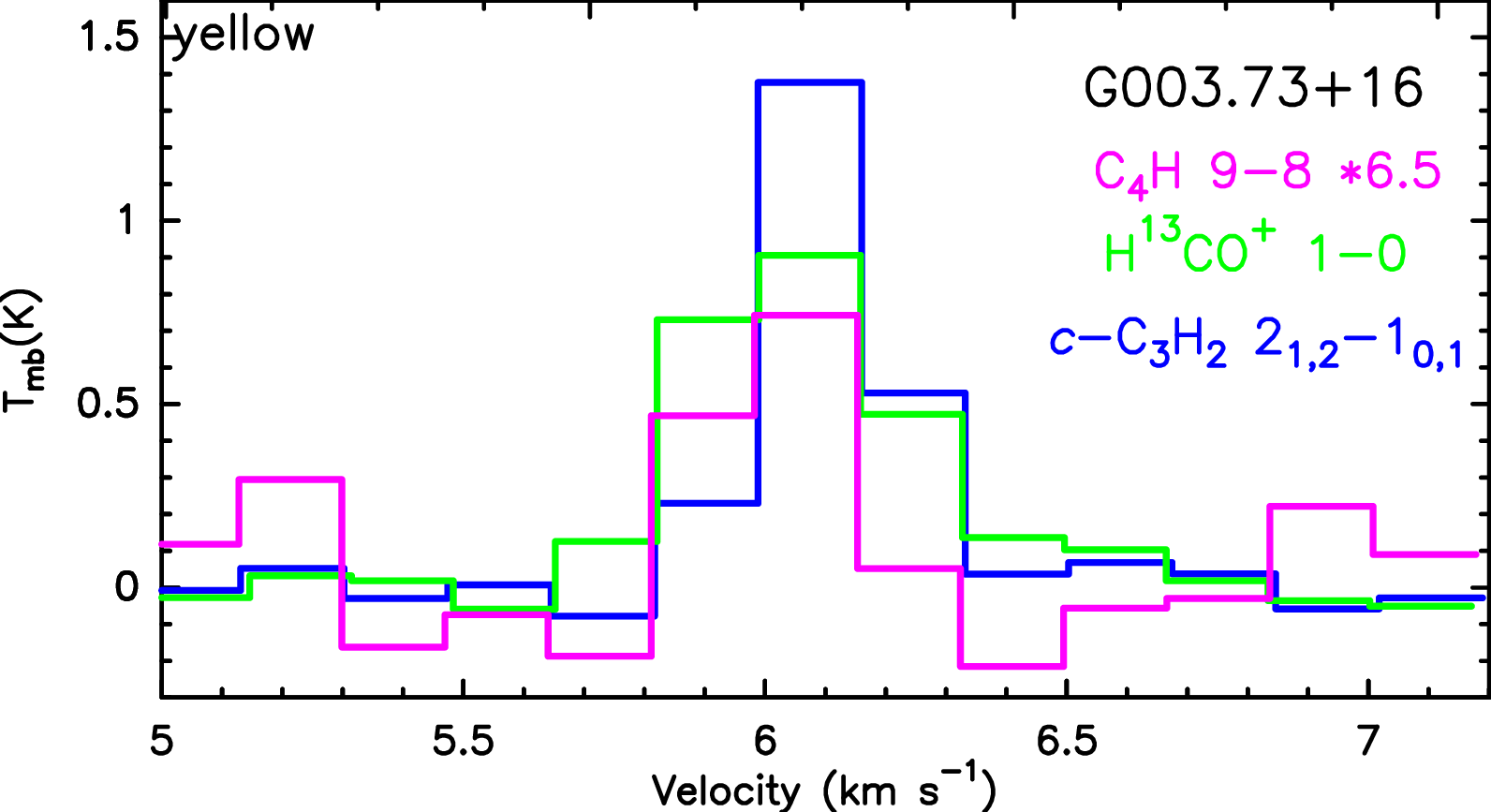} 
\includegraphics[width=0.4\columnwidth]{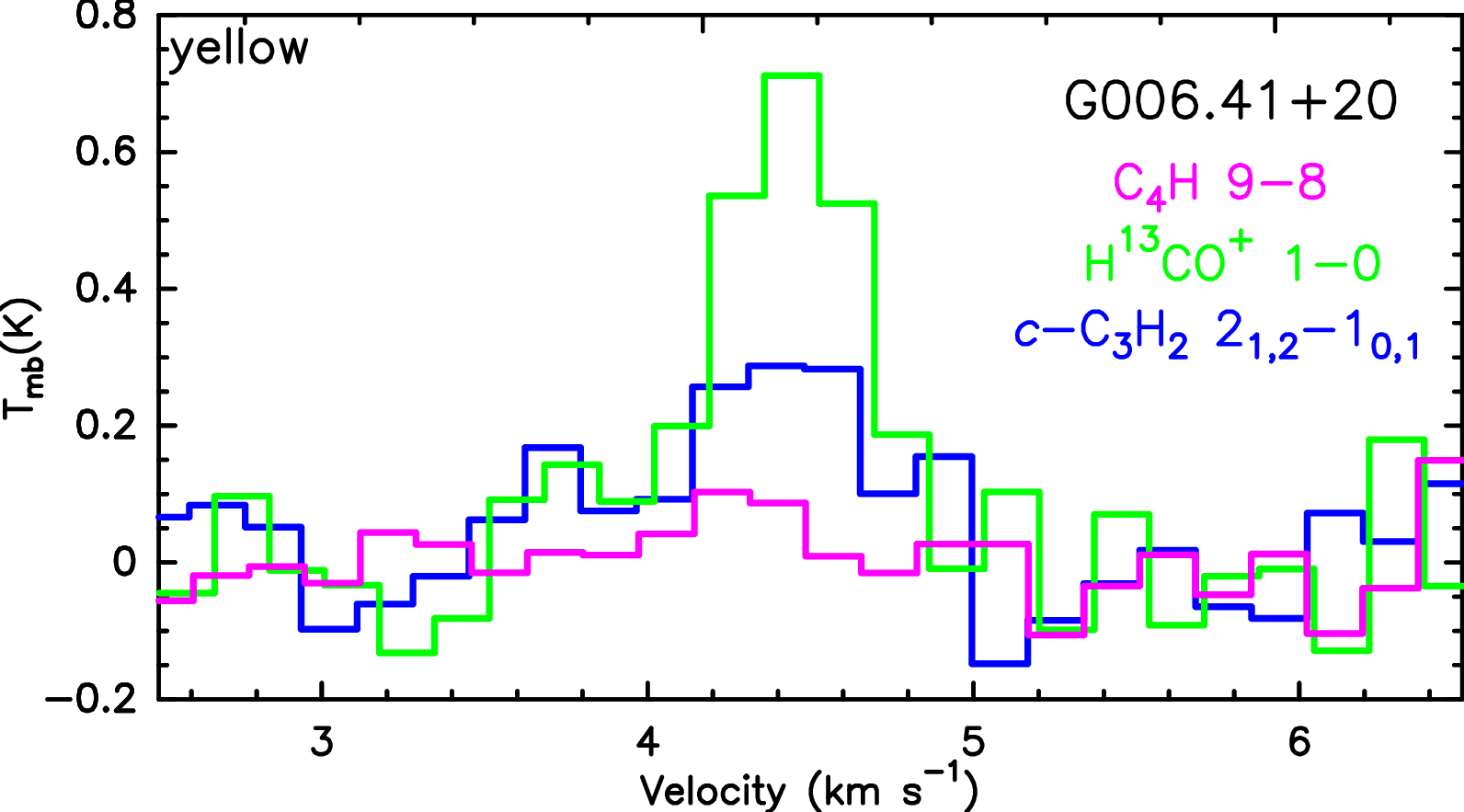} 

\centering 
\caption{Continued.}	
	
\label{appendix}
\end{figure*}

\clearpage

\begin{figure*}
\addtocounter{figure}{-1}
\centering

\centering 
\includegraphics[width=0.4\columnwidth]{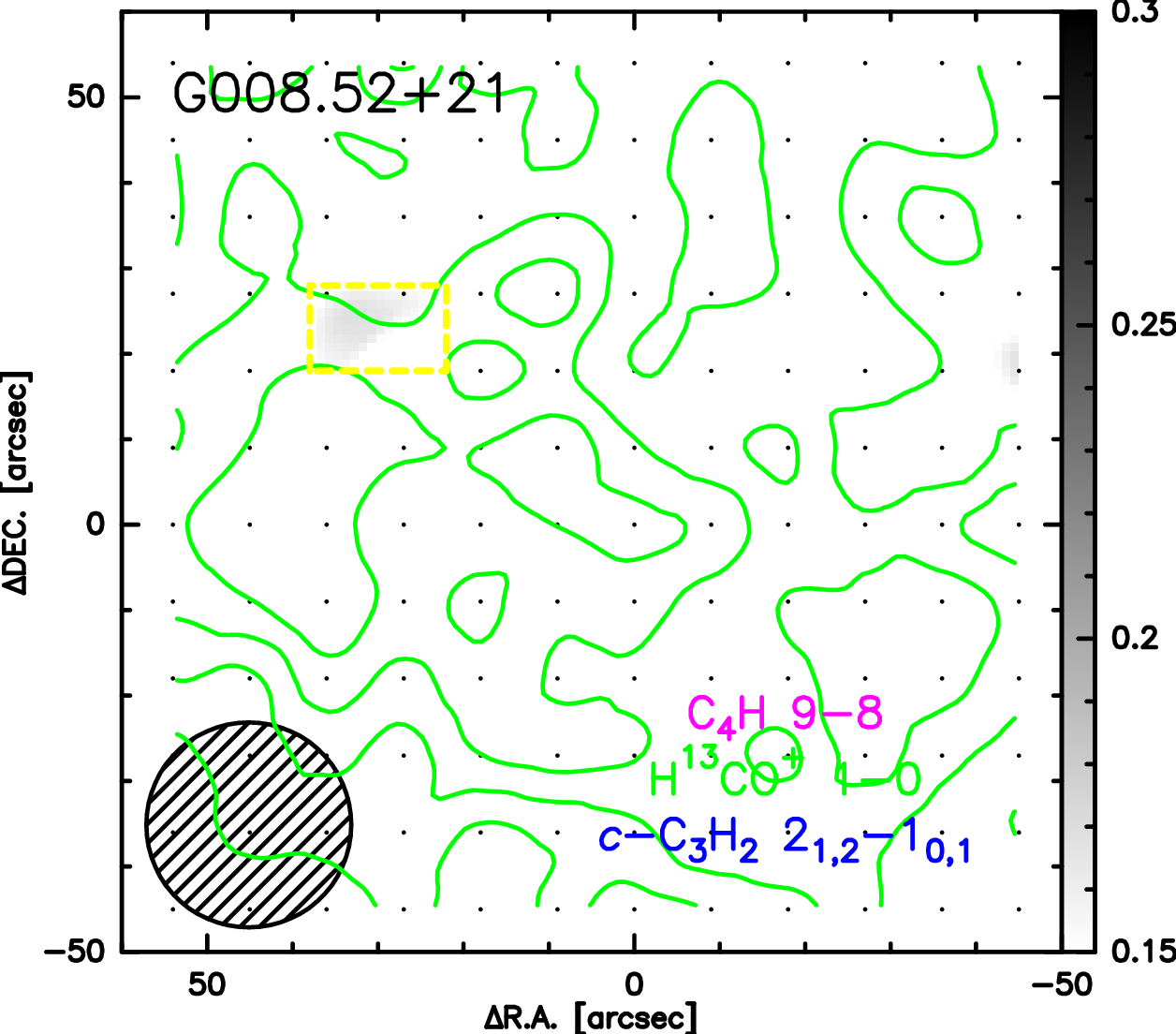} 
\centering 

\includegraphics[width=0.4\columnwidth]{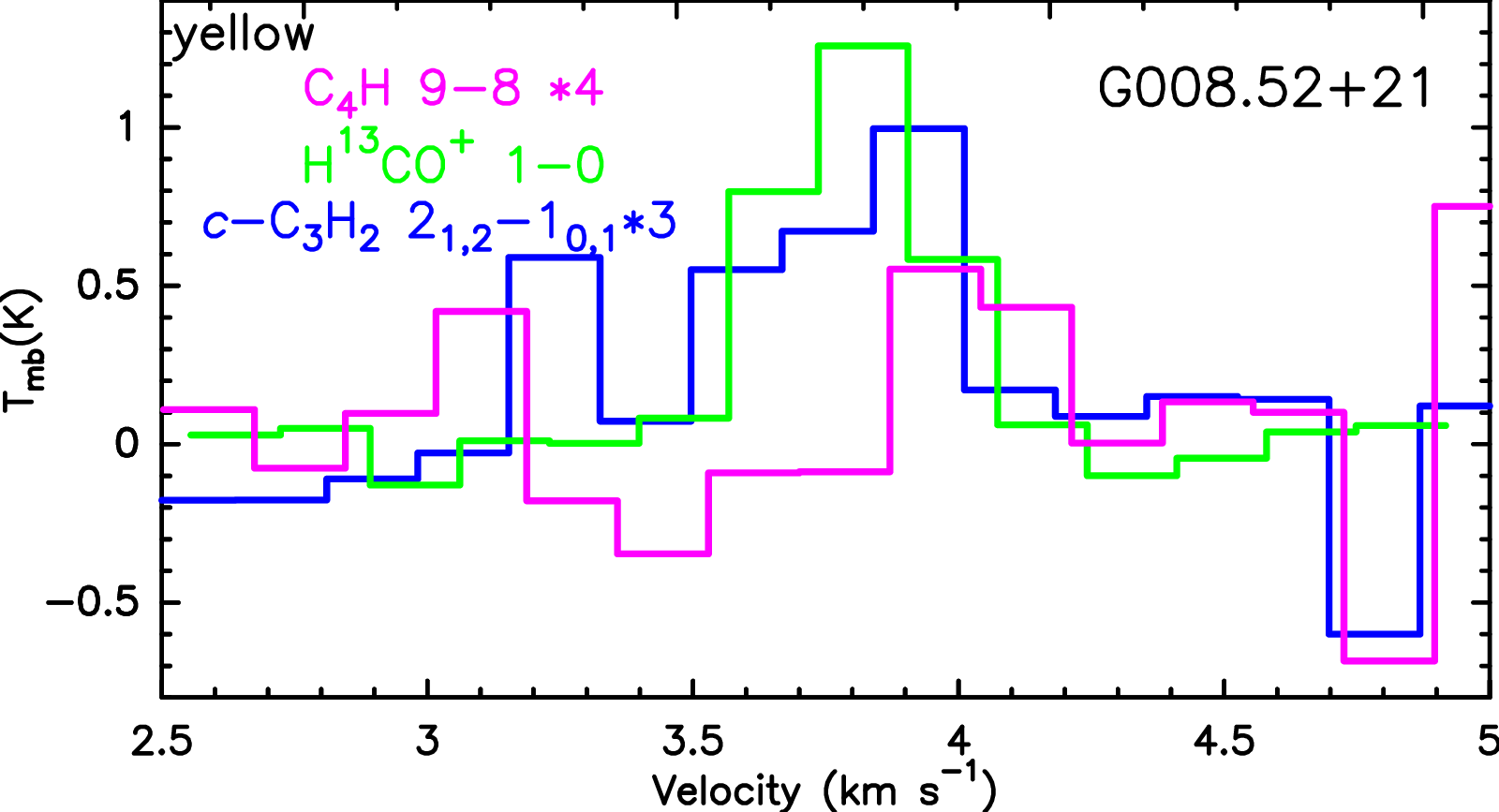} 

\centering 
\caption{Continued.}	
	
\label{appendix}
\end{figure*}

\end{document}